\documentclass[12pt]{amsart}

\usepackage{amssymb}
\usepackage{amsmath}
\usepackage{amsfonts}
\usepackage{mathrsfs}
\usepackage{graphicx}
\usepackage{caption}
\usepackage{subcaption}
\usepackage{color}
\usepackage[onehalfspacing]{setspace}
\usepackage[colorlinks=true,citecolor=blue,urlcolor=blue,pdfpagemode=UseNone,pdfstartview=FitH]{hyperref}
\usepackage{natbib}
\usepackage{enumerate}
\usepackage{dcolumn}
\usepackage{float}
\usepackage[charter,cal=cmcal]{mathdesign}

\usepackage{xargs}   
\usepackage{multirow}

\makeatletter
\def\section{\@startsection{section}{1}
	\z@{0.8\linespacing}{.6\linespacing}{\Large\centering}}

\def\subsection{\@startsection{subsection}{2}
	\z@{.6\linespacing}{.5\linespacing}{\large}}

\def\subsubsection{\@startsection{subsubsection}{3}
	\z@{.4\linespacing}{.3\linespacing}{\normalfont\bfseries}}
\makeatother

\numberwithin{equation}{section}

\newtheorem{theorem}{Theorem}[section]

\newtheorem{corollary}{Corollary}[section]

\theoremstyle{definition}

\theoremstyle{definition}
\newtheorem{assumption}{Assumption}[section]

\theoremstyle{definition}
\newtheorem{condition}{Condition}

\theoremstyle{definition}

\DeclareTextFontCommand{\bi}{%
	\fontseries\bfdefault % change series without selecting the font yet
	\itshape
}

	\title{}
	
\begin{document}
		\vspace*{5ex minus 1ex}
		\begin{center}
			\Large \textsc{Estimation and Inference on Treatment Effects  Under Treatment-Based Sampling Designs}
			\bigskip
		\end{center}
		
		\date{%
			%TCIMACRO{\TeXButton{Today}{\today}}%
			%BeginExpansion
			\today%
			%EndExpansion
		}

		\vspace*{3ex minus 1ex}
		\begin{center}
			Kyungchul Song and Zhengfei Yu\\
			\textit{University of British Columbia and University of Tsukuba}\\
			\medskip
		\end{center}
		
		\begin{abstract}
			{\footnotesize
				Causal inference in a program evaluation setting faces the problem of external validity when the treatment effect in the target population is different from the treatment effect identified from the population of which the sample is representative. This paper focuses on a situation where such discrepancy arises by a stratified sampling design based on the individual treatment status and other characteristics. In such settings, the design probability is known from the sampling design but the target population depends on the underlying population share vector which is often unknown, and except for special cases, the treatment effect parameters are not identified. In this paper, we propose a method of constructing confidence sets that are valid for a given range of population shares. When a benchmark population share vector and a corresponding estimator of a treatment effect parameter are given, we develop a method to discover the scope of external validity with familywise error rate control. Finally, we derive an optimal sampling design which minimizes the semiparametric efficiency bound given a population share associated with a target population. We provide Monte Carlo simulation results and an empirical application to demonstrate the usefulness of our proposals.
			}\bigskip\
			
			{\footnotesize \noindent \textsc{Key words.} Treatment-Based Sampling;
				Standard Stratified Sampling; Set-Identification; External Validity; Choice-Based Sampling;
				Semiparametric Efficiency; Average Treatment Effects; Optimal Sampling
				Designs; Partial Identification \bigskip\ }
			
		\end{abstract}
		
		\thanks{We thank Petra Todd who gave numerous valuable comments and advice at an early stage of this research, and Sokbae Lee for useful
comments, and Jinyong Hahn for pointing out errors in a manuscript that preceded this paper. A previous version of this paper was titled "Treatment Effects for Which Population?: Sampling Design and External Validity". All errors are ours. Song acknowledges the support from Social Sciences and Humanities Research Council of Canada. Yu acknowledges the support of JSPS KAKENHI Grant Number 19K13666.  Corresponding Address: Kyungchul Song, Vancouver School of Economics, University of British Columbia, 6000 Iona Drive, Vancouver, V6T 1L4, Canada.}
		\maketitle
		
		 \section{Introduction}
		In program evaluations, the estimated effect of a program is generally intended to provide information about an actual program's effect on a wider target population. However, if the sample in the study does not properly ``represents'' the target population, such estimates have limited use. This paper focuses on a particular source of such a problem, where the sampling process oversamples or undersamples from segments of the target population, but the shares of those segments in population are not precisely known. Indeed, as noted by \cite{Heckman/Todd:09:EJ}, the population shares are typically not available in the data set, which means that there is ambiguity about the population that the estimated effect should be targeted on.\footnote{As for the use of non-random sampling in the economics literature in program evaluations, for example, \cite{Ashenfelter/Card:85:ReStat} analyzed data from the Comprehensive Employment and Training Act (CETA) training program using a sample constructed by combining subsamples of program participants and a sample of nonparticipants drawn from the Current Population Survey (CPS). Also, the studies of \cite{Lalonde:86:AER}, \cite{Dehejia/Wahba:99:JASA, Dehejia:02:ReStat}
			and \cite{Smith/Todd:05:JOE} investigated the National Supported Work (NSW) training program where the training group consisted of individuals eligible for the program and the comparison sample were drawn from the CPS and the Panel Study of Income Dynamics (PSID) surveys. Numerous studies focused on the Job Training Partnership Act (JTPA) training program (e.g. \cite{Heckman/Ichimura/Smith/Todd:97:Eca}, \cite{Heckman/Ichimura/Todd:97:ReStud}). The participants in these data sets typically represented about 50\% in the study sample in comparison to around 3\% in the population. The eligible people in the target population often consist of drug addicts, ex-convicts, and welfare recipients, etc., and the researcher has little prior knowledge on the population share of these people.}
		
		In this paper, we propose a new inference method that accommodates such ambiguity. In many cases, while one may not know the population shares precisely, one may obtain an idea about a plausible range of the population share vector from aggregate demographic statistics from published data sets such as PSID or the U.S. Census data. Given such a range, we can write the treatment effect parameter as a function of the population share vector, and develop a robust confidence set which is valid for any target population corresponding to any population share vector in the given range. As a first result, we show how this can be done in this paper.
		
		Once a benchmark population share vector is used and an estimate is obtained, one might ask what would be the range of other population share vectors that the current estimate is ``applicable to''. This range of other population share vectors constitutes what we call \textit{the scope of external $\eta$-validity}. This scope represents the set of the population share vectors over which the benchmark treatment effect remains unchanged up to its small fraction $\eta$. When the treatment effect sensitively depends on the population share vector, the scope will be narrow, implying a small scope of external validity. Thus our second task in the paper is to discover this range from data with an appropriate measure of error controlled at a pre-specified level.
		
		A naive idea of using a confidence set for the scope of external validity suffers from a problem: when the data contains lots of noise, one may end up having a larger confidence set, claiming a greater scope of external $\eta$-validity. To remedy this problem, we propose what we call an \textit{anti-confidence set} which is a random set for an identified set whose probability of \textit{being contained in the identified set} is at least equal to a designated level. When one uses an anti-confidence set, using a test with low power forces one to claim only a small scope of external $\eta$-validity. To implement this insight, this paper adopts the step-down procedure of \cite{Romano/Shaikh:10:Eca} with the asymptotic control of its familywise error rate, and constructs an anti-confidence set for the scope of external validity.
		
		As an empirical application, we re-visit the U.S. national JTPA (Job Training Partnership Act) data and analyze the effect of the job training program. The job training program data were generated from the treatment-based sampling design, and yet the precise population shares are not available to the researcher. We first estimated the average treatment effects assuming various population shares ranging from $5\%$ to $90\%$. Then we recovered the scope of external validity to which the average treatment effect estimated assuming a benchmark share (for example, $5\%$) applies. Our result suggests a wide scope of external validity for the estimates.
		
		Given a target population for a benchmark population share, one may ask what the optimal sampling design should look like. The rationale for nonrandom sampling is often that when the participants constitute a small fraction of the population, sampling relatively more from the participants will improve the quality of inference. However, this rationale is incomplete because we also need to consider the contribution of the noise in the subsample to the variance of the estimator. We define the optimal sampling design to be one that minimizes the semiparametric efficiency bound over a range of sampling designs. We find an explicit solution for the optimal sampling design from the semiparametric efficiency bound for the treatment effect parameters under treatment-based sampling.\footnote{See \cite{Hahn/Hirano/Karlan:11:JBES} for an optimal design of social experiments in a related context.}
		
		Our paper is related to the literature of stratified sampling, program evaluation and partial identification. %Stratified sampling is one of the oldest sampling methods studied in statistics. (See e.g. \cite{Neyman:34:JRSS}.) 
		Early econometrics literature on stratified sampling assumed that the conditional
		distribution of observations given a stratum belongs to a parametric family \citep[]{Manski/Lerman:77:Eca,Manski/McFadden:81:StructuralAnalysis, Cosslett:81:StructuralAnalysis, Cosslett:81:Eca, Imbens:92:Eca, Imbens:96:JOE}. \cite{Wooldridge:99:Eca, Wooldridge:01:ET} studied \textit{M}-estimators under nonrandom sampling which do not
		rely on this assumption. Closer to this paper, \cite{Breslow/McNeney/Wellner:03:AS} and \cite{Tripathi:11:ET} investigated the problem of efficient estimation under stratified sampling schemes. %\cite{Tripathi:11:ET} considered moment-based models under various stratified sampling schemes and proved that the empirical likelihood estimators adapted to an appropriate change of measure to achieve efficiency. 
		The stratified sampling scheme studied by \cite{Tripathi:11:ET} is different from this paper's set-up because the former focuses on parametric models and assumes that the population share can be identified from an additional data source.
		Neither does this paper's framework fall into the framework of \cite{Breslow/McNeney/Wellner:03:AS} who considered variable probability sampling which is different from the standard stratified sampling studied here.
		In the program evaluations literature, there is surprisingly little research which deals with inference under treatment-based sampling. \cite{Chen/Hong/Tarozzi:08:AS} established semiparametric efficiency bounds in a general model with missing values, but their approach does not apply to our framework, because missing values arise depending on the treatment status here. \cite{Escanciano/Zhu:13:Cemmap} studied semiparametric models where the parameter of interest is conditionally identified in the sense that their moment equality restrictions admit a unique solution in terms of the parameter correponding to each fixed value of some nuisance parameters. While their general framework can potentially be applied to treatment-based sampling, we believe it is still important to study exclusively the issue of treatment-based sampling, implications for its external validity, and the problem of optimal sampling design. The results in this paper in their own context are new. \cite{Kaido/Santos:14:Eca} studied efficient estimation of a partially identified set defined by moment inequalities in a way that is amenable to convex analysis so that the identified set is essentially a function of a nuisance parameter. However, in contrast to our set-up, their identified set as a function is not necessarily a smooth function, which raises complication that does not arise in our case. In our set-up, the identified set is indexed by the population share with respect to which the treatment effect parameters vary smoothly.  \cite{Heckman/Todd:09:EJ} offered a nice, simple idea to identify and estimate the treatment effect on the treated under treatment-based sampling that is solely based on the treatment status. In contrast to \cite{Heckman/Todd:09:EJ} who focused on the case where the treatment effect parameters are point-identified, this paper accommodates more generally the set-ups where they are set-identified.
		
		This paper is organized as follows. Section \ref{sec:TSB} introduces
		treatment-based sampling data designs and discusses identification of treatment effects under treatment-based sampling. Then the section discusses inference on the treatment effects, and develops an approach to recover the scope of external validity to which the treatment effect for a benchmark population share remains applicable. Section \ref{sec:emp} applies our estimators to the U.S. national JTPA data.  Section \ref{sec:asym} establishes the asymptotic results for our methods. The final section concludes.  The Appendix presents the estimators of variances and covariances.  The online supplement collects the technical proofs. 
		
		\section{Estimating Treatment Effects under Treatment-Based Sampling}\label{sec:TSB}
		
		\subsection{Treatment-Based Sampling and Identification}\label{sec: setup}
		We consider the potential outcome framework of program evaluation. Let $D$ be a random variable
		that takes values in $\{0,1\},$ where $D=1$ means participation in the
		program and $D=0$ being left in the control group. Let $Y$ be the observed outcome defined as
		\begin{align*}
			Y = Y_1 D + Y_0 (1 - D),
		\end{align*}
		where $Y_1$ denotes the potential outcome of a person treated in the program and $Y_0$ that of a person not treated in the program. Let $X=(V,W)$ be a vector of covariates, where $W$ is a discrete random vector that is part of the sampling strata as described below.
		
		To describe treatment-based sampling, let $P$ be the \bi{target population} which is a joint distribution of $(Y,D,V,W)$. We further define
		\begin{align*}
			p_{d,w}^* = P\{D = d, W = w\},
		\end{align*}
		i.e., the proportion of individuals with $(D,W)=(d,w)$ in the target population. We call $p^*= [p^*_{d,w}]$ the \bi{population share vector} which is the vector of the target population shares $p_{d,w}^*$.
		
		We introduce a process of \bi{treatment-based sampling} as follows:\medskip
		
		(\textsc{Step 1}): A stratum $(D,W)=(d,w)$ is randomly drawn from the multinominal distribution with the \bi{design share vector} $q = [q_{d,w}]$.\medskip
		
		(\textsc{Step 2}): $(Y,V)$ is randomly drawn from the conditional distribution of $(Y,D)$ given $(D,W)=(d,w)$, where the conditional distribution is the same as that under the target population $P$.\medskip
		
		(\textsc{Step 3}): We repeat Steps 1 - 2 until our total sample size becomes $n$.\medskip
		
		In this process, the design share $q_{d,w}$ can be systematically different from the population share $p_{d,w}^*$. When the sampling strata is based only on $D=d$ (not based on $W=w$), we call the sampling the \bi{pure treatment-based sampling}.
		
		By this sampling design, there is a one-to-one correspondence between the target population $P$ (as the joint distribution of $(Y,V,D,W)$) and the population share vector $p^*$. We say that $P$ is \bi{associated with} the population share $p^*$. We denote by $Q$ the joint distribution of $(Y,V,D,W)$ determined by the design share vector $q$ and the conditional distribution of $(Y,V)$ given $(D,W)$ in Step 2 above. Hence in our treatment-based sampling set-up, the conditional distribution of $(Y,D)$ given $(D,W)$ is the same under $P$ and $Q$, but there is a significance difference between $P$ and $Q$: the sample is a random sample from $Q$, but it is not under our target population $P$. The difference between $P$ and $Q$ solely comes from the difference in the distribution of $(D,W)$ (which is $p^*$ under $P$ but $q$ under $Q$).
		
		The main objects of interest in this paper are the average treatment effect, $\tau^* _{ate},$ and the average treatment effect on the treated, $\tau^* _{tet},$ defined (under the target population) as follows: 
		\begin{equation}
			\tau_{ate}^*=\mathbf{E}\left[ Y_{1}-Y_{0}\right]\text{ and\ }\tau_{tet}^*=\mathbf{E}\left[
			Y_{1}-Y_{0}\mid D=1\right].
			\label{par}
		\end{equation}
		To make explicit their dependence on $p^*$, let us rewrite them as
		\begin{align*}
			\tau_{ate}^* = \tau_{ate}(p^*), \text{ and } \tau_{tet}^* = \tau_{tet}(p^*),
		\end{align*}
		where for $p = \{p_{d,w}\}$, 
		\begin{align}\label{par_p}
			\tau _{ate}(p) &=\sum_{w}\mathbf{E}\left[ Y_{1}-Y_{0} \mid (D,W) = (1,w)\right] p_{1,w}\\ \notag
			&\quad +\sum_{w}\mathbf{E}\left[ Y_{1}-Y_{0} \mid (D,W) = (0,w)\right] p_{0,w} \\  \notag
			\tau _{tet}(p) &=\sum_{w}\mathbf{E}\left[Y_{1}-Y_{0} \mid (D,W) = (1,w)\right] \frac{p_{1,w}}{p_1},\notag
		\end{align}
		with $p_{1}=\sum_{w}p_{1,w}$. 
		
		Let us study the identification of $\tau_{ate}^*$ and $\tau_{tet}^*$ under the standard unconfoundness condition.
		
		\begin{condition}\label{con:unconfounded}
			\noindent $(Y_{0},Y_{1})\ 
			%TCIMACRO{%
				%\TeXButton{dperp}{\mbox{$\perp\negthinspace\negthinspace\negthinspace\perp$}}}%
			%BeginExpansion
			\mbox{$\perp\negthinspace\negthinspace\negthinspace\perp$}%
			%EndExpansion
			\ D \mid X$ under $P$.
		\end{condition}
		\begin{condition}\label{con:overlap}
			\noindent There exists an $\epsilon \in (0,1/2)$ such that for all $d\in \{0,1\},\epsilon <\inf_{x}p_{d}(x)$ and $\epsilon
			<\inf_{x}q_{d}(x)$,
			where the infimum over $x$ is over the support of $X$, and
			\begin{equation*}
				p_{d}(x) \equiv P\{D=d|X=x\}\text{ and }
				q_{d}(x) \equiv Q\{D=d|X=x\}.
			\end{equation*}
		\end{condition}
		Condition \ref{con:unconfounded} is the unconfoundedness condition which requires that $(Y_{0},Y_{1})$ is conditionally independent of $D$ given $X$ under $P$. Condition \ref{con:overlap} assumes that the propensity score under $P$ (i.e., $p_{d}(x)$) and the propensity score under $Q$ (i.e., $q_d(x)$) are bounded away from
		zero on the support of $X$. This is violated when part of $X$ is only observed among the treated or untreated subsamples.\footnote{See
			\cite{Heckman/Ichimura/Todd:97:ReStud} for a discussion on this issue. See \cite{Khan/Tamer:10:Eca} for an analysis of situations where Condition \ref{con:overlap} is
			violated with $p_{d}(x)$ being arbitrarily close to 0 or 1.}
		
		Under Conditions \ref{con:unconfounded} and \ref{con:overlap}, we can identify \footnote{Note that $p_{1}(x)/p_{0}(x) =( f(v|1,w)/f(v|0,w))(p^*_{1,w}/p^*_{0,w})$, where $f(v|d,w)$ denotes the conditional density function of $V$ given $%
			(D,W)=(d,w)$, which can be identified from the data.  As a result, $\mathbf{E}\left[Y_1-Y_0\mid(D,W)=(d,w)\right]$ does not depend on $p^*_{d,w}$. } 
		\begin{align*}
			\mathbf{E}\left[
			Y_1-Y_0\mid(D,W)=(1,w)\right]&= \mathbf{E}\left[
			Y \mid(D,W)=(1,w)\right]\\
			& \quad -\frac{p^*_{0,w}}{p^*_{1,w}}\mathbf{E}\left[
			\frac{p_1(X)}{p_0(X)}Y\mid(D,W)=(0,w)\right],
		\end{align*}
		and 
		\begin{align*}
			\mathbf{E}\left[
			Y_1-Y_0\mid(D,W)=(0,w)\right]&= \frac{p^*_{1,w}}{p^*_{0,w}}\mathbf{E}\left[
			\frac{p_0(X)}{p_1(X)}Y\mid(D,W)=(1,w)\right]\\
			& \quad -\mathbf{E}\left[
			Y \mid(D,W)=(0,w)\right].
		\end{align*} 
		From this, we obtain the following identification results:
		\begin{align*}
			\tau _{ate}(p) &= \sum_{w}p_{1,w}\mathbf{E}\left[
			\frac{Y}{\bar p_1(X)}\mid(D,W)=(1,w)\right]\\
			&\quad -\sum_{w}p_{0,w}\mathbf{E}\left[
			\frac{Y}{\bar p_0(X)}\mid(D,W)=(0,w)\right]
		\end{align*}
		and
		\begin{align*}
			\tau_{tet}(p) &=\frac{1}{p_1}\sum_{w}p_{1,w}\mathbf{E}\left[
			Y \mid(D,W)=(1,w)\right]\\ \notag
			& \quad \quad -\frac{\displaystyle \sum_{w}p_{0,w}\mathbf{E}\left[
				\frac{\bar p_1(X)}{\bar p_0(X)}Y\mid(D,W)=(0,w)\right] }{\displaystyle  \sum_{w}p_{0,w}\mathbf{E}\left[
				\frac{\bar p_1(X)}{\bar p_0(X)}\mid(D,W)=(0,w)\right] },
		\end{align*}
		where  \begin{equation*}
			\bar p_{d}(x)=\frac{f(v|d,w)p_{d,w}}{f(v|1,w)p_{1,w}+f(v|0,w)p_{0,w}},  ~x=(v, w),
		\end{equation*}
		and $f(v|d,w)$ denotes the conditional density function of $V$ given $%
		(D,W)=(d,w)$.
		
		As noted by \cite{Heckman/Todd:09:EJ}, in the case of pure treatment-based sampling where sampling strata involve only treatment status $D$ (not $W$), we can identify $\tau_{tet}^*$ without knowledge of $p^*$ as follows:
		\begin{align}
			\label{id_tet_pure}
			\tau_{tet}^*&=\mathbf{E}\left[Y\mid D=1\right] -\frac{\displaystyle \mathbf{E}\left[\frac{p_1(X)}{p_0(X)}Y \mid D=0\right] }{\displaystyle \mathbf{E}\left[\frac{p_1(X)}{p_0(X)}\mid D=0\right] } \\ \notag
			&=\mathbf{E}\left[Y\mid D=1\right] -\frac{\displaystyle \mathbf{E}\left[\frac{q_1(X)}{q_0(X)}Y \mid D=0\right] }{\displaystyle \mathbf{E}\left[\frac{q_1(X)}{q_0(X)}\mid D=0\right] },
		\end{align}
		where the second equality comes from the relationship (which holds under pure treatment-based sampling)
		\begin{equation}\label{pq_relation}
			\frac{p_1(x)}{p_0(x)}=\frac{q_1(x)}{q_0(x)}\frac{p_1^*q_0}{p_0^*q_1}.
		\end{equation}
		The last difference in (\ref{id_tet_pure}) is identified without knowledge of the population share vector $p$, because the ratio $q_1(x)/q_0(x)$ is always identified from the observed sample in this case.
		
		In general, the treatment effects $\tau_{ate}^*$ and  $\tau_{tet}^*$ are identified only up to the population share $p$. Hence as functions, $\tau_{ate}(\cdot)$ and $\tau_{tet}(\cdot)$ are point-identified.
		
		\subsection{Inference under Treatment-Based Sampling}\label{sec:eff_est}
		\subsubsection{Estimation}\label{sec: estimation}
		In this section, we propose efficient estimators for $\tau_{ate}(\cdot)$ and  $\tau_{tet}(\cdot)$. For simplicity of exposition, we assume that $V_{i}$ is a continuous random vector and its support $\mathcal{V}\in \mathbf{R}^{d_1}$. It is not hard to extend the result to include discrete components. 
		
		First, we obtain a propensity score estimator:
		\begin{equation}
			\tilde{p}_{d,i}(X_{i})\equiv \frac{\tilde{\lambda}_{d,i}(X_{i})}{\tilde{%
					\lambda}_{1,i}(X_{i})+\tilde{\lambda}_{0,i}(X_{i})},  \label{ps2}
		\end{equation}%
		where, with$\ \hat{L}_{d,w,i}\equiv (p_{d,w}/\hat{q}_{d,w})1\{(D_i,W_i) = (d,w)\}$,
		$\hat{q}_{d,w}\equiv n_{d,w}/n$, and $n_{d,w} \equiv \sum_{i=1}^n1\{(D_i,W_i)=(d,w)\}$, we define%
		\begin{equation*}
			\tilde{\lambda}_{d,i}(x)=\frac{1}{n-1}\sum_{j=1,j\neq i}^{n}\hat{L}%
			_{d,w,j}K_{h}\left( V_j-v\right), ~\text{where} ~ x=(v,w),
		\end{equation*}%
		and $K_{h}(s_{1},...,s_{d_{1}})=K(s_{1}/h,...,s_{d_{1}}/h)/h^{d_{1}}$ and $%
		K(\cdot )$ is a multivariate kernel function. Then we construct the
		following estimator of $\tau _{ate}(p)$:
		\begin{equation}\label{ate_est}
			\hat{\tau}_{ate}(p)= \sum_{w}\sum_{i: (D_i,W_i) = (1,w)}\tilde{g}_{1,w,i}Y_{i}-\sum_{w}\sum_{i: (D_i,W_i) = (0,w)}\tilde{g}_{0,w,i}Y_{i},
		\end{equation}%
		where
		\begin{equation}
			\tilde{g}_{d,w,i}=\frac{p_{d,w}\tilde{1}_{n,i}}{n_{d,w}\tilde{p}_{d,i}(X_{i})},  \label{gwi}
		\end{equation} 
		and 
		\begin{equation}\label{til_ind}
			\tilde{1}%
			_{n,i}=1\left\{\min\{\tilde{\lambda}_{1,i}(X_{i}),\tilde{\lambda}%
			_{0,i}(X_{i}) \} \geq \delta _{n}\right\},
		\end{equation}
		for a positive sequence $\delta _{n}\rightarrow 0$. (For example, our choice of $\delta_n = n^{-1/2}$ shows good finite sample performance for our procedures in our simulation study.)
		
		Similarly, we construct an estimator of $\tau _{tet}(p)$ as
		follows:%
		\begin{equation}\label{tet_est}
			\hat{\tau}_{tet}(p)=\frac{1}{p_1}\sum_{w}\frac{p_{1,w}}{%
				n_{1,w}}\sum_{i: (D_i,W_i) = (1,w)}Y_{i}-\frac{\displaystyle \sum_{w%
				}\sum_{i: (D_i,W_i) = (0,w)}\tilde{g}_{0,w,i}\tilde{p}_{1,i}(X_{i})Y_{i}}{\displaystyle \sum_w \sum_{i: (D_i,W_i) = (0,w)}\tilde{g}_{0,w,i}\tilde{p}_{1,i}(X_{i})}.
		\end{equation}%
		In the case of pure treatment-based sampling, the estimator $\hat{\tau}_{tet}(p)$ is reduced
		to the following simpler form: 
		\begin{equation*}
			\frac{1}{n_1}\sum_{i: D_i = 1}Y_{i}-\frac{\displaystyle 
				\sum_{i: D_i = 0}Y_{i}\tilde{1}_{n,i}\sum_{j: D_j = 1}K_{ij}/\sum_{j:D_j = 0}K_{ij}}{\displaystyle \sum_{i:D_i = 0}\tilde{1}_{n,i}\sum_{j:D_j = 1}K_{ij} / \sum_{j:D_j = 0}K_{ij}},
		\end{equation*}%
		where $K_{ij}=K_{h}\left( V_{j}-V_{i}\right)$. This estimator does not involve the
		population share $p_{d}$. Hence one can make inference on $\tau_{tet}(p)$ without knowledge of the population share in this case.
		
		\subsubsection{Robust Confidence Intervals}\label{sec:rcs}
		Suppose that there exists a true population share vector $p^*$ under which Conditions \ref{con:unconfounded} and \ref{con:overlap} are satisfied, so that $\tau_{ate}^* = \tau_{ate}(p^*)$, and $\tau_{tet}^* = \tau_{tet}(p^*)$, but that the researcher is not sure about $p^*$; she only knows a plausible range for it. Formally, let $A$ be the set of values where the true population share vector $p^*$ is known to belong. 
		We assume that $A$ is contained in the interior of the simplex:
		\begin{equation*}
			\mathcal{S}=\left\{p:\sum_{d,w}p_{d,w}=1\text{
				and }p_{d,w}>0\text{ for all }d,w\right\}%
			\text{,}
		\end{equation*}%
		so that for all $d,w$, we have $p_{d,w}\in (0,1)$. In this set-up, let us develop confidence sets for $\tau_{ate}(p^*)$ and $\tau_{tet}(p^*)$. As is often done in the literature of inference on partially identified models, we use the approach of inverting a test.
		
		First, for each $t \in \mathbf{R}$ and each $p \in A$, let
		\begin{align}
			\label{tests}
			T_{ate}(t) = \inf_{p \in A} \frac{\sqrt{n}\left\vert \hat{\tau}%
				_{ate}(p)-t\right\vert }{\hat{\sigma}_{ate}(p)} \text{   and   }  
			T_{tet}(t) = \inf_{p \in A} \frac{\sqrt{n}\left\vert \hat{\tau}%
				_{tet}(p)-t\right\vert }{\hat{\sigma}_{tet}(p)}, 
		\end{align}%
		where $\hat{\sigma}_{ate}(p)$ and $\hat{\sigma}_{tet}(p)$ are consistent
		estimators of $\sigma _{ate}(p)$ and $\sigma _{tet}(p)$ such that 
		\begin{align}
			\label{asymp normality}
			&\sqrt{n}\left( \hat{\tau}_{ate}(p)-\tau _{ate}(p)\right) \ \ \overset{d}{%
				\rightarrow }\ \ N(0,\sigma_{ate}^{2}(p))\text{ and} \\
			&\sqrt{n}\left( \hat{\tau}_{tet}(p)-\tau _{tet}(p)\right) \ \ \overset{d}{%
				\rightarrow }\ \ N(0,\sigma _{tet}^{2}(p)).\notag
		\end{align}
		The precise forms of $\hat{\sigma}_{ate}(p)$ and $\hat{\sigma}_{tet}(p)$ are given in the Appendix.
		
		We construct confidence sets for $\tau _{ate}^*$ and $\tau _{tet}^*$:%
		\begin{align}\label{CS}
			\mathcal{C}_{ate} =\left\{ t\in \mathbf{R}:T_{ate}(t)\leq c_{1-\alpha
				/2}\right\} \text{ and } 
			\mathcal{C}_{tet} =\left\{ t\in \mathbf{R}:T_{tet}(t)\leq c_{1-\alpha
				/2}\right\},
		\end{align}%
		where $c_{1-\alpha /2}=\Phi ^{-1}(1-\alpha /2)$ and $\Phi $ is the CDF of $N(0,1)$. These confidence sets are asymptotically valid as shown in the theorem below, which follows from the weak convergence result established in Theorem \ref{th:weak convergence}.
		
		\begin{theorem}\label{thm: validity of CI}
			If Conditions \ref{con:unconfounded} -- \ref{con:overlap} and Assumptions \ref{con:bdY} -- \ref{ass:kernel} in Section \ref{sec:asym} below hold, then the confidence intervals $\mathcal{C}_{ate}$ and $\mathcal{C}_{tet}$ are asymptotically valid for $\tau _{ate}^*$ and $\tau_{tet}^*$ as $n \rightarrow \infty$, namely,
			\begin{align*}
				\liminf_{n\rightarrow \infty}\ P\left\{\tau_{ate}^*\in \mathcal{C}_{ate}\right\} \geq  1-\alpha \quad \text{and} \quad 
				\liminf_{n\rightarrow \infty}\ P\left\{\tau_{tet}^*\in\mathcal{C}_{tet}\right\} \geq 1-\alpha .
			\end{align*}
		\end{theorem}
		
		\subsection{Inference on the Scope of External Validity}\label{sec: SEV}
		
		In many applications, it may be of interest to see if there are other populations that the current estimate applies to, i.e., if the current estimate is \textit{externally valid}. To explore this question, let us first clarify the meaning of ``other populations''. We say that any given population share $p'$ \bi{satisfies the unconfoundedness condition} if Conditions \ref{con:unconfounded} and \ref{con:overlap} hold when we replace $P$ by $P'$ that is associated with $p'$. Now, suppose that we are interested in a treatment effect generically denoted by $\tau$, and identify a treatment effect $\tau(p^\circ)$ using some benchmark population share vector $p^\circ$. (One can think of $\tau$ as either $\tau_{ate}$ or $\tau_{tet}$.)  We now ask at what other values of the population share that satisfy the unconfoundedness condition, the treatment effect remains similar to the benchmark treatment effect $\tau(p^\circ)$. In particular, we ask if the benchmark treatment effect $\tau(p^\circ)$ varies little as one moves away from the assumed population share vector $p^\circ$. If this is the case, then the treatment effect $\tau(p^\circ)$ exhibits a wide scope of external validity. We will formally define the scope of external validity below. In practice, when the researcher is not sure which value of $p^\circ$ to use as a benchmark, she can try more than one values of $p^\circ$ and see how the scope of external validity varies across the different values. We illustrate this in our empirical application later.
		
		There is an interesting relation between the scope of external validity and the heterogeneity of treatment effects across strata.\footnote{We thank an anonymous referee for providing this observation.} To see this clearly, let us define: for $(d,w)$,
		\begin{align}\label{par_p2}
			\tau _{d,w} = \mathbf{E}\left[ Y_{1}-Y_{0} \mid (D,W) = (d,w)\right].
		\end{align}
		The quantity $\tau _{d,w}$ represents the average treatment effect for the subgroups with $(D,W) = (d,w)$. From (\ref{par_p}), it is clear that if these treatment effects are not heterogeneous across subgroups $(d,w)$, the average treatment effects $\tau _{ate}(p)$ and $\tau _{tet}(p)$ do not vary with the population share $p$. Hence in this case, the treatment effect estimates will have a wider scope of external validity.
		
		To make the idea precise, we fix a small number $\eta>0$ and define the \bi{scope of external $\eta$-validity} as
		\begin{align}\label{SOEV}
			A(p^\circ;\eta) = \{p \in A: |\tau(p) - \tau(p^\circ)| \le \eta |\tau(p^\circ)| \}.
		\end{align}
		Hence the set $A(p^\circ;\eta)$ is the set of population share vectors in $A$ such that as we move $p$ around in $A$, the treatment effect parameter $\tau(p)$ does not move away from $\tau(p^\circ)$ by more than $100\eta$ percent of the benchmark treatment effect $|\tau(p^\circ)|$. We say that the benchmark treatment effect $\tau(p^\circ)$  is \bi{externally $\eta$-valid} for any $p \in A(p^\circ;\eta)$.
		
		We would like to develop an inference method on the set $A(p^\circ;\eta)$. One might consider using a confidence set for $A(p^\circ;\eta)$. However, such an approach has a problem, because a wider confidence set due to larger noise in the data will translate into a wider scope of external validity. Instead, we propose using a set $\mathcal{A}_n$ such that
		\begin{equation}
			\label{FWER control}
			\liminf_{n\rightarrow\infty}P\left\{ \mathcal{A}_n \subset A(p^\circ;\eta)\right\}\geq 1-\alpha.
		\end{equation}
		The set $\mathcal{A}_n$ represents the set of population share vectors $p$ for which we have strong support from data that $\tau(p)$ is within the $\eta$ fraction of the benchmark absolute treatment effect. We call such a set $\mathcal{A}_n$ an \bi{anti-confidence set} of $A(p^\circ;\eta)$ at level $1- \alpha$. When there is a lot of noise in the data, the set $\mathcal{A}_n$ tends to be smaller, forcing the researcher to claim a smaller scope of external validity.
		
		To construct such a set, we first formulate the problem as that of multiple hypothesis testing and adapt the step-down multiple testing procedure of \cite{Romano/Shaikh:10:Eca} to our set-up. Consider the following individual hypothesis for each $p$:
		\begin{align}
			\label{hypothesis 2}
			H_0(p;\eta) &: |\tau(p) - \tau(p^\circ)| > \eta |\tau(p^\circ)|, \text{against}\\ \notag
			H_1(p;\eta) &: |\tau(p) - \tau(p^\circ)| \le \eta |\tau(p^\circ)|.
		\end{align}
		In (\ref{hypothesis 2}), it is the alternative hypothesis which states the external $\eta$-validity of $\tau(p^\circ)$ for $p$ up to the $\eta$ fraction of the benchmark treatment effect.
		
		Let $\hat \tau(p)$ and $\hat \tau(p^\circ)$ be estimators of $\tau(p)$ and $\tau(p^\circ)$. Consider the statistic
		\begin{equation}\label{hatQ}
			\hat{Q}(p) = \frac{1}{2}\max \left\{\sqrt{n}\hat \Delta(p),0 \right\},
		\end{equation}
		where\footnote{The square is taken above to facilitate the application of the delta method in the asymptotic derivation.}
		\begin{align*}
			\hat \Delta(p) =\eta^2 \hat\tau^2(p^\circ) -\left(\hat\tau(p)-\hat\tau(p^\circ)\right)^2.
		\end{align*}
		For any set $S \subset A$, let $\hat c_{1-\alpha}(S)$ be such that
		\begin{align*}
			\liminf_{n \rightarrow \infty} P\left\{\sup_{p \in S} \hat{Q}(p) \le \hat c_{1-\alpha}(S) \right\} \ge 1 - \alpha,
		\end{align*}
		whenever $H_0(p;\eta)$ holds for all $p \in S$. Using $\hat c_{1-\alpha}(S)$, we construct a set $\mathcal{C}_n  \subset A$ through the following step-down procedure, and then take
		\begin{align}
			\label{anti-conf}
			\mathcal{A}_n = A \backslash \mathcal{C}_n.
		\end{align}
		(We will present a  bootstrap procedure to construct $\hat c_{1-\alpha}(S)$ after Theorem \ref{thm:step_down} below.)
		
		The step-down procedure is as follows. First, in Step 1, we let $S_1=A\backslash\{p^\circ\}$. If $\sup_{p \in S_1} \hat{Q}(p) \leq \hat{c}_{1-\alpha}(S_1)$, set $\mathcal{C}_n=S_1$. Otherwise, set 
		\begin{equation*}
			S_2=\left\{p\in A\backslash\{p^\circ\}: \hat{Q}(p) \leq \hat{c}_{1-\alpha}(S_1)\right\}.
		\end{equation*}
		In general, in Step $k \ge 1$, if $\sup_{p\in S_k}\hat{Q}(p) \leq \hat{c}_{1-\alpha}(S_k)$, set $\mathcal{C}_n=S_k$. Otherwise, set\footnote{Note that when we define $S_k$ we exclude $p^\circ$, so that the set $\mathcal{A}_n$ includes $p^\circ$.} 
		\begin{equation*}
			S_{k+1}=\left\{p\in A\backslash\{p^\circ\}: \hat{Q}(p) \leq \hat{c}_{1-\alpha}(S_k)\right\}.
		\end{equation*}
		We continue the process until there is no change in the set $S_{k}$'s, i.e., no further hypothesis is rejected. Once we obtain $\mathcal{C}_n$, we now define the anti-confidence set $\mathcal{A}_n$ as in (\ref{anti-conf}). Then $\mathcal{A}_n$ is an anti-confidence set for $A(p^\circ;\eta)$ at level $1 - \alpha$, as shown in the theorem below.
		\begin{theorem}\label{thm:step_down}
			Suppose that either $(\tau,\hat \tau) = (\tau_{ate},\hat \tau_{ate})$ or $(\tau,\hat \tau) = (\tau_{tet},\hat \tau_{tet})$. Suppose further that Conditions
			\ref{con:unconfounded} -- \ref{con:overlap} and Assumptions \ref{con:bdY} -- \ref{ass:kernel} in Section \ref{sec:asym} below hold. Then
			\begin{align*}
				\liminf_{n\rightarrow\infty}P\left\{\mathcal{A}_n \subset A(p^\circ;\eta)\right\}\geq 1-\alpha.
			\end{align*}
		\end{theorem}
		Theorem \ref{thm:step_down} shows that the set $\mathcal{A}_n$ is indeed the anti-confidence set of the scope of external validity at level $1 - \alpha$. The proof is given in the online supplement, Section S2.
		
		To construct $\hat c_{1-\alpha}(S)$, we propose using a bootstrap procedure, following the Bonferroni approach in a spirit similar to \cite{Romano/Shaikh/Wolf:14:Eca}. First, for each $b = 1,2,...,B$, let $\hat \tau_{b}^*(p)$ be the same as $\hat \tau(p)$ except that instead of using the original sample, we use the bootstrap sample resampled with replacement. Then we construct
		\begin{align*}
			\hat \Delta_b^*(p) =\eta^2 \hat\tau_{b}^{*2}(p^\circ) -\left(\hat\tau_{b}^{*}(p)-\hat\tau_{b}^{*}(p^\circ)\right)^2.
		\end{align*}
		To describe the Bonferroni approach, we write
		\begin{equation}\label{hatQ2}
			\hat{Q}(p) = \frac{1}{2}\max \left\{\sqrt{n}\left(\hat \Delta(p) - \Delta(p) + \Delta(p) - \hat \Delta(p) + \hat \Delta(p)\right),0 \right\},
		\end{equation}
		where \begin{equation*}
			\Delta(p) =\eta^2 \tau^2(p^\circ) -\left(\tau(p)-\tau(p^\circ)\right)^2.
		\end{equation*}
		Then we fix $\beta \in (0,1)$ and $S \subset A$, and find $\hat \eta_{1-\beta}(S)$ such that
		\begin{align}
			\label{beta control}
			\liminf_{n \rightarrow \infty} P\left\{\sup_{p \in S} \sqrt{n}\left(\Delta(p) - \hat \Delta(p)\right) > \hat \eta_{1-\beta}(S) \right\} \le \beta.
		\end{align}
		For example, we can take $\hat \eta_{1-\beta}(S)$ to be the $1-\beta$ quantile from the empirical distribution of $\sup_{p \in S}\sqrt{n}(\hat \Delta(p) - \hat \Delta^*_b(p)), b=1,2,...,B$. (The choice of $\beta = 0.01$ worked well in our simulation study.) Then we construct
		\begin{equation}\label{hatQ*2}
			\tilde{Q}_b^*(p) = \frac{1}{2}\max \left\{\sqrt{n}\left(\hat \Delta_b^*(p) - \hat \Delta(p) + \hat \varphi_{1-\beta}(p,S)\right),0 \right\},
		\end{equation}
		where
		\begin{align*}
			\hat \varphi_{1-\beta}(p,S) = \min \left\{\hat \Delta(p) + \frac{\hat \eta_{1-\beta}(S)}{\sqrt{n}},0 \right\}.
		\end{align*}
		We take $\hat c_{1-\alpha}(S)$ to be the $1-\alpha + \beta$ quantile of the empirical distribution of the bootstrap quantities $\sup_{p \in S} \tilde {Q}_b^*(p), \text{  } b=1,2,...,B$.
		
		Note that the dimension of $p$ depends on the dimension of strata used in the treatment-based sampling, rather than the dimension of the covariates. Nevertheless, when the dimension of $p$ is large, the procedure can be more complicated, and the scope of external validity can be large.
		
		\subsection{Monte Carlo Simulations}\label{sec: MC}
		\subsubsection{Finite Sample Performances of the Treatment Effect Estimators }\label{sec: MC_point_est}
		
		This section conducts Monte Carlo simulations to evaluate the finite sample performance of our estimation and inference approaches described in  Sections \ref{sec: estimation} and \ref{sec:rcs}. The data generating process is as follows. Let $\varepsilon
		_{0,i},u_{1i},$ and $u_{2i}$ be independent random variables drawn from $%
		N(0,1)$, and $u_{3i}$ be a independent random variable drawn from a uniform distribution on $(-1,1)$.  The covariates $V_{i}=(V_{1i},V_{2i})$ are constructed in two ways: Spec A has $V_{1i}=1\{u_{1i}+\varepsilon _{0,i}\geq 0\}$ and $V_{2i}=1\{u_{2i}+\varepsilon _{0,i}\geq 0\}$ while Spec B has $V_{1i}=u_{3i}$ and $V_{2i}=1\{u_{2i}+\varepsilon _{0,i}\geq 0\}$. Hence in Spec A, both $V_{1i}$ and $V_{2i}$ are discrete random variables,
		while in Spec B, only $V_{2i}$ is discrete. We define an index that
		determines the participation of individuals in the program:%
		\begin{equation}\label{simu_sel}
			U_{i}=a(V_{1i}+V_{2i}-1)+r_{i}+0.5(W_i-0.5),
		\end{equation}%
		where $r_{i}\sim N(0,1)$ and the parameter $a$ captures the dependence of participation decision on $(V_{1i}, V_{2i})$. The participation indicator is 
		$
		D_{i}=1\{U_{i}\leq 0.5\}
		$, and 
		the potential outcomes are specified as follows:%
		\begin{eqnarray}\label{simu_po}
			Y_{1i} &=&e_{0i}+(c_{1i}+1/2)(V_{1i}+V_{2i})/2+0.5W_{i}+b(V_{1i}+V_{2i})+e_{1i},
			\\
			Y_{0i} &=&e_{0i}+(c_{0i}+1/2)(V_{1i}+V_{2i})/2+0.5W_{i}\notag,
		\end{eqnarray}%
		where the parameter $b$ captures the variability of the individual treatment effect for different values of $(V_{1i},V_{2i})$. Random variables $e_{0,i},$ $e_{1i},$ $c_{0i},$ and $c_{1i}$ are independently drawn from $N(0,1)$. We set $a=0.5$ and $b=3$ in (\ref{simu_sel}) and (\ref{simu_po}) so that $\tau_{ate}(p)$ varies significantly with the population share $p_1$. Table \ref{table:id_interval} presents the identified interval of $\tau_{ate}^*$ with $p_1 \in [0.01, p_1^{u}]$ for different values of $p_1^{u}$.
		
		\begin{table}
			\caption {Identified Interval of $\tau_{ate}^*$ with $p_1^* \in [0.01, p_1^{u}]$, $a=0.5, b=3$, $W_{i}\in \{0,1\}$ with $p_w^* \equiv P\{W_{i}=1\}=0.2$ : $\mathcal{I}_{ate}=[\inf_{p}\tau_{ate}(p),\sup_{p}\tau_{ate}(p)]$,
				$RL= (\sup_{p}\tau_{ate}(p)-\inf_{p}\tau_{ate}(p))/\inf_{p}\tau_{ate}(p)$. }\label{table:id_interval}
			\small
			\begin{center}
				\begin{tabular}{ccc|cc}\hline\hline
					\multicolumn{1}{c}{\multirow{2}{*}{$p_1^{u}$ }}& \multicolumn{2}{c}{Spec A}& \multicolumn{2}{c}{Spec B}\\\cline{2-5}
					\multicolumn{1}{c}{}&\multicolumn{1}{c}{$\mathcal{I}_{ate}$} & $RL$ (in $\%$)& \multicolumn{1}{c}{$\mathcal{I}_{ate}$} &$RL$ (in $\%$)\\ \hline 
					0.05 &   [4.076, 4.139] &    1.547 &    [4.753, 4.919] &    3.476\\ 
					0.10 &    [3.997, 4.139] &    3.550 &    [4.547, 4.919] &    8.173\\ 
					0.30 &    [3.681, 4.139] &   12.420 &    [3.732, 4.919] &   32.114\\ 
					0.50 &    [3.366, 4.139] &   22.951 &    [2.902, 4.919] &   69.516\\ 
					\hline
				\end{tabular}
			\end{center}
		\end{table}
		We set $W_{i}=0.5$ in the case of pure
		treatment-based sampling and $W_{i}\in \{0,1\}$ with $p_w \equiv P\{W_{i}=1\}=0.2$ in the case of nonpure treatment-based sampling. The number of replications is 10000. The sample sizes are 500 and 1000.  Let the population share vector be $p=(p_{1}, 1-p_1)$ for the pure treatment-based sampling, and $p=(p_{1}p_w, p_1(1-p_w), (1-p_1)p_w, (1-p_1)(1-p_w))$ for the nonpure treatment-based sampling. We choose $p_1 \in \{0.05,0.10,0.30,0.50\}$. For each $p$ (i.e. for each $p_{1}$, as $p_{w}$ is fixed), we consider the testing problem $H_{0} :\tau _{ate}( p)=t$ against $H_{1} : \tau _{ate}(p)\neq t$ where $t$ is a specified value.
		We examine the size property of the testing procedure that 
		rejects
		$H_{0}$ if $ T_{ate}(t, p)=\sqrt{n}\left\vert \hat{\tau}_{ate}(p)-t\right\vert / \hat{\sigma}_{ate}(p)>c_{1-\alpha /2}$
		with $c_{1-\alpha /2}$ being $(1-\alpha /2)$ quantile of the
		standard normal distribution.  A similar test is conducted for $\tau _{tet}(p)$, using the statistic 
		$
		T_{tet}(t, p)=\sqrt{n}\left\vert \hat{\tau}_{tet}(p)-t\right\vert /
		\hat{\sigma}_{tet}(p).
		$
		We focus on such t-tests for different $p$'s because our robust confidence sets (\ref{CS}) are obtained by inverting the tests based on $\inf_{p\in A}T_{ate}(t,p)$ and $\inf_{p\in A}T_{tet}(t, p)$.
		In Spec B where the variable $V_{1i}$ is continuous, we implement an (undersmoothed) rule of thumb bandwidth $h_n=2.78\times \hat\sigma_{V_1}n^{-1/3}$, where the rate is modified in order to satisfy Assumption \ref{ass:kernel}(ii).  We set $\delta_n=n^{-1/2}$ inside the indicator function in (\ref{til_ind}) so that Assumption \ref{ass:kernel}(iii) is satisfied. 
		
		%Simulation results are presented in Tables \ref{table: size} to \ref{table: mse_B}. The online supplement, we present simulation results for the DGP where the individual treatment effect $Y_{i1}-Y_{i0}$ depends on $W_i$, and the results are very similar to those presented here. 
		Table \ref{table: size} shows that the rejection probability for $T_{ate}(t,p)$ stays quite stable to the variation
		of the population shares $p$. Overall, our tests perform reasonably well in size control.
		The performance of $T_{tet}(t,p)$ is similar to that of $T_{ate}(t,p)$
		except that the rejection probability turned out to be almost the same
		across different $p_{1}$. (Hence the rejection probabilities for $%
		T_{tet}(t,p)$ in Table \ref{table: size} are presented in a single column for brevity.) There is no reason this should be \textit{a priori} so, because although the
		independence of $D_{i}$ and $W_{i}$ under $P$ renders the estimator $\hat{\tau}_{tet}(p)$ invariant to the choice of $p_{1}$, the asymptotic variance $%
		\hat{\sigma}_{tet}^{2}(p)$ can still vary with the choice of $p_{1}$.
		Nevertheless, the rejection probabilities for $T_{tet}(t,p)$ have turned out
		to be the same (up to the numerical precision allowed in the simulation)
		across different population shares $p_{1},$ perhaps because $\hat{\sigma}%
		_{tet}^{2}(p)$ does not change much when we vary $p_{1}$. %The online supplement (Section S2) reports the bias, mean absolute deviation (MAD) and mean squared error (MSE) of estimators $\hat{\tau}%_{ate}(p)$ and $\hat{\tau}_{tet}(p)$.  

		\begin{table}\caption{Rejection Frequencies for Tests Using $T_{ate}(t, p)$ and $ T_{tet}(t, p)$ under $H_{0}$: nominal size $\alpha =5\%$, $t$ equals to the true treatment effect parameter, $p=p_1$ for the pure treatment-based sampling and $p=(p_1,0.2)$ for the nonpure treatment-based sampling}\label{table: size}
			\small
			\begin{center}
				\begin{tabular}{lll|llll|l}
					\hline\hline
					& &  &  & $T_{ate}(t,p)$ &  &  & $T_{tet}(t,p)$ \\ 
					\multicolumn{1}{c}{} & \multicolumn{1}{c}{} & \multicolumn{1}{c|}{$p_1$} & \multicolumn{1}{|c}{$%
						0.05$} & \multicolumn{1}{c}{$0.10$} & \multicolumn{1}{c}{$0.30$} & 
					\multicolumn{1}{c|}{$0.50$} & \multicolumn{1}{|c}{} \\ \hline\hline
					Spec A & Pure  & $n=500$ & 0.0535 & 0.0529 & 0.0526 & 0.0514 & 0.0508
					\\ 
					& & $n=1000$ & 0.0555 & 0.0544 & 0.0555 & 0.0540 & 0.0509 \\ \hline
					& Nonpure  & $n=500$ & 0.0520 & 0.0523 & 0.0554 & 0.0554 & 
					0.0550 \\ 
					& & $n=1000$ & 0.0525 & 0.0522 & 0.0534 & 0.0522 & 0.0540 \\ \hline\hline
					Spec B & Pure & $n=500$ & 0.0715 & 0.0715 & 0.0729 & 0.0775 & 0.0618
					\\ 
					& & $n=1000$ & 0.0622 & 0.0632 & 0.0689 & 0.0714 & 0.0615 \\ \hline
					& Nonpure & $n=500$ & 0.0765 & 0.0783 & 0.0805 & 0.0821 & 
					0.0636 \\ 
					& & $n=1000$ & 0.0630 & 0.0632 & 0.0702 & 0.0753 & 0.0600 \\ \hline
				\end{tabular}
			\end{center}
		\end{table}

		\subsubsection{Finite Sample Performances of the Anti-Confidence Set}
		
		This section examines the finite sample performance of the anti-confidence set for the scope of external $\eta$-validity proposed in Section \ref{sec: SEV}. We focus on pure treatment-based sampling (fix $ W_i=0.5$) and  discrete covariates  (Spec A).  Two simulation designs are considered: Design I is the same as Spec A used in Section \ref{sec: MC_point_est}. Design II is a modified version of the simulation design in (\ref{simu_sel}) and (\ref{simu_po}) with $a=0.5$, $b=0.5$,
		$Y_{1i} =e_{0i}/2+(c_{1i}+1/2)(V_{1i}+V_{2i})/2+0.5W_{i}+b(V_{1i}+V_{2i})+ 2+ e_{1i}/2$, and $Y_{0i} = e_{0i}/2+(c_{0i}+1/2)(V_{1i}+V_{2i})/2+0.5W_{i}.$ By this construction, $\tau_{ate}(p)$ is sensitive to $p$ in Design I but is not so in Design II. For each design, we consider two values of the benchmark population share $p^{\circ}= 0.1, 0.3$, and three values of the fraction $\eta=0.10, 0.15, 0.20$. The object of interest is scope of external $\eta$-validity $A(p^{\circ}; \eta)$ defined in (\ref{SOEV}) with $A=[0.01, 0.99]$.
		We construct the anti-confidence set $\mathcal{A}_n$ that satisfies (\ref{FWER control})
		with the pre-specified familywise error rate  (FWER)  $\alpha=0.05$.
		The number of simulation is $1000$. The initial set is $S_1=[0.01, 0.99]\setminus\{p^{\circ}\}$. To compute the critical value, we apply the bootstrap approach with a Bonferonni-type correction described by (\ref{hatQ*2}). The number of bootstrap is $B=200$ and the small significance level $\beta=0.01$.
		
		Tables \ref{table: anti_con_A} summarizes the simulation results. The lower and upper bounds of the average anti-confidence set in the second-to-last column are computed by taking  average of the respective lower and upper bounds of the anti-confidence set $\mathcal{A}_n$ produced by $1000$ simulations. In Design I with $\eta=0.10$, the anti-confidence set for the scope of external $\eta$-validity degenerates to a singleton $\{p^\circ\}$. It means for this case our approach does not have enough finite-sample power to recover the scope of external $\eta$-validity. In other scenarios,  our approach produces informative anti-confidence sets. The average anti-confidence set becomes closer to the true scope of external $\eta$-validity (presented in the fourth column) when the sample size increases.  
		In Design I with  $p^\circ=0.1$ and $\eta=0.20$, the average anti-confidence 
		set for $n=1000$ is $(0.0101, 0.3862)$,  which accounts for about $65\%$ of the true scope $A(p^\circ;\eta)=[0.01, 0.59]$.  When $p^\circ=0.3$ and $\eta=0.20$, the average anti-confidence set for $n=1000$ accounts for about  $78\%$ of the true scope. 
		%n Design II with $\eta=0.10$,  the average anti-confidence set ($n=1000$) accounts for about $65\%$ ($p^\circ=0.1$) and $78\%$ ($p^\circ=0.3$) of the true scopes. 
		%In Design II with $\eta=0.15$,  the average anti-confidence set ($n=1000$) accounts for about $80\%$ ($p^\circ=0.1$) and $92\%$ ($p^\circ=0.3$) of the true scopes. 
		In Design II with $\eta=0.20$, the average anti-confidence sets ($n=1000$) account for more than $95\%$ of the true scopes.\footnote{In Design II with $(\eta, p^\circ)=(0.15, 0.3), (0.20, 0.1)$ and $(0.20, 0.3)$, the true scopes of external $\eta$-validity equal to the full set $[0.01, 0.99]$, which indicates that the treatment effect parameter is very stable with respect to the population share.}  
		In all scenarios we considered,  the empirical FWERs  are below or close to the nominal level $0.05$.
		Overall,  Table \ref{table: anti_con_A} suggests that our approach is valid and informative in recovering the scope of external $\eta$-validity.  Empirical FWERs smaller than the nominal level  do not undermine our procedure too much because the researcher wants to be cautious in claiming that the benchmark treatment effect estimate is applicable to the other populations.
		\begin{table}
			\caption{Finite Sample Performances of the Anti-confidence Set $\mathcal{A}_n$, nominal $\text{FWER} =0.05$, $\beta=0.01$} \label{table: anti_con_A}
			\small 
			\begin{center}
				\begin{tabular}{ccccccc}
					\hline\hline
					& $p^{\circ}$ & $\eta$& $A(p^{\circ}; \eta)$ & $n$  &  Ave. Anti-Confidence Set &Emp. FWER\\ 
					\hline
					Design I & 0.1&0.10 &[0.01, 0.34] & 500&   $\{0.10\}$ & 0\\
					& & &  & 1000& $\{0.10\}$ & 0\\
					& &0.15 &[0.01, 0.47] & 500&  (0.0999, 0.1001) & 0 \\
					& & &  & 1000& (0.0525, 0.1677) & 0\\
					& &0.20 &[0.01, 0.59] & 500&   (0.0697, 0.1532) &  0.001\\
					& & &  & 1000& (0.0101, 0.3862) & 0 \\
					\hline
					& 0.3&0.10 &[0.08, 0.52] & 500&   $\{0.30\}$ & 0\\
					& & &  & 1000& (0.2996, 0.3005) & 0\\
					& &0.15 &[0.01, 0.64] & 500&  (0.2968, 0.3033)  & 0.001\\
					&& &  & 1000& (0.1563, 0.4437) & 0\\
					&&0.20 &[0.01, 0.75] & 500&   (0.1477,   0.4590) & 0.008\\
					& & &  & 1000& (0.0225, 0.5966) & 0.002\\
					\hline\hline
					Design II & 0.1&0.10 &[0.01, 0.60] & 500&   (0.0827, 0.1332) & 0.002 \\
					& & &  & 1000& (0.0101, 0.3660) & 0\\
					& &0.15 &[0.01, 0.85] & 500&  (0.0100, 0.6095)  & 0.053\\
					& & &  & 1000& (0.0100, 0.6828) & 0.037\\
					& &0.20 &[0.01, 0.99] & 500&   (0.0100, 0.9091) & 0\\
					&& &  & 1000& (0.0100, 0.9501) & 0\\
					\hline
					& 0.3&0.10 &[0.01, 0.78] & 500&   (0.1196, 0.4958) &0.017 \\
					& & &  & 1000& (0.0167, 0.6172) & 0.004 \\
					&&0.15 &[0.01, 0.99] & 500&  (0.0100, 0.8685)  & 0\\
					& & & & 1000& (0.0100, 0.9104) & 0\\
					&&0.20 &[0.01, 0.99] & 500&   (0.0100, 0.9861) & 0 \\
					& &&  & 1000& (0.0100, 0.9897) & 0\\
					\hline
				\end{tabular}
			\end{center}
			
		\end{table}
		
		\section{Empirical Application}\label{sec:emp}
		
		This section applies our methods to the National Job Training Partnership (JTPA) study. We first estimate the average treatment effect for a range of population shares. Then we investigate the extent to which the treatment effect estimate based on a benchmark population share can be applicable to other populations, by constructing an anti-confidence set for the scope of external $\eta$-validity. 
		
		In the JTPA study, eligible program applicants were randomly assigned to a treatment group (which
		is allowed access to the program) and a control group (which is not allowed
		to the program) over the period of November 1987 through September 1989. The
		probability of being assigned to the treatment group was two thirds. Among the people assigned to the treatment group, about $60\%$ actually participated in the program.\footnote{Details
			about the design of the program can be found in \cite{Orr:94:JTPAStudy}.}
		Suppose that a researcher is interested in the nationwide population that consists of all the people eligible for the program, which corresponds to 
		economically disadvantaged adults or out-of-school youths.
		The population share of program participation, which is the
		ratio of those who actually participated the program over all the people eligible for the program,
		is typically unknown to the researcher.  Our outcome variable $Y$ is 30 month earnings and treatment status $D$ indicates whether the person participated in the training program. The covariate $X$ consists of indicators for high-school graduates or GED holders, African or Hispanic racial status and whether the age of the applicant is below 30. 
		\cite{Donald/Hsu/Lielie:14:JBES} found that Condition \ref{con:unconfounded} passed their test for the adult female subgroup when $(Y, D, X)$ are chosen as above. Therefore, our analysis focuses on this subgroup with $5732$ observations. We focus on the pure treatment-based sampling in which the sampling strata is based on $D$ (there is no $W$) and the unknown population share is program participation share.
		
		Table \ref{jtpa_ate} presents the point estimates for the average treatment effect $\tau_{ate}(p)$ for various values of the population share $p$. They turn out stable across different population shares, which indicates that the average treatment effect based on a particular population share $p^{\circ}$ can be applied to populations with other $p$'s, without causing much bias.  %From the perspective of  heterogeneous treatment effects,  the stability of estimated $\tau_{ate}(p)$ with respect to $p$ also reflects the relative constancy of the treatment effect across different groups.
		
		To substantiate this, we apply the anti-confidence set approach in Section \ref{sec: SEV} to gauge the scope of external $\eta$-validity to which the treatment effect estimate based on the benchmark population share $p^{\circ}$  is applicable. 
		We consider two values of $p^{\circ}$: $p^{\circ}=0.05$ representing a benchmark case where the population share of participation is small, and $p^{\circ}=0.30$ representing another case where the population share of participation is large. We consider a bunch of $\eta$'s so that one can see how the anti-confidence set $\mathcal{A}_n$ expands when $\eta$ increases. The choices of $B$ and $\beta$ are the same as the simulation exercise. %We apply the multiple testing approach  with the number of the bootstrap $B=200$ and the small significance level  $\beta=0.01$. 
		
\begin{figure}
			\centering
			\caption{The Anti-Confidence Set for the Scope of External $\eta$-Validity at Different values of $\eta$}
			\label{fig:JTPA}
			\vspace{0.5cm}
			\begin{tabular}{cc}
				(a) $p^{\circ}=0.05$ &(b)  $ p^{\circ}=0.30$\\
				\includegraphics[scale=0.35]{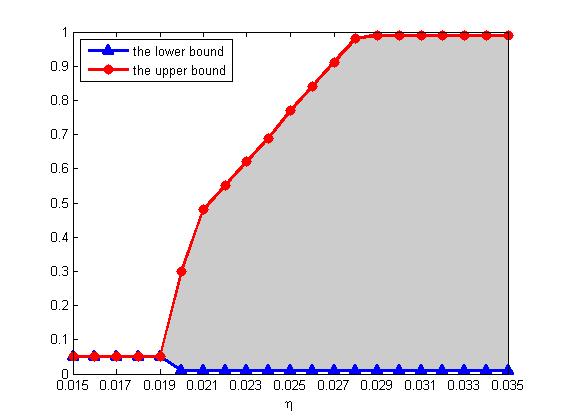} & 
				\includegraphics[scale=0.35]{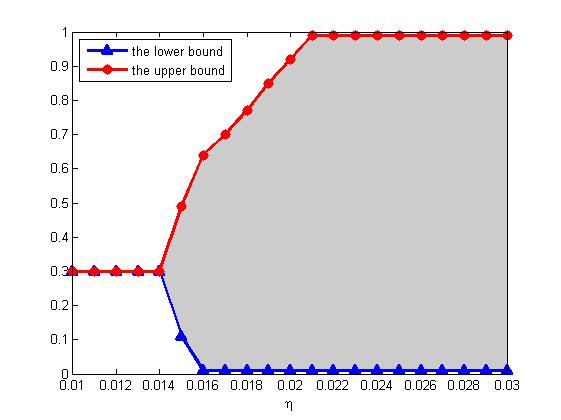}
			\end{tabular}%
		\end{figure}

		Figure \ref{fig:JTPA} illustrates how the anti-confidence set $\mathcal{A}_n$ expands with  $\eta$. Panel (a) depicts the case $p^{\circ}=0.05$ and panel (b) for $p^{\circ}=0.30$.  For $p^{\circ}=0.05$,  the lower and upper bounds of $\mathcal{A}_n$ are plotted for $\eta\in\{1.5\%,1.6\%,...,3.5\%\}$. For $p^{\circ}=0.30$,  the lower and upper bounds of $\mathcal{A}_n$ are plotted for $\eta\in\{1.0\%,1.1\%,...,3.0\%\}$. We can see that at tiny $\eta$'s ($\eta<2.0\%$ for $p^{\circ}=0.05$ or $\eta<1.5\%$ for $p^{\circ}=0.30$), the anti-confidence set shrinks to a singleton that only contains the benchmark population share. This reflects a finite-sample limitation in the power of our method.  However, the anti-confidence sets grow significantly when $\eta$ slightly increases. When $\eta\geq 2.9\%$ for $p^{\circ}=0.05$ or when $\eta\geq 2.1\%$ for $p^{\circ}=0.30$, the anti-confidence set expands to the full set $[0.01, 0.99]$. These findings suggest that in this example,  the average treatment effect estimate based on the benchmark population share applies to other populations with a broad range of the population shares. The wide scope of external validity may come from that the treatment effects are not very different across different strata.
		
		\begin{table}\caption{Point Average Treatment Effect Estimates for Various Population Shares $p$} \label{jtpa_ate}
			\small
			\begin{tabular}{c|cccccccc}
				\hline\hline
				$p$ & 0.05 & 0.10 & 0.15& 0.20 & 0.25& 0.50& 0.75& 0.90\\\hline
				$\hat\tau_{ate}(p)$ & 1888.0 & 1886.5 & 1885.0& 1883.5& 1882.0 & 1874.5 &1867.0 & 1862.5\\
				&(348.2)&(348.5) & (348.7) &(349.0)& (349.3)&(351.1)& (353.4)& (355.0)\\
				\hline
			\end{tabular}\\
			\vspace{1mm}
			\parbox{6.2in}{
				\footnotesize{Notes: All values are measured in 1990 U.S.dollar. The numbers in parentheses are standard errors.}
			}
		\end{table}

		\section{Asymptotic Theory}\label{sec:asym}
		In this section, we derive the limiting distribution theory for the \textit{stochastic processes} $\sqrt{n}(\hat \tau_{ate}(\cdot) - \tau_{ate}(\cdot))$ and $\sqrt{n}(\hat \tau_{tet}(\cdot) - \tau_{tet}(\cdot))$ on $A$. This result is crucial for showing the validity of our inference methods that were introduced in  Section \ref{sec:eff_est}. 
		We first establish the semiparametric efficiency bounds for $\tau_{ate}(\cdot )$ and  $\tau _{tet}(\cdot )$ on $A$. Then we show the week convergence of our estimators $\hat{\tau}_{ate}(\cdot )$ and $\hat{\tau}_{tet}(\cdot )$ given by (\ref{ate_est}) and (\ref{tet_est}) as well as their efficiency.
		
		\subsection{Semiparametric Efficiency Bound}
		%Under the assumptions of this paper, the treatment effect functions $\tau _{ate}(\cdot )$ and $\tau _{tet}(\cdot )$are $\sqrt{n}$-estimable infinite dimensional elements. 
		To avoid repetitive
		statements, we write $\tau (\cdot )$ to denote generically either $\tau
		_{ate}(\cdot )$ or $\tau _{tet}(\cdot )$. For any weakly regular estimator $%
		\hat{\tau}(\cdot )$ of $\tau (\cdot )$ (for the definition of weak regularity, see \cite{Bickel:93:EfficientAdaptive}), it is satisfied that%
		\begin{equation}\label{convolution_th}
			\sqrt{n}\{\hat{\tau}(\cdot )-\tau (\cdot )\}\rightsquigarrow G(\cdot
			)+\Delta (\cdot )\text{ in }l_{\infty }(A),
		\end{equation}%
		where $l_{\infty }(A)$ is the class of bounded real functions on $A,$ $%
		\rightsquigarrow $ represents weak convergence in the sense of
		Hoffman-Jorgensen, $G(\cdot )$ is a mean zero Gaussian process with
		continuous sample paths, and $\Delta (\cdot )$ is a random element that is
		independent of $G(\cdot )$. The limiting process $G(\cdot )$ is viewed as the semiparametric efficiency bound for $\tau (\cdot )$  while $\Delta (\cdot )$ is an independent noise component. An estimator $\hat{\tau}(\cdot )$ is said to be \textit{efficient} if the asymptotic distribution of $\sqrt{n}\{\hat{\tau}(\cdot )-\tau (\cdot )\}$ coincides with the distribution of $G(\cdot )$. The distribution of $G(\cdot )$ is fully characterized by its \emph{inverse information covariance kernel} denoted by $I^{-1}(p,\tilde{p})=\mathbf{E}[G(p)G(\tilde{p})], p,\tilde{p} \in A$. In order to establish the semiparametric efficiency bounds for $\tau_{ate}(\cdot )$ and 
		$\tau _{tet}(\cdot )$, we make the following additional
		assumptions. In this section,  the expectation $\mathbf{E}_p$ is under the probability associated with the population share $p$. 
		
		\begin{assumption}\label{con:bdY}
			$\mathbf{E}_{p}\mathbf{[}Y_{d}^{2}]<\infty $ for $(d,p)\in
			\{0,1\}\times A.$
		\end{assumption}
		\begin{assumption}\label{con:compact}
			$A$ is a compact set.
		\end{assumption}
		
		Let us introduce some notations.  
		Define $\beta _{d}(X)\equiv \mathbf{E}_p\left[ Y_{d}\mid X\right] $, $\tau
		(X)\equiv \mathbf{E}_p\left[ Y_{1}-Y_{0} \mid X\right] $, and 
		$e_{d}(p)\equiv (Y_{d}-\beta_{d}(X))/\bar p_{d}(X)$. For $(s,d)\in \{0,1\}\times \{0,1\}$, let
		$e_{s,d}(p)\equiv \bar p_{d}(X)(Y_{d}-\beta _{d}(X))/\bar p_{s}(X)$.
		Further define
		\begin{eqnarray*}
			R_{d,ate}(p)(X)&\equiv& t_{ate,p}(X)-\mathbf{%
				E}\left[t_{ate,p}(X)\mid (D,W) = (d,w)\right],\\
			R_{1,tet}(p)(X)& \equiv & t_{tet,p}(X)-\mathbf{E}%
			\left[t_{tet,p}(X)\mid (D,W) = (d,w)\right],
		\end{eqnarray*}
		where $t_{ate,p}(X)\equiv \tau (X)-\tau _{ate}(p)$ and $t_{tet,p}(X)\equiv \tau (X)-\tau _{tet}(p)$.
		We simply write $R_{d,ate}(p)=R_{d,ate}(p)(X)$ and $%
		R_{1,tet}(p)=R_{1,tet}(p)(X)$ below. 
		
		Theorem \ref{th:eff_bound} establishes the
		semiparametric efficiency bounds for $\tau _{ate}(\cdot )$ and $%
		\tau _{tet}(\cdot )$. The proof is given in Section S1 of the online supplement.
		
		\begin{theorem}\label{th:eff_bound}
			Under Conditions
			\ref{con:unconfounded} -- \ref{con:overlap} and Assumptions \ref{con:bdY} -- \ref{con:compact}, the following holds.
			
			\noindent \textit{(i) The inverse information covariance kernel for }$\tau
			_{ate}(\cdot )$ is equal to $I_{ate}^{-1}:A\times A\rightarrow \mathbf{R,%
			}$ \textit{defined as} 
			\begin{align*}
				I_{ate}^{-1}(p,\tilde{p}) &=\sum_{w}\left\{ \frac{p_{1,w}%
					\tilde{p}_{1,w}}{q_{1,w}}\mathbf{E}\left[ e_{1}(p)e_{1}(\tilde{p}%
				)+R_{1,ate}(p)R_{1,ate}(\tilde{p})\mid (D,W) = (1,w)\right] \right\} \\
				&+\sum_{w}\left\{ \frac{p_{0,w}\tilde{p}_{0,w}}{q_{0,w}}%
				\mathbf{E}\left[ e_{0}(p)e_{0}(\tilde{p})+R_{0,ate}(p)R_{0,ate}(\tilde{%
					p})\mid (D,W) = (0,w)\right] \right\}, \text{ } p,\tilde p \in A \text{.}
			\end{align*}%
			\textit{\noindent (ii) The inverse information covariance kernel for }$\tau
			_{tet}(\cdot )$ is equal to $I_{tet}^{-1}:A\times A\rightarrow \mathbf{R,
			}$ \textit{defined as}%
			\begin{align*}
				I_{tet}^{-1}(p,\tilde{p}) &=\frac{1}{p_{1}\tilde{p}_{1}}\sum_{w}\left\{ \frac{p_{1,w}\tilde{p}_{1,w}}{q_{1,w}}\mathbf{E}\left[ 
				e_{1,1}(p)e_{1,1}(\tilde{p})+R_{1,tet}(p)R_{1,tet}(\tilde{p})%
				\mid (D,W) = (1,w)\right] \right\} \\
				&+\frac{1}{p_{1}\tilde{p}_{1}}\sum_{w}\left\{ \frac{p_{0,w}%
					\tilde{p}_{0,w}}{q_{0,w}}\mathbf{E}\left[ e_{0,1}(p)e
				_{0,1}(\tilde{p})\mid (D,W) = (0,w)\right] \right\}, \text{ } p,\tilde p \in A.
			\end{align*}%
		\end{theorem}

		\subsection{Weak Convergence}
		In the following, we establish the asymptotic distributions of our estimators $\hat{\tau}_{ate}(\cdot )$ and $\hat{\tau}%
		_{tet}(\cdot )$ on $A$. %Let $\mathcal{W}$ be the support of $W_i$. 
		We make the following regularity assumptions.
		
		\begin{assumption}\label{ass:bd}
			For any $d$ and  $w$, the following conditions hold.
			
			\noindent (i)$\ f(v|d,w)$ and $\beta _{d}(v,w)$ are bounded and $%
			L_{1}+1$ times continuously differentiable in $v\ $with bounded derivatives
			on $\mathbf{R}^{L_{1}}$ and uniformly continuous $(L_{1}+1)$-th derivatives.
			
			\noindent (ii) $\sup_{v}\mathbf{E}\left[
			|Y_{1}|^{r}+|Y_{0}|^{r}\mid (D,V,W)=(d,v,w)\right] \mathbf{<\infty }$ and $\mathbf{E} \|V \mid (D,W)=(d,w)\|^{r}<\infty ,\ $for some $r\geq 4$.
			
			\noindent (iii) For some $\varepsilon >0$, $p_{d,w}>\varepsilon$ and $\min_{d,w}\inf_{v}f(v|d,w)>\varepsilon $.
			
			%\noindent (iv) For some $\bar{a}\geq 4,$ it is satisfied that for all $%
			%(d_{1},d_{2},w)\in \{0,1\}^{2}\times \mathcal{W}$, 
			%\begin{equation*}
			%\sup_{a\in \lbrack 0,\bar{a}]}\mathbf{E}_{d_{1},w}\left[ \sup_{p\in A}\left( 
			%\frac{f(V_{i}|d_{2},w)p_{d_{2},w}}{\sum_{d_{3}\in
					%\{0,1\}}f(V_{i}|d_{3},w)p_{d_{3},w}}\right) ^{a}\right] <%\infty .
			%\end{equation*}
			%[\textbf{KEVIN: Is not the term in the parenthesis above bonded by 1?}]
		\end{assumption}
		
		Assumption \ref{ass:bd}(i) and (ii) are regularity conditions. Assumption \ref{ass:bd}(iii) is introduced to deal with the boundary problem of kernel estimators.
		In general, the performance of kernel estimators is unstable near the
		boundary of the support of $V_{i}$. In this case, it is reasonable to trim
		part of the samples such that the realizations of $V_{i}$ appear to be
		``outliers''. For example, see \cite{Heckman/Ichimura/Todd:97:ReStud} for
		application of such trimming schemes.
		
		\begin{assumption}\label{ass:kernel}
			(i) The kernel function $K$ equals to zero outside an interior of a
			bounded set, $L_{1}+1$ times continuously differentiable with bounded
			derivatives, $\int K(s)ds=1,$ $\int
			s_{1}^{l_{1}}...s_{d_{1}}^{l_{d_{1}}}K(s)ds=0$ and $\int
			|s_{1}^{l_{1}}...s_{d_{1}}^{l_{d_{1}}}K(s)|ds<\infty $ for all nonnegative
			integers $l_{1},...,l_{d_{1}}$ such that $l_{1}+...+l_{d_{1}}\leq L_{1}$,
			where $d_{1}$ denotes the dimension of $V_{i}$.
			
			\noindent (ii)\ $n^{-1/4}h^{-d_{1}/2}\sqrt{\log n}+n^{1/2}h^{L_{1}+1}%
			\rightarrow 0,$ as $n\rightarrow \infty .$
			
			\noindent (iii)The trimming sequence $\delta _{n}$ in (\ref{til_ind}) satisfies
			that $\sqrt{n}\delta _{n}^{\gamma}\rightarrow 0$, for some $\gamma >0$.
		\end{assumption}
		
		Assumption \ref{ass:kernel}(i) is a standard assumption for higher order kernels.
		Assumptions \ref{ass:kernel}(ii) and (iii) present the conditions for the bandwidth and the trimming sequence. The condition for the trimming sequence is very weak; it requires only that it decrease at a certain polynomial rate in $n$. 
		
		Theorem \ref{th:weak convergence} establishes the asymptotic distributions of our estimators $\hat{\tau}_{ate}(\cdot)$ and $\hat{\tau}_{tet}(\cdot) $ and verifies their semiparametric efficiency. Its proof is provided in Section S3 of the online supplement.
		
		\begin{theorem}\label{th:weak convergence}
			Suppose that Conditions
			\ref{con:unconfounded} -- \ref{con:overlap} and Assumptions \ref{con:bdY} -- \ref{ass:kernel} hold. Then
			\begin{align*}
				\sqrt{n}(\hat{\tau}_{ate}(\cdot )-\tau _{ate}(\cdot ))
				\rightsquigarrow \zeta _{ate}^*(\cdot )\quad \text{and}\quad
				\sqrt{n}(\hat{\tau}_{tet}(\cdot )-\tau _{tet}(\cdot ))
				\rightsquigarrow \zeta_{tet}^*(\cdot ),
			\end{align*}%
			where $\zeta _{ate}^*$ and $\zeta _{tet}^*$ 
			are mean zero Gaussian processes with continuous sample paths that
			have covariance kernels $I_{ate}^{-1}(\cdot ,\cdot )$ and $%
			I_{tet}^{-1}(\cdot ,\cdot )$ given in Theorem \ref{th:eff_bound}.
		\end{theorem}
		
		Theorem \ref{th:weak convergence} implies the validity of the robust   confidence sets $%
		\mathcal{C}_{ate}$ and $\mathcal{C}_{tet}$ introduced in Section \ref{sec:rcs}. It is also crucial for justifying the anti-confidence set of the scope of external validity in Section \ref{sec: SEV}.

		\subsection{Optimal Treatment-Based Sampling}\label{sec:opt_sampling}
		As a by-product of Theorem \ref{th:eff_bound}, given a population share $p$ we can find out the optimal sampling design $\{q_{d,w}\}$ which minimizes the semiparametric efficiency bound for the treatment effect parameters $\tau_{ate}(p)$ and $\tau_{tet}(p)$. Corollary \ref{corr:opt_q} gives the design shares  that respectively minimize $I_{ate}^{-1}(p,p)$ and $I_{tet}^{-1}(p,p)$ characterized by Theorem \ref{th:eff_bound}. %Let $R_{d,ate}(p), R_{1,tet}(p)$ and $\sigma^2_d(X)$ be as defined above Theorem {\ref{th:eff_bound}.
			\begin{corollary}\label{corr:opt_q}
				Suppose that Conditions
				\ref{con:unconfounded} -- \ref{con:overlap} and Assumptions \ref{con:bdY} -- \ref{con:compact} hold. Then the optimal choice of $q_{d,w}$, denoted as $q_{d,w}^{ate}$ for $\tau
				_{ate}(p)$ and $q_{d,w}^{tet}$ for $\tau
				_{tet}(p)$, is given
				as follows:
				\begin{equation*}
					q_{d,w}^{ate}=\sqrt{J_{d,w}^{ate}}/\sum_{d,w}\sqrt{J_{d,w}^{ate}}~ , ~ and~~ q_{d,w}^{tet}=\sqrt{J_{d,w}^{tet}}/
					\sum_{d,w}\sqrt{J_{d,w}^{tet}},
				\end{equation*}%
				where
				\begin{align*}
					J_{d,w}^{ate} &=  p_{d,w}^{2}%
					\mathbf{E}\left[ e _{d}^{2}(p)%
					+R_{d,ate}^{2}(p)\mid (D,W)=(d,w)\right], \\
					J_{d,w}^{tet} &= p_{d,w}^{2}\mathbf{E}\left[ \frac{d}{%
						p_{1}^{2}}\left\{ e _{1,1}^{2}(p)+R_{1,tet}^{2}(p)\right\} +%
					\frac{1-d}{p_{1}^{2}}e _{0,1}^{2}(p)
					\mid (D,W)=(d,w)\right] .
				\end{align*}
			\end{corollary}
			The optimal sampling design leads to the most accurate estimator among all the efficient estimators of the treatment effects across sampling designs. Corollary  \ref{corr:opt_q} suggests that we sample from the $(d,w)$-subsample
			according to the ``noise'' proportion $(J^{ate}_{d,w})^{1/2}$  of the subsample $(d,w)$ in
			$\sum_{(d,w)\in \{0,1\}\times \mathcal{W}}\sqrt{J^{ate}_{d,w}}
			$.
			%When we have some pilot sample obtained from a two-stage sampling scheme or other data sources that can be used to draw information about $J_{d,w}$, the result here may serve as a guide for optimally choosing the size of the design shares $\{q_{d,w}\}$.
			In the case of pure treatment-based sampling, we can make precise the
			condition for treatment-based sampling to lead to inference of better
			quality than random sampling. Let $V_{TS}$ be the variance bound for $\tau_{ate}(p_1)$ when there is no $W$, and recall the definition of $J_{d}$ in Corollary \ref{corr:opt_q}, so that
			$V_{TS}= J_{1}/q_{1} + J_{0}/(1-q_{1})$. 
			Let $V_{RS}$ be the variance bound for $\tau
			_{ate}(p_1)\ $under random sampling, which is equal to $V_{TS}$ with $q_1=p_{1}$. Therefore,
			$
			V_{RS}= J_{1}/p_{1} + J_{0}/(1-p_{1}). 
			$
			Then it is not hard to see that $V_{RS}\geq V_{TS}$ if and only if
			\begin{equation}\label{range2}
				\min \left\{ p_{1},\frac{(1-p_1)J_{1}}{(1-p_1)J_{1}+p_1J_{0}}\right\} \leq q_{1}\leq \max \left\{ p_{1},\frac{(1-p_1)J_{1}}{(1-p_1)J_{1}+p_1J_{0}}\right\} .  
			\end{equation}%
			Therefore, it is not always true that sampling more from a subsample of low
			population proportion leads to a better result. The improvement happens when the design share $q_1$ lies between the population share $p_{1}$ and the value $(1-p_1) J_{1}/((1-p_1)J_{1}+p_1J_{0})$. 
			Treatment-based sampling is able to improve upon random sampling so long as the design share $q_1$ satisfies (\ref{range2}). In practice, the accuracy of the treatment effect estimate is not the only consideration one makes in designing the sampling process in the program evaluation. 
			Nevertheless, the optimal sampling design share can be a useful guidance as a benchmark design probability. 
			
			For our JTPA  application in Section \ref{sec:emp}, Table \ref{table:jtpa_range_q} presents the range of the design share $q$ that improves the estimation of $\tau_{ate}(p)$ (relative to the random sampling) and the optimal design share  $q$ that minimizes the asymptotic variance of $\tau_{ate}(p)$. The estimators of $J_1$ and $J_0$ are given in the Appendix.
			In this example the optimal design share is close to $0.5$ and is stable across different population shares, because the estimates of $J_1$ and $J_0$ are approximately equal.  In addition, when the population share is small, there is a large room for improvement in the efficiency bound through a choice of the design share.
			
			\begin{table}\caption{The Range of Design Share $q$ in the  Treatment-based Sampling That Improves the Estimation of $\tau_{ate}(p)$, JTPA data in Section \ref{sec:emp}.} \label{table:jtpa_range_q}
				\small
				\begin{center}
					\begin{tabular}{c|cccc}
						\hline\hline
						$p$ & 0.05 & 0.20 & 0.35& 0.50 \\\hline
						Efficiency-improving $q$ & [0.05, 0.951] & [0.20, 0.803] & [0.30, 0.703] &[0.50, 0.502]\\
						\hline
						Optimal $q$ & 0.5027 &   0.5021  &  0.5018  &  0.5010\\
						\hline
					\end{tabular}
				\end{center}
			\end{table}
			
			%Secondly, one may check whether how far the sampling design share $q_{d,w}$ used in the study is from the optimal sampling design, by comparing $q_{d,w}$ and an estimate of $q_{d,w}^{ate}$ in Corollary \ref{corr:opt_q}.
			
			%\begin{align*}
			%\liminf_{n\rightarrow\infty}P\left\{\mathcal{A}_n \subset A(p^\circ;\eta)\right\}\geq 1-\alpha.
			%\end{align*}
			
			\section{Conclusion}
			This paper establishes identification results for treatment effect parameters when the exact population share is unknown. We propose efficient estimators for treatment parameters that are functions of the population share vector, and construct confidence sets for treatment effects that are robust against a range of population shares. Furthermore, we develop a inference procedure for the scope of external $\eta$-validity, a set of population shares to which a benchmark treatment effect estimate can be applied. In addition, we investigate the optimal design of the treatment-based sampling.  In an empirical application, we find that the estimate of the JTPA program's impact on the earnings of adult women can be applied to populations with a broad range of program participation shares.

			\section*{Acknowledgment}
			We thank Petra Todd who gave numerous valuable comments and advice at an early stage of this research, and Sokbae Lee for useful
			comments, and Jinyong Hahn for pointing out errors in a manuscript that preceded this paper. All errors are ours. Song acknowledges the support from Social Sciences and Humanities Research Council of Canada. Yu acknowledges the support of JSPS KAKENHI Grant Number 19K13666 and 21K01419.
			
			% references
			
			\bibliographystyle{chicago}
			\bibliography{treatment_based_sampling}

			% appendix (put online supplement in a separate file)
			
			\section*{Appendix: Estimators of Variances and Covariances}\label{sec: est_V}
			\renewcommand{\theequation}{A.\arabic{equation}}
			\renewcommand{\thesection}{A}
			\renewcommand{\thelemma}{A.\arabic{lemma}}
			\setcounter{equation}{0}
			Consistent estimation of $\sigma _{ate}^{2}(p)$ can be proceeded as follows.
			First, recall $\tilde{p}_{d,i}(X_{i})$  and  $\tilde{\lambda}_{d,i}(X_{i})$ in (\ref{ps2}) and let%
			\begin{align*}
				\tilde{\beta}_{d,p}(X_{i})= \frac{\tilde{\mu}_{d}(X_{i})}{\tilde{f}%
					(X_{i})}\text{ and }\tilde{e}_{d,i}(p)= \frac{Y_{i}-\tilde{%
						\beta}_{d}(X_{i})}{\tilde{p}_{d,i}(X_{i})}\text{, }d\in \{0,1\}%
				\text{,}
			\end{align*}%
			where $\tilde{f}(X_{i})= \tilde{\lambda}_{0,i}(X_{i})+\tilde{\lambda}%
			_{1,i}(X_{i})$, and 
			\begin{align*}
				\tilde{\mu}_{d}(X_{i}) &=\frac{1}{\tilde{p}_{d,i}(X_{i})}\frac{p_{d,w}}{%
					n_{d,w}}\sum_{i: (D_i, W_i)=(d,w)}Y_{i}K_{h,i}(X_{i}).
			\end{align*}%
			We also define%
			\begin{align*}
				\tilde{R}_{d,ate,i}(p) &= \frac{\tilde{%
						\mu}_{1}(X_{i})-\tilde{\mu}_{0}(X_{i})}{\tilde{f}(X_{i})}-\hat{\tau}%
				_{ate}(p)
				-\frac{1}{n_{d,w}}\sum_{i: (D_i, W_i)=(d,w)}\left[ \frac{\tilde{%
						\mu}_{1}(X_{i})-\tilde{\mu}_{0}(X_{i})}{\tilde{f}(X_{i})}-\hat{\tau}%
				_{ate}(p)\right] .
			\end{align*}%
			Then we construct
			\begin{align*}
				\hat \sigma_{ate}(p,\tilde p) &= \sum_{w} \left\{\frac{%
					p_{1,w} \tilde p_{1,w}}{q_{1,w}n_{1,w}}\sum_{i: (D_i, W_i)=(1,w)}\left[ \tilde{e}_{1,i}(p)\tilde{e}_{1,i}(\tilde p)+%
				\tilde{R}_{1,ate,i}(p)\tilde{R}_{1,ate,i}(\tilde p)\right] \right\} \\
				&\quad +\sum_{w} \left\{\frac{p_{0,w} \tilde p_{0,w}}{q_{0,w}n_{0,w}}%
				\sum_{i: (D_i, W_i)=(0,w)}\left[ \tilde{e}_{0,i}(p)\tilde{e}_{0,i}(\tilde p)+\tilde{R}%
				_{0,ate,i}(p)\tilde{R}%
				_{0,ate,i}(\tilde p)\right] \right\}.
			\end{align*}
			In particular, when $p = \tilde p$, we simply write
			$
			\hat{\sigma}_{ate}^{2}(p) =  \hat \sigma(p,p).
			$
			Thus, we construct
			\begin{align}
				\label{Sigma p}
				\hat \Sigma_{ate}(p,\tilde p) = 
				\begin{bmatrix}
					\hat{\sigma}_{ate}^{2}(p) & \hat{\sigma}_{ate}(p,\tilde p)\\
					\hat{\sigma}_{ate}(p,\tilde p) & \hat{\sigma}_{ate}^{2}(\tilde p)
				\end{bmatrix}.
			\end{align}
			We also estimate $J^{ate}_{d,w} $, $d\in\{0,1\}$ in Corollary \ref{corr:opt_q} by
			\begin{align*}
				\hat J^{ate}_{d,w} = \frac{%
					p^2_{d,w}}{n_{d,w}}\sum_{i: (D_i, W_i)=(d,w)}\left[ \tilde{e}_{d,i}^2(p)+%
				\tilde{R}_{d,ate,i}^2(p)\right].
			\end{align*}
			
			Let us turn to estimation of the asymptotic variance of $
			\hat{\tau}_{tet}(p)$. To estimate $\sigma _{tet}^{2}(p)$, let%
			\begin{align*}
				\tilde{R}_{1,tet,i}(p) &=\frac{
					\tilde{\mu}_{1}(X_{i})-\tilde{\mu}_{0}(X_{i})}{\tilde{f}(X_{i})}-\hat{\tau}%
				_{tet}(p)
				-\frac{1}{n_{1,w}}\sum_{i: (D_i, W_i)=(1,w)}\left[ \frac{%
					\tilde{\mu}_{1}(X_{i})-\tilde{\mu}_{0}(X_{i})}{\tilde{f}(X_{i})}-\hat{\tau}%
				_{tet}(p)\right],
			\end{align*}%
			and $\tilde{e}_{s,d,i}(p)= \tilde{p}_{d,i}(X_{i})(Y_{i}-\tilde{\beta}_{d}(X_{i}))/\tilde{p}_{s,i}(X_{i})$.
			The asymptotic variance estimator we propose is:
			\begin{align*}
				\hat{\sigma}_{tet}(p,\tilde p) &=\frac{1}{p_{1} \tilde p_1}\sum_{w}\left\{ \frac{p_{1,w} \tilde p_{1,w} }{q_{1,w}n_{1,w}}\sum_{i: (D_i, W_i)=(1,w)}\left[ \tilde{e}_{1,1,i}(p)\tilde{e}_{1,1,i}(\tilde p)+\tilde{R}_{1,ate,i}(p)\tilde{R}_{1,ate,i}(\tilde p)\right] \right\} \\
				&\quad +\frac{1}{p_{1} \tilde p_1}\sum_{w}\left\{ \frac{p_{0,w} \tilde p_{0,w}}{%
					q_{0,w}n_{0,w}}\sum_{i: (D_i, W_i)=(0,w)} \tilde{e}_{0,1,i}(p)\tilde{e}_{0,1,i}(\tilde p)
				\right\} .
			\end{align*}
			Using $\sigma^2_{tet}(p)$ and $\sigma_{tet}(p,\tilde p)$, we can construct $\hat \Sigma_{ate}(p,\tilde p)$ similarly as in (\ref{Sigma p}).
			
			\pagebreak
			
			\begin{center}
				\Large \textsc{Supplemental Note to ``Estimation and Inference on Treatment Effects  Under Treatment-Based Sampling Designs''}
				\bigskip
			\end{center}
			
			\date{%
				%TCIMACRO{\TeXButton{Today}{\today}}%
				%BeginExpansion
				\today%
				%EndExpansion
			}
			
			\newtheorem{proposition}{Proposition}
			\newtheorem{remark}{Remark}

			\renewcommand{\thesection}{\arabic{section}}
			\renewcommand{\theequation}{\arabic{section}.\arabic{equation}}
			\renewcommand{\thetheorem}{\arabic{section}.\arabic{theorem}}
			\renewcommand{\theassumption}{\arabic{section}.\arabic{assumption}}
			\renewcommand{\theproposition}{\arabic{section}.\arabic{proposition}}
			\renewcommand{\thecorollary}{\arabic{section}.\arabic{corollary}}
			\renewcommand{\thelemma}{\arabic{section}.\arabic{lemma}}
			\renewcommand{\theexample}{\arabic{section}.\arabic{example}}
			\renewcommand{\theremark}{\arabic{section}.\arabic{remark}}
			\renewcommand{\thetable}{S\arabic{table}}
			\renewcommand{\thefigure}{S\arabic{figure}}
			
			% put S in front of counters for online supplement
			\renewcommand{\theequation}{S.\arabic{equation}}
			\renewcommand{\thelemma}{S\arabic{lemma}}
			\renewcommand{\thesection}{S\arabic{section}}
			\renewcommand{\thepage}{S\arabic{page}}
			
			\setcounter{equation}{0}
			\setcounter{page}{1}
			\setcounter{section}{0}
			\setcounter{subsection}{0}
			\setcounter{equation}{0}
			\setcounter{lemma}{0}

		\vspace*{3ex minus 1ex}
         \begin{center}
            	Kyungchul Song and Zhengfei Yu\\
       	\textit{University of British Columbia and University of Tsukuba}\\
     	\medskip
\end{center}
    
    The supplemental note collects auxillary results and proofs for  \cite{Song/Yu:20:WP}. %Section \ref{supsec:MC} collects additional simulation results that complement Section \ref{sec: MC_point_est}.
    Section \ref{sec: semiparametric efficiency} computes the semiparametric efficiency bounds (for Theorem \ref{th:eff_bound}).  Section \ref{supsec:fwsr} presents a proof for Theorem \ref{thm:step_down}. Section \ref{supsec:weakcon} contains a proof for Theorem \ref{th:weak convergence}. Section \ref{supsec:auxiliary} contains auxiliary results that are used in the proof of Theorem \ref{th:weak convergence}.
    Let us clarify the expectation notations used in the supplement. The notation of expectation, $\mathbf{E}$, without a subscript, is assumed to be under $P$, the target population. Expectation $\mathbf{E}_{Q}$ denotes expectation under $Q$, the design probability. The shorthand notation $\mathbf{E}_{d,w}\equiv\mathbf{E}\left[\cdot \mid (D,W)=(d,w)\right]$. %Expectation $\mathbf{E}$ depends on the population shares $p_{d,w}$. By writing $\mathbf{E}_{p}$, we sometimes make its dependence on $p_{d,w}$ explicit. 
    We use $\mathbf{E}_{p}$ to denote the expectation for the population that is associated with a generic share vector $p$.

    \section{Semiparametric Efficiency Bounds and Proofs}
    \label{sec: semiparametric efficiency}
    Suppose that $\mathbf{P}$ is a model (a collection of probability measures $%
    P $ having a density function with respect to a common $\sigma $-finite
    measure $\mu $). After identifying each probability in $\mathbf{P}$ as the
    square root of its density, we view $\mathbf{P}$ as a subset of $L_{2}(\mu )$%
    . Let $\mathcal{C}_{b}(A)$ be the collection of bounded and continuous real
    functions defined on $A\subset \mathbf{R}^{2\times |\mathcal{W}|}$ and $%
    ||\cdot ||$ be the supremum norm on $\mathcal{C}_{b}(A)$. The following
    definitions are from \cite{Bickel:93:EfficientAdaptive} (BKRW from here on).

    \bigskip

    \noindent \textsc{Definition B1 [Curve]:} $\mathbf{V}$ is a \emph{curve} in $%
    L_{2}(\mu )$ if it can be represented as the image of the open interval $%
    (-1,1)$ under a continuously Fr\'{e}chet differentiable map. That is, we can
    write
    \begin{align*}
    	\mathbf{V=\{v}(t)\in L_{2}(\mu ):|t|<1\mathbf{\},}
    \end{align*}%
    where there exists a $\mathbf{\dot{v}}\in L_{2}(\mu )$ such that $\mathbf{v}%
    (t+\Delta )=$ $\mathbf{v}(t)+\Delta \mathbf{\dot{v}}(t)+o(|\Delta |)$, as $%
    |\Delta |\rightarrow 0$, for each $t\in (-1,1).$\bigskip
    
    \noindent \textsc{Definition B2 [Tangent Set]:} The \emph{tangent set} at $%
    \mathbf{v}_{0}\in \mathbf{P}$, denoted as $\mathbf{\dot{P}}^{0}$, is the
    union of all $\mathbf{\dot{v}}$ of curves $\mathbf{V}$ $\subset $ $\mathbf{P}
    $ passing through $\mathbf{v}_{0},$ where $\mathbf{v}_{0}=\mathbf{v}(0)$.
    The closed linear span of $\mathbf{\dot{P}}^{0}$ is the \emph{tangent space}%
    , denoted as $\mathbf{\dot{P}.}$\bigskip
    
    \noindent \textsc{Definition B3 [Pathwise Differentiability]:} A parameter $%
    \tau :$ $\mathbf{P\rightarrow }$ $\mathcal{C}_{b}(A)$ is \emph{pathwise
    	differentiable} at $\mathbf{v}_{0}$ if there exists a bounded linear
    function $\dot{\tau}(\mathbf{v}_{0})(\cdot )= \dot{\tau}(\cdot ):%
    \mathbf{\dot{P}}\rightarrow \mathcal{C}_{b}(A)$ such that for any curve $%
    \mathbf{V}\subset $ $\mathbf{P}$ with tangent $s\in \mathbf{\dot{P}}^{0}$,
    we have 
    \begin{align*}
    	\left\Vert \frac{\tau (\mathbf{v}(t))-\tau (\mathbf{v}_{0})}{t}-\dot{\tau}%
    	(s)\right\Vert =o(1),
    \end{align*}%
    as $t\rightarrow 0$.\bigskip
    
    \noindent \textsc{Proof of Theorem \ref{th:eff_bound}:} Let $f(y,v,d,w)$ be the density of $%
    (Y,V,D,W)$ with respect to a $\sigma $-finite measure $\mu $ under $P\in 
    \mathcal{P}$, where $\mathcal{P}$ is the collection of potential
    distributions for $(Y,V,D,W)$. Let $f(y,v|d,w)$ be the conditional density
    of $(Y,V)$ given $(D,W)=(d,w),$ and $\mathbf{P}_{d,w}$ denotes the
    collection of conditional densities $f(\cdot ,\cdot |d,w)$ of $(Y,V)$ given $%
    (D,W)=(d,w)$ with $P$ running in $\mathcal{P}$. Let $\mathbf{Q}=
    \{f_{Y,V|D,W}(\cdot |\cdot )q_{d.w}:f_{Y,V|D,W}\in \mathbf{P}_{d,w},(d,w)\in
    \{0,1\}\times \mathcal{W}\}$. Let $\mathbf{v}_{0}\in \mathbf{Q}$ be the true
    density and $Q$ the associated probability measure. We use subscript $Q$ for
    densities and expectations associated with $\mathbf{v}_{0}$. This subscript
    is not needed for the conditional densities (and conditional expectations)
    given $(D,W)=(d,w)$ or given $(D,W,V)=(d,w,v)$ because they remain the same
    both under $P$ and under $Q.$ Use notations $\int \cdot d\mu (w),$ $\int
    \cdot d\mu (v)$, $\int \cdot d\mu (y)$, etc., to denote the integrations
    with respect to the marginals of $\mu $ for the coordinates of $w,v,y$, etc.
    
    Since $A$ is compact, the space $(\mathcal{C}_{b}(A),||\cdot ||)$ equipped
    with the supremum norm $||\cdot ||$ is a Banach space. With a slight abuse of
    notation, we view the treatment effect parameters $\tau _{ate}(\cdot)$ and $\tau
    _{tet}(\cdot)$ as maps from $\mathbf{Q}$ into$\ \mathcal{C}_{b}(A),$\textbf{\ }so
    that, for example, $\tau _{ate}(\mathbf{v}),$ $\mathbf{v\in Q}$, is an
    element in $\mathcal{C}_{b}(A)$ but $\tau _{ate}(\mathbf{v})(p)\in \mathbf{R}
    $.\bigskip
    
    \noindent (i) First consider the semiparametric efficiency bound for $\tau
    _{ate}(\cdot ).$ The proof is composed of three steps:\medskip
    
    \noindent \emph{Step 1. Calculate the tangent space. }Following \cite{Hahn:98:Eca},
    under Condition \ref{con:unconfounded} we write the density $f(y,v,d,w)\ $as%
    \begin{align*}
    	f(y,v,d,w)=\left[ f_{1}(y|v,w)p(v,w)\right] ^{d}\left[ f_{0}(y|v,w)\left(
    	1-p(v,w)\right) \right] ^{1-d}f(v,w),
    \end{align*}%
    where%
    \begin{align*}
    	f_{1}(y|v,w) &=f(y|1,v,w), \\
    	f_{0}(y|v,w) &=f(y|0,v,w),\text{ }p(v,w)= P\left\{
    	D=1|V=v,W=w\right\} ,
    \end{align*}%
    and $f(y|d,v,w)$ denotes the conditional density of $Y$ given $%
    (D,V,W)=(d,v,w)$. Consider a curve $\mathbf{v}(t)$ identified with $%
    f^{t}(y,v,d,w)$ ($|t|$ $<1$), we have%
    \begin{align}
    	f^{t}(y,v,d,w)=\left[ f_{1}^{t}(y|v,w)p^{t}(v,w)\right] ^{d}\left[
    	f_{0}^{t}(y|v,w)\left( 1-p^{t}(v,w)\right) \right] ^{1-d}f^{t}(v,w),
    	\label{submodel}
    \end{align}%
    such that $f^{0}(y,v,d,w)=f(y,v,d,w)$. Since $%
    f_{Q}(y,v,d,w)=f(y,v,d,w)q_{d,w}/p_{d,w}$, the density under $Q$, $%
    f_{Q}(y,v,d,w)$ can be written as 
    \begin{align*}
    	f_{Q}(y,v,d,w)=\left[ f_{1}(y|v,w)p(v,w)\right] ^{d}\left[
    	f_{0}(y|v,w)\left( 1-p(v,w)\right) \right] ^{1-d}f(v,w)q_{d,w}/p_{d,w},
    \end{align*}%
    and consider a curve $Q^{t}$ identified with $f^{t}(y,v,d,w)q_{d,w}/p_{d,w}$. The score of the above curve is
    \begin{align*}
    	s^{t}(y,v,d,w)&=ds_{1}^{t}(y|v,w)+(1-d)s_{0}^{t}(y|v,w)\\
    	& \quad +\frac{\partial
    		p^{t}(v,w)/\partial t}{p^{t}(v,w)\left( 1-p^{t}(v,w)\right) }\left[ d-p(v,w)%
    	\right] +s^{t}(v,w),
    \end{align*}%
    where $s_{1}^{t}(y|v,w)$, $s_{0}^{t}(y|v,w)$ and $s^{t}(v,w)$ are the scores
    of $f_{1}^{t}(y|v,w),f_{0}^{t}(y|v,w)$ and $f^{t}(v,w)$ respectively. Also
    let $s(y,v,d,w)= s^{0}(y,v,d,w)$ (the score evaluated at the $t=0$).
    Now we can calculate the \emph{tangent set} at $\mathbf{v}_{0}\in \mathbf{Q}$
    as
    \begin{align*}
    	\mathbf{\dot{Q}}^{0}\mathcal{=}\left\{ 
    	\begin{array}{c}
    		dh_{1}(y|v,w)+(1-d)h_{0}(y|v,w)+a(v,w)(d-p(v,w))+h(v,w) \\ 
    		:h_{1},h_{0},a,h\in L_{2}(Q),\int h_{1}(y|v,w)f_{1}(y|v,w)=0, \\ 
    		\int h_{0}(y|v,w)f_{0}(y|v,w)=0,\text{ and}\int h(v,w)f(v,w)=0%
    	\end{array}%
    	\right\} ,
    \end{align*}%
    where we recall that $Q$ in $L_{2}(Q)$ is the probability measure associated
    with $\mathbf{v}_{0}$. Observe that $\mathbf{\dot{Q}}^{0}$ is linear and
    closed, so it is the \emph{tangent space} which we denote by $\mathbf{\dot{Q}%
    }$.\medskip
    
    \noindent \emph{Step 2. Prove the pathwise differentiability of }$\tau
    _{ate} $ \emph{and compute its derivative. }As for the pathwise
    differentiability, for given $\mathbf{v}_{0}\in \mathbf{Q}$, let $\mathbf{V}%
    \subset \mathbf{Q}$ be a curve passing through $\mathbf{v}_{0}$,
    parametrized by $t\in (-1,1)$. Then the weighted average treatment effect
    under a point in this curve $\mathbf{v}(t)$, say, $\tau _{ate}(\mathbf{v}%
    (t)) $ at $p\in A$ is written as%
    \begin{align*}
    	&\sum_{w}\int \int y\left\{
    	f^{t}(y|v,1,w)-f^{t}(y|v,0,w)\right\} d\mu (y)f^{t}(v,w)d\mu (v)\\
    	&=\sum_{d,w}\int \left\{ \int
    	yf_{1}^{t}(y|v,w)d\mu (y)-\int yf_{0}^{t}(y|v,w)d\mu (y)\right\}
    	p_{d,w}f^{t}(v|d,w)d\mu (v),
    \end{align*}%
    for $p\in A$. The first order derivative of $\tau _{ate}(\mathbf{v}(t))(p)$
    with respect to $t$ at $t=0$ is equal to 
    \begin{align*}
    	\mathbf{E}_{p}\left[ \mathbf{E}%
    	\left[ Ys_{1}(Y|X)|X\right] -\mathbf{E}\left[ Ys_{0}(Y|X)|X\right]\right] 
    	+\mathbf{E}_{p}[s(V|D,W)\{\tau (X)-\tau
    	_{ate}(p)\}],
    \end{align*}%
    where $\tau (X)= \mathbf{E}_{p}\left[ Y_{1}-Y_{0}|X\right] .$ Let 
    \begin{align*}
    	\dot{\psi}_{ate,P}(y,v,d,w)=\frac{%
    		d(y-\beta _{1}(v,w))}{p_{1}(v,w)}-\frac{(1-d)(y-\beta _{0}(v,w))}{p_{0}(v,w)} +R_{d,ate}(v,w).
    \end{align*}%
    (Recall $R_{d,ate}(p)(v,w)= t_{ate,p}(v,w)-\mathbf{E}%
    _{d,w}[t_{ate,p}(X)]$.)\ We can write%
    \begin{align*}
    	\frac{\partial \tau _{ate}(\mathbf{v}(t))(p)}{\partial t}=\sum_{(d,w)\in
    		\{0,1\}\times \mathcal{W}}\mathbf{E}_{d,w}\left[ \dot{\psi}%
    	_{ate,P}(Y,V,D,W)s(Y,V,D,W)\right] p_{d,w}\mathbf{.}
    \end{align*}%
    Define $\dot{\psi}_{ate,Q}(y,v,d,w)(p)=\dot{\psi}%
    _{ate,P}(y,v,d,w)p_{d,w}/q_{d,w}$ and rewrite%
    \begin{align}
    	\frac{\partial \tau _{ate}(\mathbf{v}(t))(p)}{\partial t}=\mathbf{E}_{Q}%
    	\left[ \dot{\psi}_{ate,Q}(Y,V,D,W)(p)s(Y,V,D,W)\right] \mathbf{.}
    	\label{derv5}
    \end{align}%
    Define an operator $\dot{\tau}_{ate}:\mathbf{\dot{Q}}\mathcal{%
    	\longrightarrow }$ $\mathcal{C}_{b}(A)$ as 
    \begin{align*}
    	\dot{\tau}_{ate}(s)(p)= \mathbf{E}_{Q}\left[ \dot{\psi}%
    	_{ate,Q}(Y,V,D,W)(p)s(Y,V,D,W)\right] \text{, }s\in \mathbf{\dot{Q}}\text{,\ 
    	}p\in A\text{.}
    \end{align*}%
    Since (\ref{derv5}) holds for all $p\in A$ and $\dot{\psi}%
    _{ate,Q}(Y,V,D,W)(p)$ is continuous in $p$ on the compact set $A$, we have:%
    \begin{align}
    	\sup_{p\in A}\left\vert \tau _{ate}(\mathbf{v}(t))(p)-\tau _{ate}(\mathbf{v}%
    	_{0})(p)-t\dot{\tau}_{ate}(s)(p)\right\vert =o(t),\text{ as }t\rightarrow 0,
    	\label{pathdif}
    \end{align}%
    \newline
    for all curves $f_{Q}^{t}(y,v,d,w)=f_{Q}(y,v,d,w)+ts(y,v,d,w)+o(t).$ Under
    Conditions C2-C4, 
    \begin{align*}
    	\sup_{p\in A}\mathbf{E}_{Q}[\dot{\psi}_{ate,Q}^{2}(Y,V,D,W)(p)]<\infty .
    \end{align*}%
    Then there exists a finite $M_{1}$ such that 
    \begin{align*}
    	\sup_{p\in A}\mathbf{E}_{Q}\left[ \dot{\psi}_{ate,Q}(Y,V,D,W)(p)s(Y,V,D,W)%
    	\right] \leq M_{1}\sqrt{\mathbf{E}_{Q}\left[ s^{2}(Y,V,D,W)\right] },
    \end{align*}%
    for all $s\in \mathbf{\dot{Q}}$, which implies that $\dot{\tau}_{ate}$ is
    bounded. Also obviously $\dot{\tau}_{ate}$ is linear. Therefore $\tau _{ate}$%
    \textbf{\ }is pathwise differentiable at $\mathbf{v}_{0}$ with derivative $%
    \dot{\tau}_{ate}.$\medskip
    
    \noindent \emph{Step 3. Calculate the efficient influence function, inverse
    	information covariance functional and the semiparametric efficiency bound. }%
    For a generic element $b^*\in \left( \mathcal{C}_{b}(A)\right) ^{\ast
    } $ (the dual space of $\mathcal{C}_{b}(A)$), we have 
    \begin{align*}
    	b^*\dot{\tau}_{ate}(s)=\mathbf{E}_{Q}\left[ \left( b^*\dot{\psi}%
    	_{ate,Q}(Y,V,D,W)\right) s(Y,V,D,W)\right] .
    \end{align*}%
    Notice that $\dot{\psi}_{ate,Q}\in \mathbf{\dot{Q}}$, so the linearity of
    expectation and the dual operator $b^*$ lead to $b^*\dot{\psi}%
    _{ate,Q}\in \mathbf{\dot{Q}}$. Then the projection of $b^*\dot{\psi}%
    _{ate,Q}$ onto $\mathbf{\dot{Q}}$ is itself and we obtain the efficient
    influence operator (see BKRW p.178 for its definition) of $\tau _{ate}$ as $%
    \tilde{I}_{ate}:\left( \mathcal{C}_{b}(A)\right) ^*\longrightarrow $\ $%
    \mathbf{\dot{Q},}$ where $\tilde{I}_{ate}(b^*)=b^*\dot{\psi}%
    _{ate,Q}$. In particular, for the evaluation map $b^*=\pi _{p}$
    defined by $\pi _{p}(b)= b(p)$ for all $b\in \mathcal{C}_{b}(A)$, the
    efficient influence operator becomes $\tilde{I}_{ate}(\pi _{p})=\dot{\psi}%
    _{ate,Q}(\cdot ,\cdot ,\cdot ,\cdot )(p).$ Following BKRW p.184, the inverse
    information covariance functional for $\tau _{ate}$, $I_{ate}^{-1}:A\times
    A\longrightarrow \mathbf{R}$ is given by%
    \begin{align}
    	I_{ate}^{-1}(p,\tilde{p})=\mathbf{E}_{Q}[\dot{\psi}_{ate,Q}(Y,V,D,W)(p)\dot{%
    		\psi}_{ate,Q}(Y,V,D,W)(\tilde{p})].  \label{I_wate}
    \end{align}%
    By Theorem 5.2 BKRW(Convolution Theorem), an efficient weakly regular
    estimator $\hat{\tau}_{ate}$ of $\tau _{ate}$ weakly converges to a mean
    zero Gaussian process $\zeta ^*(\cdot )$ with the inverse information
    covariance functional $I_{ate}^{-1}(p,\tilde{p})$ characterized by (\ref%
    {I_wate}). As a special case, the variance bound for any weakly regular
    estimator of the real parameter $\tau _{ate}(\mathbf{v}_{0})(p)$ can be
    written as:%
    \begin{align*}
    	\sum_{(d,w)\in \mathcal{\{}0,1\mathcal{\}}\times \mathcal{W}}\mathbf{E}%
    	_{d,w}[\dot{\psi}_{ate,Q}^{2}(Y,V,D,W)]q_{d,w}=\sum_{(d,w)\in \mathcal{\{}0,1%
    		\mathcal{\}}\times \mathcal{W}}\frac{p_{d,w}^{2}}{q_{d,w}}\mathbf{E}_{d,w}[%
    	\dot{\psi}_{ate,P}^{2}(Y,V,D,W)].
    \end{align*}%
    
    \noindent (ii) Let us turn to the semiparametric efficiency bound for $\tau
    _{tet}(\cdot)$. The tangent space remains the same as that in (i). To establish
    the semiparametric efficiency bound, the only needed change is the
    computation of the efficient influence operator. Similarly as before, for
    given $\mathbf{v}_{0}\in \mathbf{Q}$, let $\mathbf{V}\subset \mathbf{Q}$ be
    a curve passing through $\mathbf{v}_{0}$, parametrized by $t\in (-1,1)$. The
    weighted average treatment effect on the treated under a point in this curve 
    $\mathbf{v}(t)$, say, $\tau _{tet}(\mathbf{v}(t))$ at $p\in A$ is written as%
    \begin{align*}
    	\tau _{tet}(\mathbf{v}(t))(p)=\sum_{w}\int \int
    	y\left\{ f^{t}(y|v,w,1)-f^{t}(y|v,w,0)\right\} d\mu
    	(y)f^{t}(v|w,1)p_{w|1}d\mu (v),
    \end{align*}%
    where $p_{w|1}=p_{1,w}/\{\Sigma _{w}p_{1,w}\}.$ The first
    order derivative of $\tau _{tet}(\mathbf{v}(t))(p)$ with respect to $t$ is
    equal to%
    \begin{align*}
    	& \mathbf{E}_{1,p}\left[ \mathbf{E}\left[ Ys_{1}(Y|X)|X,D=1\right] -%
    	\mathbf{E}\left[ Ys_{0}(Y|X)|X,D=0\right] \right] \\
    	& \quad +\mathbf{E}_{1,p}\left[ s(V|D,W)\{\tau (X)-\tau _{tet}\}\right].
    \end{align*}%
    We take 
    \begin{align*}
    	\dot{\psi}_{tet,P}(y,v,d,w) &=
    	d(y-\beta _{1}(v,w))/p_{1} -p_{1}(v,w)(1-d)(y-\beta _{0}(v,w))/\{p_{0}(v,w)p_{1}\}%
    	\\
    	&\quad -dR_{1,tet}(p)(v,w)/p_{1}.
    \end{align*}%
    (Recall $R_{1,tet}(p)(v,w)= t_{tet,p}(v,w)-\mathbf{E}%
    _{1,w}[t_{tet,p}(X)]$.) The remainder of the proof follows the argument in
    the proof of (i): we construct
    \begin{align*}
    	\dot{\psi}_{tet,Q}(y,v,d,w)(p)= \dot{
    		\psi}_{tet,P}(y,v,d,w)p_{d,w}/q_{d,w}.
    \end{align*}
    Under Conditions \ref{con:unconfounded} to \ref{con:compact}, we can
    verify the pathwise differentiability of $\tau _{tet}:\mathbf{Q}%
    \longrightarrow \mathcal{C}_{b}(A)$ at $\mathbf{v}_{0}$. Write the efficient
    influence operator as $\tilde{I}_{tet}(b^*)=b^*\dot{\psi}%
    _{tet,Q} $ and compute the inverse information covariance functional as%
    \begin{align*}
    	I_{tet}^{-1}(p,\tilde{p})=\mathbf{E}_{Q}[\dot{\psi}_{tet,Q}(Y,V,D,W)(p)\dot{%
    		\psi}_{tet,Q}(Y,V,D,W)(\tilde{p})].
    \end{align*}
    
    Let us turn to the situation with pure treatment-based sampling, where
    parameter $\tau _{tet}(p)$ does not depend on $p$. Thus for each $\mathbf{v}%
    \in \mathbf{Q}$, $\tau _{tet}(\mathbf{v})$ is a constant real map on $A$. We
    simply write $\tau _{tet}$ suppressing the argument $p$. In this special
    case of pure treatment-based sampling, the functional $I_{tet}^{-1}(p,\tilde{%
    	p})$ no longer depends on $(p,\tilde{p})$. In particular, write 
    \begin{align*}
    	\tau _{tet}(\mathbf{v}(t))(p)=\int \int y\left\{
    	f^{t}(y|x,1)-f^{t}(y|x,0)\right\} d\mu (y)f^{t}(x|1)d\mu (x).
    \end{align*}%
    The first order derivative of $\tau _{tet}^{t}(p)$ with respect to $t$ is
    equal to 
    \begin{align*}
    	&\mathbf{E}_{1}\left[\mathbf{E}\left[ Ys_{1}(Y|X)|X,D=1\right]-
    	\mathbf{E}\left[ Ys_{0}(Y|X)|X,D=0\right] \right]  +\mathbf{E}_{1}\left[ s(X|D)\{\tau (X)-\tau _{tet}(p)\}\right].
    \end{align*}%
    Therefore, we take%
    \begin{align*}
    	\dot{\psi}_{tet,P}(y,x,d)= \frac{d(y-\beta
    		_{1}(x)-\{\tau (x)-\tau _{tet}\})}{p_{1}}-\frac{p_{1}(x)(1-d)(y-\beta
    		_{0}(x))}{p_{0}(x)p_{1}},
    \end{align*}%
    because $\mathbf{E}_{1,p}\left[ \tau (X)-\tau _{tet}(p)\right] =0.$ Let $%
    \dot{\psi}_{tet,Q}(y,x,d)(p)= \dot{\psi}_{tet,P}(y,x,d)p_{d}/q_{d}.$
    Now the inverse information covariance functional becomes%
    \begin{align}
    	I_{tet}^{-1}(p,\tilde{p}) &=\sum_{d\in \{0,1\}}q_{d}\mathbf{E}_{d}\left[ 
    	\dot{\psi}_{tet,Q}(Y,X,D)(p)\dot{\psi}_{tet,Q}(Y,X,D)(\tilde{p})\right]
    	\label{I} \\
    	&=\frac{1}{q_{1}}\mathbf{E}_{1}\left[(Y_{1}-\beta _{1}(X)-\{\tau (X)-\tau _{tet}\})^{2}\right]  \notag  +\frac{1}{q_{0}}\frac{p_{0}\tilde{p}_{0}}{p_{1}%
    		\tilde{p}_{1}}\mathbf{E}_{0}\left[ \frac{p_{1}(X)\tilde{p}%
    		_{1}(X)(Y_{0}-\beta _{0}(X))^{2}}{p_{0}(X)\tilde{p}_{0}(X)}\right] .
    	\notag
    \end{align}%
    Note that by Bayes' rule,%
    \begin{align*}
    	\frac{p_{0}p_{1}(X)}{p_{1}p_{0}(X)}=\frac{p_{0}f(X|1)p_{1}}{p_{0}f(X|0)p_{1}}%
    	=\frac{f(X|1)}{f(X|0)}=\frac{\tilde{p}_{0}f(X|1)\tilde{p}_{1}}{\tilde{p}%
    		_{0}f(X|0)\tilde{p}_{1}}=\frac{\tilde{p}_{0}\tilde{p}_{1}(X)}{p_{1}\tilde{p}%
    		_{0}(X)}.
    \end{align*}%
    We rewrite the last term in (\ref{I}) as%
    \begin{align*}
    	\frac{1}{q_{0}}\mathbf{E}_{0}\left[ \frac{f^{2}(X|1)%
    	}{f^{2}(X|0)}(Y_{0}-\beta _{0}(X))^{2}\right] .
    \end{align*}%
    Thus the semiparametric efficiency bound does not depend on $p=
    \{p_{d}\}$. $\mathbf{\blacksquare }$\bigskip
    \section{Proof of Familywise Error Rate Control}\label{supsec:fwsr}
    \noindent \textsc{Proof of Theorem \ref{thm:step_down}:} 
    Choose any $S \subset A$ such that $\eta |\tau_{ate}(p^\circ)| \le |\tau_{ate}(p)-\tau_{ate}(p^\circ)|$ for all $p \in S$. Write
    \begin{align*}
    	\sup_{p \in S} \hat Q(p) \le \sup_{p \in K_0} \hat Q(p) + \sup_{p \in S\backslash K_0} \hat Q(p). 
    \end{align*}
    The last term vanishes as $n \rightarrow \infty$, by the definition of $\hat Q$ and $K_0$. Thus we see that for each $t \in \mathbf{R}$,
    \begin{align*}
    	\limsup_{n \rightarrow \infty} P\left\{\sup_{p \in S} \hat Q(p) > t \right\} \le
    	\limsup_{n \rightarrow \infty} P\left\{\sup_{p \in K_0} \xi(p) > t \right\},
    \end{align*}
    by Theorem \ref{th:weak convergence}, the continuous mapping theorem, and the Delta method. Hence, by the condition of $\hat c_{1-\alpha}(S)$ that $\hat c_{1-\alpha}(S) = c_{1-\alpha}(S) +o_P(1)$, we have
    \begin{align*}
    	\limsup_{n \rightarrow \infty} P\left\{\sup_{p \in S} \hat Q(p) > \hat c_{1-\alpha}(S)  \right\} \le \alpha.
    \end{align*}
    Furthermore, $\hat c_{1-\alpha}(S)$ is increasing in the set $S$. By Theorem 2.1 of \cite{Romano/Shaikh:10:Eca}, we obtain the desired result. $\blacksquare$
    
    \section{Efficient Estimation and Proofs}\label{supsec:weakcon}
    
    For the proof of Theorem \ref{th:weak convergence}, we first establish the asymptotic linear
    representations for $\hat{\tau}_{ate}(\cdot )$, and $\hat{\tau}%
    _{tet}(\cdot )$. We introduce some notations. 
    %(Throughout this supplemental note, we suppress $p$ in $\mathbf{E}_{p}$ and $\mathbf{E}_{d,p}$ from the notation and write $\mathbf{E}$ and $\mathbf{E}_{d}$ simply.)
    First, define mean-deviated quantities:%
    \begin{align}
    	\xi _{d,ate}(V_{i},w) &= \tau (V_{i},w)-\tau _{ate}(p)%
    	-\mathbf{E}_{d,w}\left[\left( \tau (V_{i},w)-\tau
    	_{ate}(p)\right) \right] ,  \label{def4} \\
    	\xi _{1,tet}(V_{i},w) &= \tau (V_{i},w)-\tau _{tet}(p)%
    	-\mathbf{E}_{1,w}\left[ \left( \tau (V_{i},w)-\tau
    	_{tet}(p)\right) \right] ,  \notag
    \end{align}%
    where $
    \tau (X)= \mathbf{E}\left[ Y_{1}|X\right] -\mathbf{E}\left[ Y_{0}|X%
    \right]$.
    Also, define
    $\varepsilon _{d,w,i}= Y_{di}-\beta _{d}(V_{i},w)$.
    
    Lemma A1 below establishes the asymptotic linear representations for $%
    \hat{\tau}_{ate}(\cdot )$ and  $\hat{\tau}_{tet}(\cdot )$. For that purpose, we define%
    \begin{align}
    	Z_{i}(p) &=\sum_{w}\left\{ \frac{%
    		L_{1,w,i}(p)\varepsilon _{1,w,i}}{p_{1}(V_{i},w)}-\frac{%
    		L_{0,w,i}(p)\varepsilon _{0,w,i}}{p_{0,w}(V_{i},w)}\right\}\label{Zi} \\
    	&\quad +\sum_{w}\left( \xi
    	_{1,ate}(V_{i},w)L_{1,w,i}(p)+\xi _{0,ate}(V_{i},w)L_{0,w,i}(p)\right), \text{ and }
    	\notag\\
    	\tilde{Z}_{i}(p) &=\sum_{w\in \mathcal{%
    			W}}\left\{ L_{1,w,i}(p)\varepsilon _{1,w,i}-\frac{%
    		L_{0,w,i}(p)p_{1}(V_{i},w)\varepsilon _{0,w,i}}{p_{0,w}(V_{i},w)}%
    	\right\}+\sum_{w}\xi_{1,tet}(V_{i},w)L_{1,w,i}(p),\label{Z2i} 
    \end{align}%
    where $L_{d,w,i}(p)= (p_{d,w}/q_{d,w}) 1\{\left( D_{i},W_{i}\right)
    =\left( d,w\right) \}$. From here on, we suppress the argument notation and
    write $L_{d,w,i}(p)$ simply as $L_{d,w,i}$.\medskip
    
    \noindent \textsc{Lemma A1}\textbf{: }\textit{Suppose that Condition \ref{con:unconfounded} 
    	and Assumptions \ref{ass:bd} and \ref{ass:kernel} } \textit{hold}. \textit{Then\ uniformly over\ }$p\in A,$%
    \textit{\ }%
    \begin{align}
    	\sqrt{n}\left( \hat{\tau}_{ate}(p)-\tau _{ate}(p)\right) &=\frac{1}{%
    		\sqrt{n}}\sum_{i=1}^{n}Z_{i}(p)+o_{P}(1),\mathit{\ }\text{\textit{and}}
    	\label{convergences} \\
    	\sqrt{n}\left( \hat{\tau}_{tet}(p)-\tau _{tet}(p)\right) &=\frac{1}{%
    		\sqrt{n}}\sum_{i=1}^{n}\tilde{Z}_{i}(p)+o_{P}(1).  \notag
    \end{align}%

    The proof of Lemma A1 is given in Section \ref{supsec:auxiliary} of this note.
    
    \medskip
    
    \noindent \textsc{Proof of Theorem \ref{th:weak convergence}: }We focus on $\hat{\tau}_{ate}(\cdot )$ only. The proof for the case of $\hat{\tau}_{tet}(\cdot )$ is similar. By Lemma A1, it suffices to prove that 
    \begin{align}
    	\frac{1}{\sqrt{n}}\sum_{i=1}^{n}Z_{i}(\cdot )\rightsquigarrow \zeta
    	_{ate}^*(\cdot ).  \label{proc}
    \end{align}%
    Since $\mathbf{E}_{Q}[Z_{i}(p)]=0$ and $\mathbf{E}_{Q}[Z_{i}^{2}(p)]<\infty $
    for all $p$, for every finite subset $\{p_{1},\ldots ,p_{K}\}\subset A,$ the
    Central Limit Theorem yields that $(Z_{i}(p_{1}),\ldots ,Z_{i}(p_{K}))$
    converges in distribution to a normal distribution with mean zero and
    covariance matrix $\Sigma = \lbrack \sigma _{kl}],$ where%
    \begin{align*}
    	\sigma _{kl}=\sum_{d,w}\frac{%
    		p_{k,d,w}p_{l,d,w}}{q_{d,w}}\mathbf{E}_{d,w}\left[
    	e_{d}(p_{k})e_{d}(p_{l})+R_{d,ate}(p_{k})R_{d,ate}(p_{l})\right] .
    \end{align*}
    
    Now we verify the stochastic equicontinuity of the process $(1/\sqrt{n}%
    )\sum_{i=1}^{n}Z_{i}(\cdot )$. Note that $Z_{i}(p)$ is differentiable with
    respect to $p$. By the mean-value theorem, 
    \begin{align*}
    	\left\vert \frac{1}{\sqrt{n}}\sum_{i=1}^{n}Z_{i}(p)-\frac{1}{\sqrt{n}}%
    	\sum_{i=1}^{n}Z_{i}(\tilde{p})\right\vert \leq \left( \sup_{p\in
    		A}\sum_{d,w}\left\vert \frac{1}{\sqrt{n}}%
    	\sum_{i=1}^{n}\frac{\partial Z_{i}(p)}{\partial p_{d,w}}\right\vert \right)
    	||p-\tilde{p}||,
    \end{align*}%
    for any pair of $p,\tilde{p}\in A$. Therefore, the stochastic equicontinuity
    follows once we show that%
    \begin{align}
    	\sup_{p\in A}\sum_{d,w}\left\vert \frac{1}{%
    		\sqrt{n}}\sum_{i=1}^{n}\frac{\partial Z_{i}(p)}{\partial p_{d,w}}\right\vert
    	=O_{p}(1).  \label{cv3}
    \end{align}%
    (See e.g., Theorem 21.10 of \cite{Davidson:94:StochasticLimitTheory}, p.339).
    It suffices to show that 
    \begin{align}
    	\frac{1}{\sqrt{n}}\sum_{i=1}^{n}Q_{i,d,w}(\cdot ),\text{ where } Q_{i,d,w}(p)=
    	\partial Z_{i}(p)/\partial p_{d,w}  \label{two_procs}
    \end{align}%
    weakly converge in $l_{\infty }(A)$. This can be shown by establishing the
    convergence of the finite dimensional distributions using the Central Limit
    Theorem, and stochastic equicontinuity of the processes which follows by
    showing the first order derivatives of the the process in (\ref{two_procs}%
    ) are stochastically bounded uniformly over $p\in A$. Details are omitted. $%
    \mathbf{\blacksquare }$
    
    \section{Further Auxiliary Results}\label{supsec:auxiliary}
    
    This section presents the proof of Lemma A1. We begin with
    Lemmas B1, B2 and B3 that will be used in the proof. 
    First introduce some definitions: for $d=0,1,$ 
    \begin{align*}
    	\hat{p}_{d,i}(V_{i},w)= \frac{\hat{\lambda}_{d,i}(V_{i},w)}{\hat{\lambda%
    		}_{1,i}(v,w)+\hat{\lambda}_{0,i}(V_{i},w)},
    \end{align*}%
    where $\hat{\lambda}_{d,i}(V_{i},w)= \frac{1}{n}\sum_{j=1,j\neq
    	i}^{n}L_{d,w,j}K_{h}\left( V_{1j}-V_{1i}\right) 1\{V_{2j}=V_{2i}\}$. Also,
    define 
    \begin{align*}
    	\hat{1}_{n,i}= 1\left\{ \hat{\lambda}_{1,i}(V_{i},w)\wedge \hat{\lambda}%
    	_{0,i}(V_{i},w)\geq \delta _{n}:d\in \{0,1\}\right\} ,
    \end{align*}%
    where $\delta _{n}$ is a sequence that appears in Assumption
    \ref{ass:kernel}(iii). In addition, let $L_{w,i}=L_{1,w,i}+L_{0,w,i}$.
    \medskip
    
    \noindent \textsc{Lemma B1:} \textit{Suppose that} \emph{Assumptions 1-4} 
    \textit{hold.} $\tilde{p}%
    _{1,i}(V_{i},w)$ \textit{is defined below (\ref{ps2}) of \cite{Song/Yu:20:WP}. Then, for each} $w$, \textit{uniformly over }$%
    p\in A,$%
    \begin{align}
    	\max_{1\leq i\leq n}\hat{1}_{n,i}\left\vert p_{1}(V_{i},w)-\hat{p}%
    	_{1,i}(V_{i},w)\right\vert &=O_{P}(\varepsilon _{n})\text{ \textit{and}} 
    	\notag \\
    	\max_{1\leq i\leq n}\tilde{1}_{n,i}\left\vert p_{1}(V_{i},w)-\tilde{p}%
    	_{1,i}(V_{i},w)\right\vert &=O_{P}(\varepsilon _{n}),  \label{LemmaA1}
    \end{align}%
    \textit{where }$\varepsilon _{n}=n^{-1/2}h^{-d_{1}/2}\sqrt{\log n}%
    +h^{L_{1}+1}$.\medskip
    
    \noindent \textsc{Proof:} Consider the first statement. For simplicity, we
    assume that $V=V_{1}$ and define $\mathbf{E}_{Q,w,i}[L_{1,w,i}]=\mathbf{E}%
    _{Q}[L_{1,w,i}|V_{i},W_{i}=w]$ and $\mathbf{E}_{Q,w,i}[L_{w,i}]=\mathbf{E}%
    _{Q}[L_{w,i}|V_{i},W_{i}=w]$. Recall that $q_{1}(v,w)$ is the propensity
    score under $Q$, i.e., $q_{1}(v,w)=Q\left\{
    D_{i}=1|(V_{i},W_{i})=(v,w)\right\} $. By Bayes' rule,
    \begin{align}
    	f(V_{i}|1,w)=q_{1,w}(V_{i})f_{Q}(V_{i})/q_{1,w}=q_{1}(V_{i},w)q_{w}(V_{i})f_{Q}(V_{i})/q_{1,w},
    	\label{eq0}
    \end{align}%
    where $q_{1,w}(V_{i})=\mathbf{E}_{Q}[1\{(D_{i},W_{i})=(d,w)\}|V_{i}],$ $%
    q_{w}(V_{i})=\mathbf{E}_{Q}[1\{W_{i}=w\}|V_{i}]$ and $f_{Q}(\cdot )$ is the
    density of $V_{i}$ under $Q$. Hence%
    \begin{align}\label{eq1}
    	p_{1}(V_{i},w) &=\frac{f(V_{i}|1,w)p_{1,w}}{%
    		f(V_{i}|1,w)p_{1,w}+f(V_{i}|0,w)p_{0,w}}   =\frac{(q_{1}(V_{i},w)/q_{1,w})p_{1,w}}{\sum_{d\in \{0,1\}%
    		}(q_{d}(V_{i},w)/q_{d,w})p_{d,w}}=\frac{\mathbf{E}_{Q,w,i}[L_{1,w,i}]}{%
    		\mathbf{E}_{Q,w,i}[L_{w,i}]}.  
    \end{align}%
    Let $K_{ji}=K_{h}\left( V_{1j}-V_{1i}\right) $ for brevity. Also let%
    \begin{align*}
    	\mathbf{\hat{E}}_{Q,w,i}[L_{1,w,i}] =\frac{\frac{1}{n-1}\sum_{j=1,j\neq
    			i}^{n}L_{1,w,j}K_{ji}}{\frac{1}{n-1}\sum_{j=1,j\neq i}^{n}1\left\{
    		W_{j}=w\right\} K_{ji}}\text{~and~} 
    	\mathbf{\hat{E}}_{Q,w,i}[L_{w,i}] =\frac{\frac{1}{n-1}\sum_{j=1,j\neq
    			i}^{n}L_{w,j}K_{ji}}{\frac{1}{n-1}\sum_{j=1,j\neq i}^{n}1\left\{
    		W_{j}=w\right\} K_{ji}}.
    \end{align*}%
    By applying Theorem 6 of \cite{Hansen:08:ET}, we find that uniformly over $i\in
    \{1,...,n\},$%
    \begin{align}\label{rac}
    	\mathbf{E}_{Q,w,i}[L_{1,w,i}]-\mathbf{\hat{E}}_{Q,w,i}[L_{1,w,i}]
    	=O_{P}(\varepsilon _{n}), \text{~and~} 
    	\mathbf{E}_{Q,w,i}[L_{w,i}]-\mathbf{\hat{E}}_{Q,w,i}[L_{w,i}]
    	=O_{P}(\varepsilon _{n}). 
    \end{align}%
    Furthermore, (\ref{rac}) holds uniformly for all $p\in A$, because\textbf{\ }%
    \begin{align*}
    	\mathbf{E}_{Q,w,i}[L_{1,w,i}]-\mathbf{\hat{E}}_{Q,w,i}[L_{1,w,i}]
    	=-\frac{p_{1,w}}{q_{1,w}}\left\{ 
    	\begin{array}{c}
    		\sum_{j=1,j\neq i}^{n}\mathbf{1}\{(D_{i},W_{i})=(1,w)\}K_{ji}/\sum_{j=1,j%
    			\neq i}^{n}1\{W_{j}=w\}K_{ji} \\ 
    		-\mathbf{E}_{Q,w,i}[\mathbf{1}\{(D_{i},W_{i})=(1,w)\}]%
    	\end{array}%
    	\right\} .
    \end{align*}%
    The term in the bracket is $O_{P}(\varepsilon _{n})$\ by Theorem 6 of Hansen
    (2008), and this convergence is uniformly for all $p$\ since it does not
    depend on $p.$ Observe that%
    \begin{align}
    	&\hat{1}_{n,i}\left[ p_{1}(V_{i},w)-\hat{p}_{1,i}(V_{i},w)\right]  \notag \\
    	&=\hat{1}_{n,i}\frac{\mathbf{E}_{Q,w,i}[L_{1,w,i}]-\mathbf{\hat{E}}%
    		_{Q,w,i}[L_{1,w,i}]}{\mathbf{E}_{Q,w,i}[L_{w,i}]}  \notag \\
    	&\quad +\hat{1}_{n,i}\frac{\mathbf{\hat{E}}_{Q,w,i}[L_{1,w,i}]\left\{ \mathbf{%
    			\hat{E}}_{Q,w,i}[L_{w,i}]-\mathbf{E}_{Q,w,i}[L_{w,i}]\right\} }{\left( 
    		\mathbf{E}_{Q,w,i}[L_{w,i}]\right) ^{2}}+o_{P}(\varepsilon _{n}).
    	\label{decomp21}
    \end{align}%
    Using Bayes' rule, we deduce that%
    \begin{align*}
    	\mathbf{E}_{Q,w,i}[L_{w,i}] &=\frac{p_{1,w}}{q_{1,w}}P_{Q}\left\{
    	D_{i}=1|V_{i},W_{i}=w\right\} +\frac{p_{0,w}}{q_{0,w}}P_{Q}\left\{
    	D_{i}=0|V_{i},W_{i}=w\right\} \\
    	&=\frac{p_{1,w}}{q_{1,w}}\frac{f_{Q}(V_{i}|w,1)q_{1,w}}{f_{Q}(V_{i},w)}+%
    	\frac{p_{0,w}}{q_{0,w}}\frac{f_{Q}(V_{i}|w,0)q_{0,w}}{f_{Q}(V_{i},w)} \\
    	&=\frac{p_{1,w}f(V_{i}|w,1)}{f_{Q}(V_{i},w)}+\frac{%
    		p_{0,w}f(V_{i}|w,0)}{f_{Q}(V_{i},w)} \\
    	&=\frac{f(V_{i},w)}{f_{Q}(V_{i},w)}=\frac{f(V_{i},w)}{%
    		q_{1,w}f(V_{i}|w,1)+q_{0,w}f(V_{i}|w,0)}.
    \end{align*}%
    Therefore,%
    \begin{align} 
    	\label{dec5}
    	&\mathbf{E}_{Q}\left[\sup_{p\in A}\left( \mathbf{E}_{Q,w,i}\left[
    	L_{w,i}\right] \right) ^{-a}\right] \\  \notag
    	&=\mathbf{E}_{Q}\left[\sup_{p\in A}\left( \frac{%
    		q_{1,w}f(V_{i}|w,1)+q_{0,w}f(V_{i}|w,0)}{f(V_{i},w)}\right) ^{a}%
    	\right]  \\ \notag
    	&\leq 2^{a-1}\sum_{d,w}q_{d,w}\left\{ \mathbf{E}_{d,w}\left[\sup
    	_{p\in A}\left( \frac{f(V_{i}|w,1)}{f(V_{i},w)}\right) ^{a}\right] 
    	+ \mathbf{E}_{d,w}\left[ \sup_{p\in A}\left( \frac{f(V_{i}|w,0)}{%
    		f(V_{i},w)}\right) ^{a}\right] \right\} <\infty, 
    \end{align}%
    for $a \geq 1$. The last inequality
    comes from Assumption \ref{ass:bd} (i) and (iii). Combining this with (\ref{rac}) and (\ref%
    {decomp21}), we have%
    \begin{align*}
    	\hat{1}_{n,i}\left\{ p_{1}(V_{i},w)-\hat{p}_{1,i}(V_{i},w)\right\}
    	=O_{P}(\varepsilon _{n}),
    \end{align*}%
    uniformly over $p\in A$ and over $1\leq i\leq n.$ Hence we obtain the first
    statement of (\ref{LemmaA1}).
    
    For the second statement of (\ref{LemmaA1}), let 
    \begin{align*}
    	\mathbf{\hat{E}}_{Q,w,i}[\hat{L}_{1,w,i}] &=\frac{\frac{1}{n-1}%
    		\sum_{j=1,j\neq i}^{n}\hat{L}_{1,w,j}K_{ji}}{\frac{1}{n-1}\sum_{j=1,j\neq
    			i}^{n}1\left\{ W_{j}=w\right\} K_{ji}}\text{ and} \\
    	\mathbf{\hat{E}}_{Q,w,i}[\hat{L}_{w,i}] &=\frac{\frac{1}{n-1}%
    		\sum_{j=1,j\neq i}^{n}\hat{L}_{w,j}K_{ji}}{\frac{1}{n-1}\sum_{j=1,j\neq
    			i}^{n}1\left\{ W_{j}=w\right\} K_{ji}}.
    \end{align*}%
    Observe that 
    \begin{align*}
    	\left\vert \mathbf{\hat{E}}_{Q,w,i}[\hat{L}_{1,w,i}]-\mathbf{\hat{E}}%
    	_{Q,w,i}[L_{1,w,i}]\right\vert \leq \left\vert \frac{p_{d,w}}{q_{d,w}}-\frac{%
    		p_{d,w}}{\hat{q}_{d,w}}\right\vert \cdot \left\vert \frac{\sum_{j\in
    			S_{d,w}\backslash \{i\}}K_{ji}}{\sum_{j\in \mathcal{S}_{w}\backslash
    			\{i\}}K_{ji}}\right\vert =o_{P}(\varepsilon _{n}).
    \end{align*}%
    Hence the argument in the proof of first statement can be applied to prove
    the second statement of (\ref{LemmaA1}). $\mathbf{\blacksquare }$\medskip
    
    \noindent \textsc{Lemma B2 :} \textit{Suppose that} $S_{i}=\varphi
    (Y_{i},X_{i},D_{i})$\textit{, for a given real-valued map }$\varphi $ 
    \textit{such that for each }$w$, 
    \begin{equation*}
    	\sup_{v\in 
    		\mathcal{V}(w)}\mathbf{E}_{Q}\left[ |S_{i}|^{2}|(V_{i},W_{i})=(v,w)\right]
    	<\infty
    \end{equation*}
    \textit{and} $\mathbf{E}_{Q}[S_{i}|V_{1i}=\cdot ,W=w]$ \textit{is%
    } $L_{1}+1$ \textit{times continuously differentiable with bounded
    	derivatives and uniformly continuous }$(L_{1}+1)$-\textit{th derivatives}.
    
    \noindent (i) \textit{Suppose that Condition \ref{con:unconfounded}, Assumptions \ref{ass:bd} and \ref{ass:kernel}} 
    \textit{hold}. \textit{Then}, \textit{for} $d=0,1,$%
    \begin{align*}
    	&\frac{1}{n}\sum_{i=1}^{n}S_{i}\hat{1}_{n,i}\left( p_{d}(V_{i},w)-\hat{p}%
    	_{d,i}(V_{i},w)\right) \\
    	&=-\frac{1}{n}\sum_{i=1}^{n}\frac{\mathbf{E}_{Q,w,i}[S_{i}]\mathcal{J}%
    		_{d,w,i}}{\mathbf{E}_{Q,w,i}[L_{w,i}]}+\frac{1}{n}\sum_{i=1}^{n}\frac{%
    		\mathbf{E}_{Q,w,i}[S_{i}]p_{d}(V_{i},w)\mathcal{J}_{w,i}}{\mathbf{E}%
    		_{Q,w,i}[L_{w,i}]}+o_{P}(n^{-1/2}),
    \end{align*}%
    \textit{uniformly for }$p\in A,$ \textit{where} $\mathcal{J}_{d,w,i}=
    L_{d,w,i}-\mathbf{E}_{Q,w,i}\left[ L_{d,w,i}\right] $ \textit{and} $\mathcal{%
    	J}_{w,i}=\mathcal{J}_{1,w,i}+\mathcal{J}_{0,w,i}.$\medskip
    
    \noindent (ii) \textit{Suppose that Condition \ref{con:unconfounded}, Assumptions \ref{ass:bd} and \ref{ass:kernel}} 
    \textit{hold}. \textit{Then, for }$d=0,1,$%
    \begin{align*}
    	&\frac{1}{n}\sum_{i=1}^{n}S_{i}\hat{1}_{n,i}\left( \hat{p}_{d,i}(V_{i},w)-%
    	\tilde{p}_{d,i}(V_{i},w)\right) \\
    	&=\mathbf{E}_{Q,w}\left[ p_{1-d}(V_{i},w)p_{d}(V_{i},w)S_{i}\right] \left( 
    	\frac{\hat{q}_{d,w}-q_{d,w}}{q_{d,w}}-\frac{\hat{q}_{1-d,w}-q_{1-d,w}}{%
    		q_{1-d,w}}\right) +o_{P}(n^{-1/2}),
    \end{align*}%
    \textit{uniformly for }$p\in A.$\medskip
    
    \noindent \textsc{Proof:}\textbf{\ }(i) By adding and subtracting the sum:%
    \begin{align*}
    	\frac{1}{n}\sum_{i=1}^{n}S_{i}\hat{1}_{n,i}\frac{\frac{1}{n-1}%
    		\sum_{j=1,j\neq i}^{n}L_{1,w,j}K_{ji}}{\mathbf{E}%
    		_{Q,w,i}[L_{w,i}]f_{Q}(V_{i},w)},
    \end{align*}
    and noting (\ref{eq1}), we write%
    \begin{align}
    	\label{dev2}
    	&\frac{1}{n}\sum_{i=1}^{n}S_{i}\hat{1}_{n,i}\left( p_{1}(V_{i},w)-\hat{p}%
    	_{1,i}(V_{i},w)\right) \\
    	&=\frac{1}{n}\sum_{i=1}^{n}S_{i}\hat{1}_{n,i}\left\{ \frac{\mathbf{E}%
    		_{Q,w,i}[L_{1,w,i}]}{\mathbf{E}_{Q,w,i}[L_{w,i}]}-\frac{\sum_{j=1,j\neq
    			i}^{n}L_{1,w,j}K_{ji}}{\sum_{j=1,j\neq i}^{n}L_{w,j}K_{ji}}\right\}  
    	\\ \notag
    	&=\frac{1}{n}\sum_{i=1}^{n}\frac{S_{i}\hat{1}_{n,i}}{\mathbf{E}%
    		_{Q,w,i}[L_{w,i}]}\left\{ \mathbf{E}_{Q,w,i}[L_{1,w,i}]-\frac{\frac{1}{n-1}%
    		\sum_{j=1,j\neq i}^{n}L_{1,w,j}K_{ji}}{f_{Q}(V_{i},w)}\right\} \\ \notag
    	&+\frac{1}{n}\sum_{i=1}^{n}S_{i}\hat{1}_{n,i}\left\{ \frac{\frac{1}{n-1}%
    		\sum_{j=1,j\neq i}^{n}L_{1,w,j}K_{ji}}{\mathbf{E}%
    		_{Q,w,i}[L_{w,i}]f_{Q}(V_{i},w)}-\frac{\sum_{j=1,j\neq i}^{n}L_{1,w,j}K_{ji}%
    	}{\sum_{j=1,j\neq i}^{n}L_{w,j}K_{ji}}\right\}. 
    \end{align}
    We write the last sum as%
    \begin{align*}
    	&\frac{1}{n}\sum_{i=1}^{n}S_{i}\frac{\hat{1}_{n,i}\sum_{j=1,j\neq
    			i}^{n}L_{1,w,j}K_{ji}}{\mathbf{E}_{Q,w,i}[L_{w,i}]\sum_{j=1,j\neq
    			i}^{n}L_{w,j}K_{ji}}\left\{ \frac{\frac{1}{n-1}\sum_{j=1,j\neq
    			i}^{n}L_{w,j}K_{ji}}{f_{Q}(V_{i},w)}-\mathbf{E}_{Q,w,i}[L_{w,i}]\right\} \\
    	&=-\frac{1}{n}\sum_{i=1}^{n}\frac{S_{i}\hat{1}_{n,i}\mathbf{E}%
    		_{Q,w,i}[L_{1,w,j}]}{\mathbf{E}_{Q,w,i}[L_{w,i}]^{2}}\left\{ \mathbf{E}%
    	_{Q,w,i}[L_{w,i}]-\frac{\frac{1}{n-1}\sum_{j=1,j\neq i}^{n}L_{w,j}K_{ji}}{%
    		f_{Q}(V_{i},w)}\right\} +o_{P}(n^{-1/2}) \\
    	&=-\frac{1}{n}\sum_{i=1}^{n}\frac{S_{i}\hat{1}_{n,i}p_{1}(V_{i},w)}{\mathbf{%
    			E}_{Q,w,i}[L_{w,i}]}\left\{ \mathbf{E}_{Q,w,i}[L_{w,i}]-\frac{\frac{1}{n-1}%
    		\sum_{j=1,j\neq i}^{n}L_{w,j}K_{ji}}{f_{Q}(V_{i},w)}\right\}
    	+o_{P}(n^{-1/2}).
    \end{align*}%
    uniformly for all $p\in A.$ The first equality uses Lemma B1 and the second
    uses (\ref{eq1}). Let 
    \begin{align*}
    	K_{n,i}= \mathbf{E}_{Q,w,i}[L_{w,i}]-\frac{\frac{1}{n-1}\sum_{j=1,j\neq
    			i}^{n}L_{w,j}K_{ji}}{f_{Q}(V_{i},w)},
    \end{align*}%
    and write the last sum as%
    \begin{align*}
    	&-\frac{1}{n}\sum_{i=1}^{n}\frac{S_{i}\hat{1}_{n,i}p_{1}(V_{i},w)K_{n,i}}{%
    		\mathbf{E}_{Q,w,i}[L_{w,i}]} \\
    	&=-\frac{1}{n}\sum_{i=1}^{n}\frac{S_{i}p_{1}(V_{i},w)K_{n,i}}{\mathbf{E}%
    		_{Q,w,i}[L_{w,i}]}-\frac{1}{n}\sum_{i=1}^{n}\frac{S_{i}\left\{ 1-\hat{1}%
    		_{n,i}\right\} p_{1}(V_{i},w)K_{n}}{\mathbf{E}_{Q,w,i}[L_{w,i}]}.
    \end{align*}%
    Observe that%
    \begin{align}
    	1-\hat{1}_{n,i}\leq 1\left\{ \hat{\lambda}_{1,i}(V_{i},w)<\delta
    	_{n}\right\} +1\left\{ \hat{\lambda}_{0,i}(V_{i},w)<\delta _{n}\right\} .
    	\label{arg}
    \end{align}%
    We write the first indicator on the right hand side as 
    \begin{align}
    	1\left\{ \frac{\mathbf{\hat{E}}_{Q,w,i}[L_{1,w,i}]}{n-1}\sum_{j=1,j\neq
    		i}^{n}1\{W_{j}=w\}K_{ji}<\delta _{n}\right\} \leq 1\left\{ \mathbf{E}%
    	_{Q,w,i}[L_{1,w,i}]<\kappa _{n}\right\} ,  \label{bdd}
    \end{align}%
    where $\kappa _{n}=(\delta _{n}+R_{1n})/c$ (with $c>0$ such that min$_{w\in 
    	\mathcal{W}}$inf$_{v\in \mathcal{V}(w)}f_{Q}(v,w)>c$ (see Assumption \ref{ass:bd}(iii))
    and%
    \begin{align}
    	R_{1n}= \max_{1\leq i\leq n}\left\vert \frac{\mathbf{\hat{E}}%
    		_{Q,w,i}[L_{1,w,i}]}{n-1}\sum_{j=1,j\neq i}^{n}K_{ji}-\mathbf{E}%
    	_{Q,w,i}[L_{1,w,i}]\cdot f_{Q}(V_{i},w)\right\vert .  \label{LemmaB2_R1}
    \end{align}%
    Note that from (\ref{rac}), we have$\ R_{1n}=O_{P}(\varepsilon _{n})$. Thus
    we can take a nonstochastic sequence $\kappa _{n}^{\prime }$ and $\eta >0$
    such that $\kappa _{n}^{\prime \eta }=o(n^{-1/2})$ and $\max \{\gamma
    ,2\}\leq \eta$, using Assumptions \ref{ass:kernel} (ii) and (iii). (Here $%
    \gamma $ is the constant in Assumptions \ref{ass:kernel} (iii).)
    Replacing $\kappa _{n}$ in (\ref{bdd})\ by this $\kappa _{n}^{\prime }$, we
    find that with probability approaching one, we have%
    \begin{align}
    	&\sup_{p\in A}\left\vert \frac{1}{n}\sum_{i=1}^{n}\frac{S_{i}\left\{ 1-\hat{%
    			1}_{n,i}\right\} p_{1}(V_{i},w)K_{n,i}}{\mathbf{E}_{Q,w,i}[L_{w,i}]}%
    	\right\vert  \label{LemmaB2_sum1} \\
    	&\leq \sup_{p\in A}\left\{ \frac{K_{n}}{n}\sum_{i=1}^{n}\left\vert \frac{%
    		S_{i}p_{1}(V_{i},w)}{\mathbf{E}_{Q,w,i}[L_{w,i}]}\right\vert \left( 1\left\{ 
    	\mathbf{E}_{Q,w,i}[L_{1,w,i}]\leq \kappa _{n}^{\prime }\right\} +1\left\{ 
    	\mathbf{E}_{Q,w,i}[L_{0,w,i}]\leq \kappa _{n}^{\prime }\right\} \right)
    	\right\} ,  \notag
    \end{align}%
    where $K_{n}=\max_{1\leq i\leq n}$ $\left\vert K_{n,i}\right\vert $. It is
    not hard to see that $\sup_{p\in A}K_{n}=O_{P}(1)$, because 
    \begin{align*}
    	\sup_{p\in A}\max_{1\leq i\leq n}\left\vert K_{n,i}\right\vert \leq
    	\sup_{p\in A}\max_{w}\sup_{v\in \mathcal{V}(w)}\frac{%
    		2f(v,w)}{f_{Q}(v,w)}+O_{P}(\varepsilon _{n})=O_{P}(1)
    \end{align*}%
    and min$_{w}$inf$_{v\in \mathcal{V}(w)}f_{Q}(v,w)>c$ for some
    positive constant $c>0$, using Assumption \ref{ass:bd} (iii). Note that the expectation $%
    \mathbf{E}_{Q}$ of (\ref{LemmaB2_sum1}) is bounded by (for some $C>0$)%
    \begin{align*}
    	C\kappa _{n}^{\prime \eta }\mathbf{E}_{Q}\left[ \mathbf{E}_{Q,w,i}^{-\eta
    	}[L_{w,i}]\right] =O\left( \kappa _{n}^{\prime \eta }\right) =o(n^{-1/2}),
    \end{align*}%
    uniformly over $p\in A$, using (\ref{dec5}). Hence we conclude that%
    \begin{align}
    	\frac{1}{n}\sum_{i=1}^{n}\frac{S_{i}\hat{1}_{n,i}p_{1}(V_{i},w)K_{n,i}}{%
    		\mathbf{E}_{Q,w,i}[L_{w,i}]}=\frac{1}{n}\sum_{i=1}^{n}\frac{%
    		S_{i}p_{1}(V_{i},w)K_{n,i}}{\mathbf{E}_{Q,w,i}[L_{w,i}]}+o_{P}(n^{-1/2}),
    	\label{state}
    \end{align}%
    uniformly over $p\in A.$ Applying the similar argument to the second to the
    last sum of (\ref{dev2}) to eliminate $\hat{1}_{n,i}$, we finally write 
    \begin{align*}
    	&\frac{1}{n}\sum_{i=1}^{n}S_{i}\hat{1}_{n,i}\left( p_{1}(V_{i},w)-\hat{p}%
    	_{1,i}(V_{i},w)\right) \\
    	&=\frac{1}{n}\sum_{i=1}^{n}\frac{S_{i}}{\mathbf{E}_{Q,w,i}[L_{w,i}]}\left\{ 
    	\mathbf{E}_{Q,w,i}[L_{1,w,i}]-\frac{\frac{1}{n-1}\sum_{j=1,j\neq
    			i}^{n}L_{1,w,j}K_{ji}}{f_{Q}(V_{i},w)}\right\} \\
    	& \quad -\frac{1}{n}\sum_{i=1}^{n}\frac{S_{i}p_{1}(V_{i},w)}{\mathbf{E}%
    		_{Q,w,i}[L_{w,i}]}\left\{ \mathbf{E}_{Q,w,i}[L_{w,i}]-\frac{\frac{1}{n-1}%
    		\sum_{j=1,j\neq i}^{n}L_{w,j}K_{ji}}{f_{Q}(V_{i},w)}\right\}
    	+o_{P}(n^{-1/2}),
    \end{align*}%
    uniformly over $p\in A.$ By Lemma D1 below, the difference of the last two
    terms is asymptotically equivalent to (up to $o_{P}(n^{-1/2}),$ uniformly
    over $p\in A.$)\ 
    \begin{align*}
    	&\frac{1}{n}\sum_{i=1}^{n}\left\{ \mathbf{E}\left[ \frac{\mathbf{E}%
    		_{Q,w,i}[S_{i}]\mathbf{E}_{Q,w,i}[L_{1,w,i}]}{\mathbf{E}_{Q,w,i}[L_{w,i}]}%
    	\right] -\frac{\mathbf{E}_{Q,w,i}[S_{i}]}{\mathbf{E}_{Q,w,i}[L_{w,i}]}%
    	L_{1,w,i}\right\} \\
    	&-\frac{1}{n}\sum_{i=1}^{n}\left\{ \mathbf{E}\left[ \frac{\mathbf{E}%
    		_{Q,w,i}[S_{i}]p_{1}(V_{i},w)\mathbf{E}_{Q,w,i}[L_{w,i}]}{\mathbf{E}%
    		_{Q,w,i}[L_{w,i}]}\right] -\frac{\mathbf{E}%
    		_{Q,w,i}[S_{i}]p_{1}(V_{i},w)L_{w,i}}{\mathbf{E}_{Q,w,i}[L_{w,i}]}\right\} \\
    	&=-\frac{1}{n}\sum_{i=1}^{n}\frac{\mathbf{E}_{Q,w,i}[S_{i}]\mathcal{J}%
    		_{1,w,i}}{\mathbf{E}_{Q,w,i}[L_{w,i}]}+\frac{1}{n}\sum_{i=1}^{n}\frac{%
    		\mathbf{E}_{Q,w,i}[S_{i}]p_{1}(V_{i},w)\mathcal{J}_{w,i}}{\mathbf{E}%
    		_{Q,w,i}[L_{w,i}]}+R_{n},
    \end{align*}%
    using the definitions of $\mathcal{J}_{1,w,i}$ and $\mathcal{J}_{w,i},$ where%
    \begin{align*}
    	R_{n} &=\frac{1}{n}\sum_{i=1}^{n}\left\{ \mathbf{E}\left[ \frac{\mathbf{E}%
    		_{Q,w,i}[S_{i}]\mathbf{E}_{Q,w,i}[L_{1,w,i}]}{\mathbf{E}_{Q,w,i}[L_{w,i}]}%
    	\right] -\frac{\mathbf{E}_{Q,w,i}[S_{i}]\mathbf{E}_{Q,w,i}\left[ L_{1,w,i}%
    		\right] }{\mathbf{E}_{Q,w,i}[L_{w,i}]}\right\} \\
    	& \quad -\frac{1}{n}\sum_{i=1}^{n}\left\{ \mathbf{E}\left[ \frac{\mathbf{E}%
    		_{Q,w,i}[S_{i}]p_{1}(V_{i},w)\mathbf{E}_{Q,w,i}[L_{w,i}]}{\mathbf{E}%
    		_{Q,w,i}[L_{w,i}]}\right] -\frac{\mathbf{E}_{Q,w,i}[S_{i}]p_{1}(V_{i},w)%
    		\mathbf{E}_{Q,w,i}[L_{w,i}]}{\mathbf{E}_{Q,w,i}[L_{w,i}]}\right\} \\
    	&=\frac{1}{n}\sum_{i=1}^{n}\left\{ \mathbf{E}\left[ \mathbf{E}%
    	_{Q,w,i}[S_{i}]p_{1}(V_{i},w)\right] -\mathbf{E}%
    	_{Q,w,i}[S_{i}]p_{1}(V_{i},w)\right\} \\
    	&\quad -\frac{1}{n}\sum_{i=1}^{n}\left\{ \mathbf{E}\left[ \mathbf{E}%
    	_{Q,w,i}[S_{i}]p_{1}(V_{i},w)\right] -\mathbf{E}%
    	_{Q,w,i}[S_{i}]p_{1}(V_{i},w)\right\} \\
    	&=0,
    \end{align*}%
    using (\ref{eq1}).\medskip
    
    \noindent (ii) We focus on the case of $d=1$. The case for $d=0$ can be
    dealt with precisely in the same way. First, we let $1_{n,i}=1\left\{ 
    \mathbf{E}_{Q,w,i}\left[ L_{w,i}\right] \geq \delta _{n}\right\} ,$ and write%
    \begin{align}
    	&\frac{1}{n}\sum_{i=1}^{n}S_{i}\hat{1}_{n,i}\left( \hat{p}_{1,i}(V_{i},w)-%
    	\tilde{p}_{1,i}(V_{i},w)\right)  \label{dec3} \\
    	&=\frac{1}{n}\sum_{i=1}^{n}S_{i}\hat{1}_{n,i}1_{n,i}\left( \hat{p}%
    	_{1,i}(V_{i},w)-\tilde{p}_{1,i}(V_{i},w)\right)   +\frac{1}{n}\sum_{i=1}^{n}S_{i}\hat{1}_{n,i}\left( 1-1_{n,i}\right) \left( 
    	\hat{p}_{1,i}(V_{i},w)-\tilde{p}_{1,i}(V_{i},w)\right) .  \notag
    \end{align}%
    By Lemma B1, $\max_{1\leq i\leq n}\left\vert \hat{p}_{1,i}(V_{i},w)-\tilde{p}%
    _{1,i}(V_{i},w)\right\vert \hat{1}_{n,i}=O_{P}(\varepsilon _{n})$ uniformly
    over $p\in A.$ Furthermore, since $S_{i}$'s are i.i.d. under $Q$, and the
    absolute conditional moment given $(V_{i},W_{i})=(v,w)$ is bounded uniformly
    over $v\in \mathcal{V}(w)$ and over $w$, we find that%
    \begin{align}\label{dec4} 
    	\mathbf{E}_{Q}\left[ \frac{1}{n}\sum_{i=1}^{n}\left\vert S_{i}\right\vert
    	\left\vert 1-1_{n,i}\right\vert \right] &\leq C\mathbf{E}_{Q}\left[
    	1\left\{ \mathbf{E}_{Q,w,i}\left[ L_{w,i}\right] \leq \delta _{n}\right\} %
    	\right]  \leq C\delta _{n}^{a}\mathbf{E}_{Q}\left[ \mathbf{E}_{Q,w,i}^{-a}\left[
    	L_{w,i}\right] \right],
    \end{align}%
    by Markov's inequality, for some $a\geq\gamma$. By (\ref{dec5}%
    ), the last expectation is finite. Since $\delta _{n}^{\gamma }=o(n^{-1/2})$
    (Assumption \ref{ass:kernel}(iii)), we conclude that%
    \begin{align*}
    	\frac{1}{n}\sum_{i=1}^{n}S_{i}\hat{1}_{n,i}\left( \hat{p}_{1,i}(V_{i},w)-
    	\tilde{p}_{1,i}(V_{i},w)\right) &= \frac{1}{n}\sum_{i=1}^{n}S_{i}\hat{1}%
    	_{n,i}1_{n,i}\left( \hat{p}_{1,i}(V_{i},w)-\tilde{p}_{1,i}(V_{i},w)\right)+o_{P}(n^{-1/2}),
    \end{align*}%
    uniformly for $p\in A.$
    
    As for the leading sum on the right hand side (\ref{dec3}), note that%
    \begin{align*}
    	&\frac{1}{n}\sum_{i=1}^{n}S_{i}\hat{1}_{n,i}1_{n,i}\left( \hat{p}%
    	_{1,i}(V_{i},w)-\tilde{p}_{1,i}(V_{i},w)\right) \\
    	&=\frac{1}{n}\sum_{i=1}^{n}S_{i}\hat{1}_{n,i}1_{n,i}\left\{ \frac{%
    		\sum_{j=1,j\neq i}^{n}L_{1,w,j}K_{ji}}{\sum_{j=1,j\neq i}^{n}L_{w,j}K_{ji}}-%
    	\frac{\sum_{j=1,j\neq i}^{n}\hat{L}_{1,w,j}K_{ji}}{\sum_{j=1,j\neq i}^{n}%
    		\hat{L}_{w,j}K_{ji}}\right\} \\
    	&=\frac{1}{n}\sum_{i=1}^{n}S_{i}\hat{1}_{n,i}1_{n,i}\frac{\sum_{j=1,j\neq
    			i}^{n}\left\{ L_{1,w,j}-\hat{L}_{1,w,j}\right\} K_{ji}}{\sum_{j=1,j\neq
    			i}^{n}L_{w,j}K_{ji}} \\
    	&\quad +\frac{1}{n}\sum_{i=1}^{n}S_{i}\hat{1}_{n,i}1_{n,i}\sum_{j=1,j\neq i}^{n}%
    	\hat{L}_{1,w,j}K_{ji}\left\{ \frac{1}{\sum_{j=1,j\neq i}^{n}L_{w,j}K_{ji}}-%
    	\frac{1}{\sum_{j=1,j\neq i}^{n}\hat{L}_{w,j}K_{ji}}\right\} .
    \end{align*}%
    Now, note that as for the second term,%
    \begin{align*}
    	&\frac{1}{n}\sum_{i=1}^{n}S_{i}\hat{1}_{n,i}1_{n,i}\sum_{j=1,j\neq i}^{n}%
    	\hat{L}_{1,w,j}K_{ji}\left\{ \frac{1}{\sum_{j=1,j\neq i}^{n}L_{w,j}K_{ji}}-%
    	\frac{1}{\sum_{j=1,j\neq i}^{n}\hat{L}_{w,j}K_{ji}}\right\} \\
    	&=\frac{1}{n}\sum_{i=1}^{n}S_{i}\hat{1}_{n,i}1_{n,i}\frac{\sum_{j=1,j\neq
    			i}^{n}\hat{L}_{1,w,j}K_{ji}}{\sum_{j=1,j\neq i}^{n}\hat{L}_{w,j}K_{ji}}%
    	\left\{ \frac{\sum_{j=1,j\neq i}^{n}\left\{ \hat{L}_{w,j}-L_{w,j}\right\}
    		K_{ji}}{\sum_{j=1,j\neq i}^{n}L_{w,j}K_{ji}}\right\} .
    \end{align*}%
    Using Lemma B1, we can write the last sum as%
    \begin{align*}
    	\frac{1}{n}\sum_{i=1}^{n}S_{i}\hat{1}_{n,i}1_{n,i}\frac{\sum_{j=1,j\neq
    			i}^{n}L_{1,w,j}K_{ji}}{\sum_{j=1,j\neq i}^{n}L_{w,j}K_{ji}}\left\{ \frac{%
    		\sum_{j=1,j\neq i}^{n}\left\{ \hat{L}_{w,j}-L_{w,j}\right\} K_{ji}}{%
    		\sum_{j=1,j\neq i}^{n}L_{w,j}K_{ji}}\right\} +o_{P}(n^{-1/2}),
    \end{align*}%
    uniformly for $p\in A.$ Therefore, we can write%
    \begin{align}
    	&\frac{1}{n}\sum_{i=1}^{n}S_{i}\hat{1}_{n,i}1_{n,i}\{\hat{p}_{1,i}(V_{i},w)-%
    	\tilde{p}_{1,i}(V_{i},w)\}  \label{dev3} \\
    	&=-\frac{1}{n}\sum_{i=1}^{n}S_{i}\hat{1}_{n,i}1_{n,i}\frac{\sum_{j=1,j\neq
    			i}^{n}L_{0,w,j}K_{ji}}{\sum_{j=1,j\neq i}^{n}L_{w,j}K_{ji}}\frac{%
    		\sum_{j=1,j\neq i}^{n}\left\{ \hat{L}_{1,w,j}-L_{1,w,j}\right\} K_{ji}}{%
    		\sum_{j=1,j\neq i}^{n}L_{w,j}K_{ji}}  \notag \\
    	&\quad +\frac{1}{n}\sum_{i=1}^{n}S_{i}\hat{1}_{n,i}1_{n,i}\frac{\sum_{j=1,j\neq
    			i}^{n}L_{1,w,j}K_{ji}}{\sum_{j=1,j\neq i}^{n}L_{w,j}K_{ji}}\frac{%
    		\sum_{j=1,j\neq i}^{n}\left\{ \hat{L}_{0,w,j}-L_{0,w,j}\right\} K_{ji}}{%
    		\sum_{j=1,j\neq i}^{n}L_{w,j}K_{ji}}+o_{P}(n^{-1/2})  \notag \\
    	&=-\frac{1}{n}\sum_{i=1}^{n}S_{i}\hat{1}_{n,i}1_{n,i}\frac{%
    		p_{0}(V_{i},w)\sum_{j=1,j\neq i}^{n}\left\{ \hat{L}_{1,w,j}-L_{1,w,j}\right%
    		\} K_{ji}}{\sum_{j=1,j\neq i}^{n}L_{w,j}K_{ji}}  \notag \\
    	&\quad +\frac{1}{n}\sum_{i=1}^{n}S_{i}\hat{1}_{n,i}1_{n,i}\frac{%
    		p_{1}(V_{i},w)\sum_{j=1,j\neq i}^{n}\left\{ \hat{L}_{0,w,j}-L_{0,w,j}\right%
    		\} K_{ji}}{\sum_{j=1,j\neq i}^{n}L_{w,j}K_{ji}}+o_{P}(n^{-1/2}),  \notag
    \end{align}%
    uniformly over $p\in A.$ Here the uniformity over $p\in A$ follows from 
    \textbf{\ }%
    \begin{align*}
    	&\frac{\sum_{j=1,j\neq i}^{n}L_{0,w,j}K_{ji}}{\sum_{j=1,j\neq
    			i}^{n}L_{w,j}K_{ji}} \\
    	&=\left( 1+\frac{p_{1,w}q_{0,w}\sum_{j=1,j\neq
    			i}^{n}1\{(D_{j},W_{j})=(1,w)\}K_{ji}/\sum_{j=1,j\neq i}^{n}1\{W_{i}=w\}K_{ji}%
    	}{p_{0,w}q_{1,w}\sum_{j=1,j\neq
    			i}^{n}1\{(D_{j},W_{j})=(0,w)\}K_{ji}/\sum_{j=1,j\neq i}^{n}1\{W_{i}=w\}K_{ji}%
    	}\right) ^{-1},
    \end{align*}%
    where $\sum_{j=1,j\neq i}^{n}1\{(D_{j},W_{j})=(d,w)\}K_{ji}/\sum_{j=1,j\neq
    	i}^{n}1\{W_{i}=w\}K_{ji}$\ converges to $q_{d}(V_{i},w)$\ and does not
    depends on $p$.
    
    We write%
    \begin{align}
    	&\hat{1}_{n,i}1_{n,i}\frac{\sum_{j=1,j\neq i}^{n}\{\hat{L}%
    		_{1,w,j}-L_{1,w,j}\}K_{ji}}{\sum_{j=1,j\neq i}^{n}L_{w,j}K_{ji}}
    	\label{dev1} \\
    	&=\left( \frac{p_{1,w}}{\hat{q}_{1,w}}-\frac{p_{1,w}}{q_{1,w}}\right) \hat{1%
    	}_{n,i}1_{n,i}\frac{\sum_{j=1,j\neq i}^{n}1\{(D_{j},W_{j})=(1,w)\}K_{ji}}{%
    		\sum_{j=1,j\neq i}^{n}L_{w,j}K_{ji}}.  \notag
    \end{align}%
    As for the last term, we note that%
    \begin{align*}
    	&\hat{1}_{n,i}1_{n,i}\frac{\sum_{j=1,j\neq
    			i}^{n}1\{(D_{i},W_{i})=(1,w)\}K_{ji}}{\sum_{j=1,j\neq i}^{n}L_{w,j}K_{ji}} \\
    	&=\hat{1}_{n,i}1_{n,i}\frac{\sum_{j=1,j\neq
    			i}^{n}1\{(D_{i},W_{i})=(1,w)\}K_{ji}/\sum_{j=1,j\neq i}^{n}1\{W_{i}=w\}K_{ji}%
    	}{\sum_{j=1,j\neq i}^{n}L_{w,j}K_{ji}/\sum_{j=1,j\neq
    			i}^{n}1\{W_{i}=w\}K_{ji}} \\
    	&=\frac{q_{1}(V_{i},w)}{%
    		q_{1}(V_{i},w)p_{1,w}/q_{1,w}+q_{0}(V_{i},w)p_{0,w}/q_{0,w}}+o_{P}(n^{-1/4}),
    \end{align*}%
    uniformly over $p\in A,$ (using the fact that $O_{p}(\varepsilon
    _{n})=o_{P}(n^{-1/4})$ by Assumption 2(ii)). Hence the first term in (\ref%
    {dev1}) is written as%
    \begin{align*}
    	&\left( \frac{p_{1,w}}{\hat{q}_{1,w}}-\frac{p_{1,w}}{q_{1,w}}\right) \frac{%
    		q_{1}(V_{i},w)}{q_{1}(V_{i},w)p_{1,w}/q_{1,w}+q_{0}(V_{i},w)p_{0,w}/q_{0,w}}%
    	+o_{P}(n^{-1/2}) \\
    	&=\left( \frac{q_{1,w}-\hat{q}_{1,w}}{q_{1,w}}\right) \frac{%
    		q_{1}(V_{i},w)p_{1,w}/q_{1,w}}{%
    		q_{1}(V_{i},w)p_{1,w}/q_{1,w}+q_{0}(V_{i},w)p_{0,w}/q_{0,w}}+o_{P}(n^{-1/2})
    	\\
    	&=\left( \frac{q_{1,w}-\hat{q}_{1,w}}{q_{1,w}}\right)
    	p_{1}(V_{i},w)+o_{P}(n^{-1/2}),
    \end{align*}%
    where we used (\ref{eq1}) for the last equality.
    
    Similarly, we find that%
    \begin{align*}
    	\hat{1}_{n,i}1_{n,i}\frac{p_{1}(V_{i},w)\sum_{j=1}^{n}\left\{ \hat{L}%
    		_{0,w,j}-L_{0,w,j}\right\} K_{ji}}{\sum_{j=1}^{n}L_{w,j}K_{ji}}=\left( \frac{%
    		q_{1,w}-\hat{q}_{1,w}}{q_{1,w}}\right) p_{1}(V_{i},w)+o_{P}(n^{-1/2}),
    \end{align*}%
    uniformly over $p\in A$. Applying these results back to the last two sums in
    (\ref{dev3}), we conclude that 
    \begin{align*}
    	&\frac{1}{n}\sum_{i=1}^{n}S_{i}\hat{1}_{n,i}1_{n,i}\{\hat{p}_{1,i}(V_{i},w)-%
    	\tilde{p}_{1,i}(V_{i},w)\} \\
    	&=\frac{1}{n}\sum_{i=1}^{n}S_{i}p_{0}(V_{i},w)p_{1}(V_{i},w)\left( \frac{%
    		\hat{q}_{1,w}-q_{1,w}}{q_{1,w}}-\frac{\hat{q}_{0,w}-q_{0,w}}{q_{0,w}}\right)
    	+o_{P}(n^{-1/2}),
    \end{align*}%
    uniformly over $p\in A.$ Finally, we write the last sum as%
    \begin{align*}
    	\mathbf{E}_{Q}\left[ p_{0}(V_{i},w)p_{1}(V_{i},w)S_{i}\right] \left( \frac{%
    		\hat{q}_{1,w}-q_{1,w}}{q_{1,w}}-\frac{\hat{q}_{0,w}-q_{0,w}}{q_{0,w}}\right)
    	+o_{P}(n^{-1/2}),
    \end{align*}%
    uniformly over $p\in A$ and this completes the proof. $\mathbf{\blacksquare }
    $\medskip
    
    \noindent \textsc{Lemma B3 }\textit{Suppose that Condition
    	\ref{con:unconfounded}, Assumptions \ref{ass:bd} and \ref{ass:kernel} hold, and} \textit{let} $\varepsilon
    _{d,w,i}=Y_{di}-\beta _{d}(V_{i},w)$. \textit{Then the following statements hold.}
    
    (i)
    \begin{align*}
    	&\frac{p_{1,w}}{q_{1,w}n}\sum_{i\in S_{1,w}}\hat{1}_{n,i}\frac{%
    		Y_{i}}{\hat{p}_{1,i}(V_{i},w)}-\frac{p_{0,w}}{q_{0,w}n}\sum_{i\in
    		S_{0,w}}\hat{1}_{n,i}\frac{Y_{i}}{\hat{p}_{0,i}(V_{i},w)} \\
    	&=\frac{1}{n}\sum_{i=1}^{n}\frac{L_{1,w,i}\varepsilon _{1,w,i}}{%
    		p_{1}(V_{i},w)}-\frac{1}{n}\sum_{i=1}^{n}\frac{L_{0,w,i}%
    		\varepsilon _{0,w,i}}{p_{0}(V_{i},w)} +\frac{1}{n}\sum_{i=1}^{n}\tau (V_{i},w)L_{w,i}+o_{P}(n^{-1/2}),
    \end{align*}%
    \textit{uniformly over} $p\in A.$
    
    \noindent (ii)
    \begin{align*}
    	&\frac{p_{1,w}}{n_{1,w}}\sum_{i\in S_{1,w}}\tilde{1}_{n,i}\frac{%
    		Y_{i}}{\tilde{p}_{1,i}(V_{i},w)}-\frac{p_{0,w}}{n_{0,w}}\sum_{i\in
    		S_{0,w}}\tilde{1}_{n,i}\frac{Y_{i}}{\tilde{p}_{0,i}(V_{i},w)} \\
    	&=\frac{1}{n}\sum_{i=1}^{n}\frac{L_{1,w,i}\varepsilon _{1,w,i}}{%
    		p_{1}(V_{i},w)}-\frac{1}{n}\sum_{i=1}^{n}\frac{L_{0,w,i}%
    		\varepsilon _{0,w,i}}{p_{0}(V_{i},w)} \\
    	&\quad +\frac{1}{n}\sum_{i=1}^{n}\left\{ \tau (V_{i},w)-\mathbf{E}_{1,w}%
    	\left[ \tau (V_{i},w)\right] \right\} L_{1,w,i} \\
    	&\quad +\frac{1}{n}\sum_{i=1}^{n}\left\{ \tau (V_{i},w)-\mathbf{E}_{0,w}%
    	\left[ \tau (V_{i},w)\right] \right\} L_{0,w,i} \\
    	&\quad +\mathbf{E}_{1,w}\left[ \tau (V_{i},w)\right] p_{1,w}+\mathbf{E}%
    	_{0,w}\left[ \tau (V_{i},w)\right] p_{0,w}+o_{P}(n^{-1/2}),
    \end{align*}%
    \textit{uniformly over }$p\in A.$\medskip
    
    \noindent \textsc{Proof :}\textbf{\ }(i) We first write%
    \begin{align*}
    	&\frac{p_{1,w}}{q_{1,w}n}\sum_{i\in S_{1,w}}\hat{1}_{n,i}\frac{%
    		Y_{i}}{\hat{p}_{1,i}(V_{i},w)}-\frac{p_{0,w}}{q_{0,w}n}\sum_{i\in
    		S_{0,w}}\hat{1}_{n,i}\frac{Y_{i}}{\hat{p}_{0,i}(V_{i},w)} \\
    	&=\frac{1}{n}\sum_{i=1}^{n}\hat{1}_{n,i}\frac{Y_{i}L_{1,w,i}}{%
    		\hat{p}_{1,i}(V_{i},w)}-\frac{1}{n}\sum_{i=1}^{n}\hat{1}_{n,i}\frac{%
    		Y_{i}L_{0,w,i}}{\hat{p}_{0,i}(V_{i},w)}=A_{1n}-A_{2n}.
    \end{align*}%
    We first write%
    \begin{align*}
    	A_{1n}=\frac{1}{n}\sum_{i=1}^{n}\frac{Y_{i}\hat{1}_{n,i}L_{1,w,i}}{%
    		p_{1}(V_{i},w)}+\tilde{A}_{1n},
    \end{align*}%
    where%
    \begin{align*}
    	\tilde{A}_{1n}=\frac{1}{n}\sum_{i=1}^{n}Y_{i}L_{1,w,i}\hat{1}%
    	_{n,i}\left( \frac{1}{\hat{p}_{1,i}(V_{i},w)}-\frac{1}{p_{1}(V_{i},w)}%
    	\right) .
    \end{align*}%
    As for $\tilde{A}_{1n}$, note that%
    \begin{align*}
    	&\frac{1}{n}\sum_{i=1}^{n}Y_{i}L_{1,w,i}\hat{1}_{n,i}\left( \frac{%
    		p_{1}(V_{i},w)-\hat{p}_{1,i}(V_{i},w)}{\hat{p}_{1,i}(V_{i},w)p_{1}(V_{i},w)}%
    	\right) \\
    	&=\frac{1}{n}\sum_{i=1}^{n}Y_{i}L_{1,w,i}\hat{1}_{n,i}\left( 
    	\frac{p_{1}(V_{i},w)-\hat{p}_{1,i}(V_{i},w)}{p_{1}^{2}(V_{i},w)}\right) \\
    	&\quad +\frac{1}{n}\sum_{i=1}^{n}Y_{i}L_{1,w,i}\hat{1}_{n,i}\frac{%
    		p_{1}(V_{i},w)-\hat{p}_{1,i}(V_{i},w)}{p_{1}(V_{i},w)}\left( \frac{1}{\hat{p}%
    		_{1,i}(V_{i},w)}-\frac{1}{p_{1}(V_{i},w)}\right) .
    \end{align*}%
    The supremum (over $p$) of the absolute value of the last sum has an upper
    bound with leading term%
    \begin{align}
    	\frac{1}{n}\sup_{p\in A}\sum_{i=1}^{n}\left\vert
    	Y_{i}L_{1,w,i}\right\vert \hat{1}_{n,i}\frac{\left( p_{1}(V_{i},w)-%
    		\hat{p}_{1,i}(V_{i},w)\right) ^{2}}{p_{1}(V_{i},w)^{3}}.  \label{sum}
    \end{align}%
    On the other hand, observe that from (\ref{eq1}), for any $q \geq 1$, 
    \begin{align}
    	&\mathbf{E}_{Q}\left[ \sup_{p\in A}p_{1}^{-q}(V_{i},w)\right]  
    	= \sum_{d,w}\mathbf{E}_{d,w}\left[\sup_{p\in A}\left\{ \frac{f(V_{i}|1,w)p_{1,w}+f(V_{i}|0,w)p_{0,w}}{%
    		f(V_{i}|1,w)p_{1,w}}\right\} ^{q}\right] q_{d,w}.  \label{bd}
    \end{align}%
    The last term is bounded due to Assumption \ref{ass:bd} (i) and (iii). Furthermore, observe that
    for some $C>0,$ 
    \begin{align*}
    	\sup_{v\in \mathcal{V}(w)}\mathbf{E}_{Q}\left[ \sup_{p\in A}\left\vert
    	Y_{i}L_{1,w,i}\right\vert ^{2}|(V_{i},W_{i})=(v,w)\right] 
    	&\leq &C\sup_{v\in \mathcal{V}(w)}\mathbf{E}_{Q}\left[
    	Y_{i}^{2}|(V_{i},W_{i})=(v,w)\right] .
    \end{align*}%
    The last term is bounded due to Assumption \ref{ass:bd} (ii). Hence by Lemma B1, we find
    that the sum in (\ref{sum}) is $o_{P}(n^{-1/2})$ (by the fact that $%
    \varepsilon _{n}^{2}=o_{P}(n^{-1/2})$). We conclude that%
    \begin{align}
    	\tilde{A}_{1n}=\frac{1}{n}\sum_{i=1}^{n}\frac{Y_{i}L_{1,w,i}}{%
    		p_{1}^{2}(V_{i},w)}\hat{1}_{n,i}\left( p_{1}(V_{i},w)-\hat{p}%
    	_{1,i}(V_{i},w)\right) +o_{P}(n^{-1/2}),  \label{sm}
    \end{align}%
    uniformly over $p\in A.$ Let $S_{i}=$ $%
    Y_{i}L_{1,w,i}/p_{1}^{2}(V_{i},w)\hat{1}_{n,i}.$ Then, for some $%
    C>0,$%
    \begin{align*}
    	\sup_{v\in \mathcal{V}(w)}\mathbf{E}_{Q}\left[ S_{i}^{2}|(V_{i},W_{i})=(v,w)%
    	\right] \leq C\sup_{v\in \mathcal{V}(w)}\mathbf{E}%
    	_{Q}[Y_{i}^{2}|(V_{i},W_{i})=(v,w)].
    \end{align*}%
    The last term is bounded due to Assumption \ref{ass:bd}(ii). As we saw in (\ref{bd}), the last
    term is bounded. We apply Lemma B2(i) to obtain that the leading sum in (\ref%
    {sm}) is asymptotically equivalent to (up to $o_{P}(n^{-1/2})$)%
    \begin{align}
    	-\frac{1}{n}\sum_{i=1}^{n}\frac{\mathbf{E}_{Q,w,i}[Y_{i}L_{1,w,i}]%
    		\mathcal{J}_{1,w,i}}{p_{1}^{2}(V_{i},w)\mathbf{E}_{Q,w,i}[L_{w,i}]}+\frac{1}{%
    		n}\sum_{i=1}^{n}\frac{\mathbf{E}_{Q,w,i}[Y_{i}L_{1,w,i}]\mathcal{J}%
    		_{w,i}}{p_{1}(V_{i},w)\mathbf{E}_{Q,w,i}[L_{w,i}]},  \label{decomp1}
    \end{align}%
    where $\mathcal{J}_{1,w,i}$ and $\mathcal{J}_{w,i}$ are as defined in Lemma
    B2. Using the fact that 
    \begin{align*}
    	\mathbf{E}_{Q,w,i}[Y_{i}L_{1,w,i}] &=\mathbf{E}%
    	[Y_{1i}|V_{i},(D_{i},W_{i})=(1,w)]q_{1}(V_{i},w)p_{1,w}/q_{1,w} \\
    	&=\beta _{1}(V_{i},w)q_{1}(V_{i},w)p_{1,w}/q_{1,w}.
    \end{align*}%
    and $q_{1}(V_{i},w)p_{1,w}/\{\mathbf{E}_{Q,w,i}[L_{w,i}]q_{1,w}%
    \}=p_{1}(V_{i},w)\ $from (\ref{eq1}), we write%
    \begin{align}
    	\frac{\mathbf{E}_{Q,w,i}[Y_{i}L_{1,w,i}]}{\mathbf{E}_{Q,w,i}[L_{w,i}]}=\beta
    	_{1}(V_{i},w)p_{1}(V_{i},w),\ \text{(using Condition \ref{con:unconfounded}, )}  \label{eq21}
    \end{align}%
    Using this, we write the first term in (\ref{decomp1}) as%
    \begin{align*}
    	-\frac{1}{n}\sum_{i=1}^{n}\frac{\beta _{1}(V_{i},w)\mathcal{J}%
    		_{1,w,i}}{p_{1}(V_{i},w)},
    \end{align*}%
    and the second term as 
    \begin{align*}
    	\frac{1}{n}\sum_{i=1}^{n}\beta _{1}(V_{i},w)\mathcal{J}_{w,i}=%
    	\frac{1}{n}\sum_{i=1}^{n}\beta _{1}(V_{i},w)\left\{ \mathcal{J}%
    	_{1,w,i}+\mathcal{J}_{0,w,i}\right\} .
    \end{align*}%
    Hence the difference in (\ref{decomp1}) is equal to 
    \begin{align*}
    	-\frac{1}{n}\sum_{i=1}^{n}\frac{\beta _{1}(V_{i},w)p_{0}(V_{i},w)}{%
    		p_{1}(V_{i},w)}\mathcal{J}_{1,w,i}+\frac{1}{n}\sum_{i=1}^{n}\beta
    	_{1}(V_{i},w)\mathcal{J}_{0,w,i}.
    \end{align*}%
    Therefore, we conclude that%
    \begin{align*}
    	&\tilde{A}_{1n}=-\frac{1}{n}\sum_{i=1}^{n}\frac{\beta
    		_{1}(V_{i},w)p_{0}(V_{i},w)}{p_{1}(V_{i},w)}\mathcal{J}_{1,w,i} 
    	+\frac{1}{n}\sum_{i=1}^{n}\beta _{1}(V_{i},w)\mathcal{J}%
    	_{0,w,i}+o_{P}(n^{-1/2}).
    \end{align*}%
    uniformly over $p\in A.$
    
    We turn to $A_{2n}$, which can be written as%
    \begin{align*}
    	A_{2n}=\frac{1}{n}\sum_{i=1}^{n}\frac{\hat{1}_{n,i}Y_{i}L_{0,w,i}}{%
    		p_{0}(V_{i},w)}+\tilde{A}_{2n}+o_{P}(n^{-1/2}),
    \end{align*}%
    where%
    \begin{align*}
    	\tilde{A}_{2n}=\frac{1}{n}\sum_{i=1}^{n}\hat{1}%
    	_{n,i}Y_{i}L_{0,w,i}\left( \frac{1}{\hat{p}_{0,i}(V_{i},w)}-\frac{1}{%
    		p_{0}(V_{i},w)}\right) .
    \end{align*}%
    Similarly as before, we write%
    \begin{align*}
    	\tilde{A}_{2n} &=\frac{1}{n}\sum_{i=1}^{n}\beta _{0}(V_{i},w)%
    	\mathcal{J}_{1,w,i} 
    	-\frac{1}{n}\sum_{i=1}^{n}\frac{\beta _{0}(V_{i},w)p_{1}(V_{i},w)%
    	}{p_{0}(V_{i},w)}\mathcal{J}_{0,w,i}+o_{P}(n^{-1/2}),
    \end{align*}%
    uniformly over $p\in A.$ Using the arguments employed to show (\ref{state})
    and combining the two results for $\tilde{A}_{1n}$ and $\tilde{A}_{2n}$, we
    deduce that%
    \begin{align*}
    	\tilde{A}_{1n}-\tilde{A}_{2n} &=-\frac{1}{n}\sum_{i=1}^{n}\left( 
    	\frac{\beta _{1}(V_{i},w)p_{0}(V_{i},w)}{p_{1}(V_{i},w)}+\beta
    	_{0}(V_{i},w)\right) \mathcal{J}_{1,w,i} \\
    	&\quad +\frac{1}{n}\sum_{i=1}^{n}\left( \beta _{1}(V_{i},w)+\frac{\beta
    		_{0}(V_{i},w)p_{1}(V_{i},w)}{p_{0}(V_{i},w)}\right) \mathcal{J}%
    	_{0,w,i}+o_{P}(n^{-1/2}) \\
    	&=-\frac{1}{n}\sum_{i=1}^{n}\left( \frac{\beta _{1}(V_{i},w)-\tau
    		(V_{i},w)p_{1}(V_{i},w)}{p_{1}(V_{i},w)}\right) \mathcal{J}_{1,w,i} \\
    	&\quad +\frac{1}{n}\sum_{i=1}^{n}\left( \frac{\tau
    		(V_{i},w)p_{0}(V_{i},w)+\beta _{0}(V_{i},w)}{p_{0}(V_{i},w)}\right) \mathcal{%
    		J}_{0,w,i}+o_{P}(n^{-1/2}),
    \end{align*}%
    using the fact that $\tau (X)=\beta _{1}(X)-\beta _{0}(X).$
    
    Therefore,%
    \begin{align*}
    	&\frac{p_{1,w}}{q_{1,w}n}\sum_{i\in S_{1,w}}\hat{1}_{n,i}\frac{%
    		Y_{i}}{\hat{p}_{1,i}(V_{i},w)}-\frac{p_{0,w}}{q_{0,w}n}\sum_{i\in
    		S_{0,w}}\hat{1}_{n,i}\frac{Y_{i}}{\hat{p}_{0,i}(V_{i},w)} \\
    	&=\frac{1}{n}\sum_{i=1}^{n}\frac{L_{1,w,i}\varepsilon _{1,w,i}}{%
    		p_{1}(V_{i},w)}-\frac{1}{n}\sum_{i=1}^{n}\frac{L_{0,w,i}%
    		\varepsilon _{0,w,i}}{p_{0}(V_{i},w)} \\
    	&\quad +\frac{1}{n}\sum_{i=1}^{n}\frac{L_{1,w,i}\beta _{1}(V_{i},w)}{%
    		p_{1}(V_{i},w)}-\frac{1}{n}\sum_{i=1}^{n}\frac{L_{0,w,i}\beta
    		_{0}(V_{i},w)}{p_{0}(V_{i},w)} \\
    	&-\frac{1}{n}\sum_{i=1}^{n}\left( \frac{\beta _{1}(V_{i},w)-\tau
    		(V_{i},w)p_{1}(V_{i},w)}{p_{1}(V_{i},w)}\right) \mathcal{J}_{1,w,i} \\
    	&\quad +\frac{1}{n}\sum_{i=1}^{n}\left( \frac{\tau
    		(V_{i},w)p_{0}(V_{i},w)+\beta _{0}(V_{i},w)}{p_{0}(V_{i},w)}\right) \mathcal{%
    		J}_{0,w,i}+o_{P}(n^{-1/2}).
    \end{align*}%
    By rearranging the terms, we rewrite%
    \begin{align*}
    	&\frac{p_{1,w}}{q_{1,w}n}\sum_{i\in S_{1,w}}\hat{1}_{n,i}\frac{%
    		Y_{i}}{\hat{p}_{1,i}(V_{i},w)}-\frac{p_{0,w}}{q_{0,w}n}\sum_{i\in
    		S_{0,w}}\hat{1}_{n,i}\frac{Y_{i}}{\hat{p}_{0,i}(V_{i},w)} \\
    	&=\frac{1}{n}\sum_{i=1}^{n}\frac{L_{1,w,i}\varepsilon _{1,w,i}}{%
    		p_{1}(V_{i},w)}-\frac{1}{n}\sum_{i=1}^{n}\frac{L_{0,w,i}%
    		\varepsilon _{0,w,i}}{p_{0}(V_{i},w)} \\
    	&\quad +\frac{1}{n}\sum_{i=1}^{n}\tau (V_{i},w)L_{1,w,i}+\frac{1}{n}%
    	\sum_{i=1}^{n}\tau (V_{i},w)L_{0,w,i} \\
    	&\quad +\frac{1}{n}\sum_{i=1}^{n}\left( \frac{\beta _{1}(V_{i},w)-\tau
    		(V_{i},w)p_{1}(V_{i},w)}{p_{1}(V_{i},w)}\right) \left( \mathbf{E}%
    	_{Q,w,i}[L_{1,w,i}]\right) \\
    	&\quad -\frac{1}{n}\sum_{i=1}^{n}\left( \frac{\tau
    		(V_{i},w)p_{0}(V_{i},w)+\beta _{0}(V_{i},w)}{p_{0}(V_{i},w)}\right) \left( 
    	\mathbf{E}_{Q,w,i}[L_{0,w,i}]\right) +o_{P}(n^{-1/2}),
    \end{align*}%
    uniformly over $p\in A.$ As for the last two terms, observe that%
    \begin{align*}
    	H_{n,i} &=\left\{ \frac{\beta _{1}(V_{i},w)}{p_{1}(V_{i},w)}-\tau
    	(V_{i},w)\right\} \mathbf{E}_{Q,w,i}[L_{1,w,i}]-\left\{ \frac{\beta
    		_{0}(V_{i},w)}{p_{0}(V_{i},w)}+\tau (V_{i},w)\right\} \mathbf{E}%
    	_{Q,w,i}[L_{0,w,i}] \\
    	&=\left\{ \frac{\beta _{1}(V_{i},w)}{p_{1}(V_{i},w)}-\tau (V_{i},w)\right\} 
    	\frac{q_{1}(V_{i},w)p_{1,w}}{q_{1,w}}-\left\{ \frac{\beta _{0}(V_{i},w)}{%
    		p_{0}(V_{i},w)}+\tau (V_{i},w)\right\} \frac{q_{0}(V_{i},w)p_{0,w}}{q_{0,w}}.
    \end{align*}%
    However, by Bayes' rule (see (\ref{eq0})), 
    \begin{align}
    	\frac{p_{1,w}q_{1}(V_{i},w)}{q_{1,w}}=\frac{%
    		p_{1,w}q_{1}(V_{i},w)f_{Q}(V_{i},w)}{q_{1,w}f_{Q}(V_{i},w)}=\frac{%
    		p_{1,w}f(V_{i}|1,w)}{f_{Q}(V_{i},w)}=\frac{p_{1}(V_{i},w)f(V_{i},w)}{%
    		f_{Q}(V_{i},w)}.  \label{deve3}
    \end{align}%
    Therefore, 
    \begin{align*}
    	H_{n,i}=\frac{f(V_{i},w)}{f_{Q}(V_{i},w)}\left\{ \left\{ \frac{\beta
    		_{1}(V_{i},w)}{p_{1}(V_{i},w)}-\tau (V_{i},w)\right\} p_{1}(V_{i},w)-\left\{ 
    	\frac{\beta _{0}(V_{i},w)}{p_{0}(V_{i},w)}+\tau (V_{i},w)\right\}
    	p_{0}(V_{i},w)\right\}
    \end{align*}%
    from which it follows that $H_{n,i}=0$ by the definition of $\tau (V_{i},w).$
    Hence we obtain the wanted result.\medskip
    
    \noindent (ii)\ We write%
    \begin{align}
    	&\frac{1}{n}\sum_{i=1}^{n}\frac{Y_{i}\tilde{1}_{n,i}\hat{L}%
    		_{1,w,i}}{\tilde{p}_{1,i}(V_{i},w)}-\frac{1}{n}\sum_{i=1}^{n}\frac{%
    		Y_{i}\tilde{1}_{n,i}\hat{L}_{0,w,i}}{\tilde{p}_{0,i}(V_{i},w)}
    	\label{devel} \\
    	&=\frac{1}{n}\sum_{i=1}^{n}\frac{Y_{i}\tilde{1}_{n,i}L_{1,w,i}}{%
    		\tilde{p}_{1,i}(V_{i},w)}-\frac{1}{n}\sum_{i=1}^{n}\frac{Y_{i}%
    		\tilde{1}_{n,i}L_{0,w,i}}{\tilde{p}_{0,i}(V_{i},w)}  \notag \\
    	&\quad +\frac{1}{n}\sum_{i=1}^{n}\frac{Y_{i}\tilde{1}_{n,i}\{\hat{L}%
    		_{1,w,i}-L_{1,w,i}\}}{\tilde{p}_{1,i}(V_{i},w)}-\frac{1}{n}\sum_{i=1}^{n}%
    	\frac{Y_{i}\tilde{1}_{n,i}\{\hat{L}_{0,w,i}-L_{0,w,i}\}}{\tilde{p}%
    		_{0,i}(V_{i},w)}.  \notag
    \end{align}%
    We write the first difference as%
    \begin{align*}
    	&\left\{ \frac{1}{n}\sum_{i=1}^{n}\frac{Y_{i}\hat{1}%
    		_{n,i}L_{1,w,i}}{\hat{p}_{1,i}(V_{i},w)}-\frac{1}{n}\sum_{i=1}^{n}\frac{%
    		Y_{i}\hat{1}_{n,i}L_{0,w,i}}{\hat{p}_{0,i}(V_{i},w)}\right\} \\
    	&\quad +\left\{ \frac{1}{n}\sum_{i=1}^{n}\frac{Y_{i}\hat{1}%
    		_{n,i}L_{1,w,i}}{p_{1}^{2}(V_{i},w)}A_{i}-\frac{1}{n}\sum_{i=1}^{n}\frac{%
    		Y_{i}\hat{1}_{n,i}L_{0,w,i}}{p_{0}^{2}(V_{i},w)}B_{i}\right\}
    	+o_{P}(n^{-1/2}) \\
    	&=J_{1n}+J_{2n}+o_{P}(n^{-1/2}),\text{ say,}
    \end{align*}%
    uniformly over $p\in A$, where 
    \begin{align*}
    	A_{i} &=\hat{p}_{1,i}(V_{i},w)-\tilde{p}_{1,i}(V_{i},w),\text{ and} \\
    	B_{i} &=\hat{p}_{0,i}(V_{i},w)-\tilde{p}_{0,i}(V_{i},w).
    \end{align*}%
    Note that the normalized sums with trimming factor $\tilde{1}_{n,i}$ can be
    replaced by the same sums but with $\hat{1}_{n,i}$ (with the resulting
    discrepancy confined to $o_{P}(n^{-1/2}),$ uniformly for $p\in A$), because 
    \begin{align}
    	1-\hat{1}_{n,i} &=o_{P}(n^{-1/2}),\text{ and}  \label{cv4} \\
    	1-\tilde{1}_{n,i} &=o_{P}(n^{-1/2})\text{,}  \notag
    \end{align}%
    uniformly over $p\in A$. The first line was shown in the proof of Lemma B2.
    (See arguments below (\ref{arg}).) Similar arguments apply to the second
    line so that 
    \begin{align*}
    	1-\tilde{1}_{n,i}\leq 1\left\{ \tilde{\lambda}_{1,i}(V_{i},w)<\delta
    	_{n}\right\} +1\left\{ \tilde{\lambda}_{0,i}(V_{i},w)<\delta _{n}\right\} .
    \end{align*}%
    We write the first indicator on the right hand side as 
    \begin{align}
    	1\left\{ \frac{\mathbf{\hat{E}}_{Q,w,i}[\tilde{L}_{1,w,i}]}{n-1}%
    	\sum_{j=1,j\neq i}^{n}1\{W_{i}=w\}K_{h,ji}<\delta _{n}\right\} \leq 1\left\{ 
    	\mathbf{E}_{Q,w,i}[L_{1,w,i}]<\kappa _{2n}\right\} ,  \label{bdd2}
    \end{align}%
    where $\kappa _{2n}=(\delta _{n}+R_{1n}+R_{2n})/c$ (with $c>0$ such that min$%
    _{w}$inf$_{v\in \mathcal{V}(w)}f_{Q}(v,w)>c$ (see Assumption
    1(iii)), $R_{1n}$ is as defined in (\ref{LemmaB2_R1}) and%
    \begin{align*}
    	R_{2n} &=\max_{1\leq i\leq n}\left\vert \frac{\mathbf{\hat{E}}%
    		_{Q,w,i}[L_{1,w,i}]-\mathbf{\hat{E}}_{Q,w,i}[\hat{L}_{1,w,i}]}{n-1}%
    	\sum_{j=1,j\neq i}^{n}1\{W_{i}=w\}K_{ji}\right\vert \\
    	&\leq \left\vert \frac{p_{d,w}}{q_{d,w}}-\frac{p_{d,w}}{\hat{q}_{d,w}}%
    	\right\vert \cdot \max_{1\leq i\leq n}\left\vert \frac{1}{n-1}%
    	\sum_{j=1,j\neq i}^{n}1\{W_{i}=w\}K_{ji}\right\vert =o_{P}(\varepsilon _{n}).
    \end{align*}%
    Recall that$\ R_{1n}=O_{P}(\varepsilon _{n})$. Thus as before, we can take a
    nonstochastic sequence $\kappa _{2n}^{\prime }$ and $\eta >0$ such that $%
    \kappa _{2n}^{\prime \eta }=o(n^{-1/2})$ and $\max \{\gamma ,2\}\leq \eta
    $, using Assumptions \ref{ass:kernel}(ii) and (iii). Replacing $\kappa _{2n}$
    in (\ref{bdd2})\ by this $\kappa _{2n}^{\prime }$, we find that with
    probability approaching one,%
    \begin{align*}
    	\left\vert 1-\tilde{1}_{n,i}\right\vert \leq 1\left\{ \mathbf{E}%
    	_{Q,w,i}[L_{1,w,i}]\leq \kappa _{2n}^{\prime }\right\} +1\left\{ \mathbf{E}%
    	_{Q,w,i}[L_{0,w,i}]\leq \kappa _{2n}^{\prime }\right\}
    \end{align*}%
    Note that the expectation $\mathbf{E}_{Q}$ of the last term is bounded by
    (for some $C>0$)%
    \begin{align*}
    	C\kappa _{n}^{\prime \eta }\mathbf{E}_{Q}\left[ \mathbf{E}_{Q,w,i}^{-\eta
    	}[L_{w,i}]\right] =O\left( \kappa _{2n}^{\prime \eta }\right) =o(n^{-1/2}),
    \end{align*}%
    uniformly over $p\in A$. Thus we obtain the second convergence in (\ref{cv4}%
    ).
    
    As for $J_{2n},$ by applying Lemma B2(ii), we have%
    \begin{align*}
    	J_{2n} &=\mathbf{E}_{Q}\left[ \frac{Y_{i}L_{1,w,i}p_{0}(V_{i},w)}{%
    		p_{1}(V_{i},w)}\right] \left( \frac{\hat{q}_{1,w}-q_{1,w}}{q_{1,w}}-\frac{%
    		\hat{q}_{0,w}-q_{0,w}}{q_{0,w}}\right) \\
    	&\quad -\mathbf{E}_{Q}\left[ \frac{Y_{i}L_{0,w,i}p_{1}(V_{i},w)}{%
    		p_{0}(V_{i},w)}\right] \left( \frac{\hat{q}_{0,w}-q_{0,w}}{q_{0,w}}-\frac{%
    		\hat{q}_{1,w}-q_{1,w}}{q_{1,w}}\right) +o_{P}(n^{-1/2}) \\
    	&=\mathbf{E}_{Q}\left[ Y_{i}\left\{ \frac{L_{1,w,i}p_{0}(V_{i},w)%
    	}{p_{1}(V_{i},w)}+\frac{L_{0,w,i}p_{1}(V_{i},w)}{p_{0}(V_{i},w)}\right\} %
    	\right] \frac{\hat{q}_{1,w}-q_{1,w}}{q_{1,w}} \\
    	&\quad -\mathbf{E}_{Q}\left[ Y_{i}\left\{ \frac{L_{1,w,i}p_{0}(V_{i},w)%
    	}{p_{1}(V_{i},w)}+\frac{L_{0,w,i}p_{1}(V_{i},w)}{p_{0}(V_{i},w)}\right\} %
    	\right] \frac{\hat{q}_{0,w}-q_{0,w}}{q_{0,w}} \\
    	&\quad +o_{P}(n^{-1/2}),
    \end{align*}%
    uniformly for $p\in A.$ On the other hand, as for the last difference in (%
    \ref{devel}), we have
    
    \begin{align*}
    	\frac{1}{n}\sum_{i=1}^{n}\frac{Y_{i}\tilde{1}_{n,i}\{\hat{L}%
    		_{1,w,i}-L_{1,w,i}\}}{\tilde{p}_{1,i}(V_{i},w)} 
    	&=\frac{1}{n}\sum_{i=1}^{n}\frac{Y_{i}\{\hat{L}%
    		_{1,w,i}-L_{1,w,i}\}}{p_{1}(V_{i},w)}+o_{P}(n^{-1/2}) \\
    	&=-\frac{1}{n}\sum_{i=1}^{n}\frac{Y_{i}L_{1,w,i}}{p_{1}(V_{i},w)}%
    	\frac{\hat{q}_{1,w}-q_{1,w}}{q_{1,w}}+o_{P}(n^{-1/2}) \\
    	&=-\mathbf{E}_{Q}\left[ \frac{Y_{i}L_{1,w,i}}{p_{1}(V_{i},w)}%
    	\right] \frac{\hat{q}_{1,w}-q_{1,w}}{q_{1,w}}+o_{P}(n^{-1/2}),
    \end{align*}%
    uniformly for $p\in A.$ Here uniformity again follows from the fact that $%
    p_{1,w}$ and $p_{0,w}$ can be factored out from the converging random
    sequence. In particular, \textbf{\ }%
    \begin{align*}
    	\frac{1}{n}\sum_{i=1}^{n}\frac{Y_{i}L_{1,w,i}}{p_{1}(V_{i},w)}
    	=\frac{p_{1,w}}{q_{1,w}}\frac{1}{n}\sum_{i=1}^{n}Y_{i}\mathcal{I}%
    	_{1,w,i}+\frac{p_{0,w}}{q_{1,w}}\frac{1}{n}\sum_{i=1}^{n}Y_{i}%
    	\mathcal{I}_{1,w,i}\frac{f(V_{i}|0,w)}{f(V_{i}|1,w)},
    \end{align*}%
    where $\mathcal{I}_{1,w,i}= $ $1\{(D_{i},W_{i})=(1,w)\}$. The CLT can
    be applied to terms that do not depend on $p$. Similarly,%
    \begin{align*}
    	\frac{1}{n}\sum_{i=1}^{n}\frac{Y_{i}\tilde{1}_{n,i}\{\hat{L}%
    		_{0,w,i}-L_{0,w,i}\}}{\tilde{p}_{0,i}(V_{i},w)} 
    	&=\frac{1}{n}\sum_{i=1}^{n}\frac{Y_{i}\{\hat{L}%
    		_{0,w,i}-L_{0,w,i}\}}{p_{0}(V_{i},w)}+o_{P}(n^{-1/2}) \\
    	&=-\frac{1}{n}\sum_{i=1}^{n}\frac{Y_{i}L_{0,w,i}}{p_{0}(V_{i},w)}%
    	\frac{\hat{q}_{0,w}-q_{0,w}}{q_{0,w}}+o_{P}(n^{-1/2}) \\
    	&=-\mathbf{E}_{Q}\left[ \frac{Y_{i}L_{0,w,i}}{p_{0}(V_{i},w)}%
    	\right] \frac{\hat{q}_{0,w}-q_{0,w}}{q_{0,w}}+o_{P}(n^{-1/2}),
    \end{align*}%
    uniformly for $p\in A.$ Combining these results, we conclude that%
    \begin{align}
    	&\frac{1}{n}\sum_{i=1}^{n}\frac{Y_{i}\tilde{1}_{n,i}\hat{L}%
    		_{1,w,i}}{\tilde{p}_{1,i}(V_{i},w)}-\frac{1}{n}\sum_{i=1}^{n}\frac{%
    		Y_{i}\tilde{1}_{n,i}\hat{L}_{0,w,i}}{\tilde{p}_{0,i}(V_{i},w)}
    	\label{develop} \\
    	&=\frac{1}{n}\sum_{i=1}^{n}\frac{Y_{i}\hat{1}_{n,i}L_{1,w,i}}{%
    		\hat{p}_{1,i}(V_{i},w)}-\frac{1}{n}\sum_{i=1}^{n}\frac{Y_{i}\hat{1}%
    		_{n,i}L_{0,w,i}}{\hat{p}_{0,i}(V_{i},w)}  \notag \\
    	&\quad +\mathbf{E}_{Q}\left[ Y_{i}\left\{ -L_{1,w,i}+\frac{%
    		L_{0,w,i}p_{1}(V_{i},w)}{p_{0}(V_{i},w)}\right\} \right] \frac{\hat{q}%
    		_{1,w}-q_{1,w}}{q_{1,w}}  \notag \\
    	&\quad -\mathbf{E}_{Q}\left[ Y_{i}\left\{ \frac{L_{1,w,i}p_{0}(V_{i},w)%
    	}{p_{1}(V_{i},w)}-L_{0,w,i}\right\} \right] \frac{\hat{q}_{0,w}-q_{0,w}}{%
    		q_{0,w}}+o_{P}(n^{-1/2}).  \notag
    \end{align}%
    uniformly for $p\in A.$ The last difference is written as 
    \begin{align*}
    	&\mathbf{E}_{Q}\left[ \left\{ -Y_{1i}L_{1,w,i}+Y_{0i}\frac{%
    		L_{0,w,i}p_{1}(V_{i},w)}{p_{0}(V_{i},w)}\right\} \right] \frac{\hat{q}%
    		_{1,w}-q_{1,w}}{q_{1,w}}  -\mathbf{E}_{Q}\left[ \left\{ Y_{1i}\frac{L_{1,w,i}p_{0}(V_{i},w)%
    	}{p_{1}(V_{i},w)}-Y_{0i}L_{0,w,i}\right\} \right] \frac{\hat{q}_{0,w}-q_{0,w}%
    	}{q_{0,w}} \\
    	&=\mathbf{E}_{Q}\left[ \left\{
    	-\{Y_{1i}-Y_{0i}\}L_{1,w,i}\right\} \right] \frac{\hat{q}_{1,w}-q_{1,w}}{%
    		q_{1,w}} +\mathbf{E}_{Q}\left[ Y_{0i}\left\{ \frac{L_{0,w,i}p_{1}(V_{i},w)%
    	}{p_{0}(V_{i},w)}-L_{1,w,i}\right\} \right] \frac{\hat{q}_{1,w}-q_{1,w}}{%
    		q_{1,w}} \\
    	&\quad -\mathbf{E}_{Q}\left[ \{Y_{1i}-Y_{0i}\}L_{0,w,i}\right] \frac{%
    		\hat{q}_{0,w}-q_{0,w}}{q_{0,w}} +\mathbf{E}_{Q}\left[ Y_{1i}\left\{ L_{0,w,i}-\frac{%
    		L_{1,w,i}p_{0}(V_{i},w)}{p_{1}(V_{i},w)}\right\} \right] \frac{\hat{q}%
    		_{0,w}-q_{0,w}}{q_{0,w}}.
    \end{align*}%
    The second and the fourth expectations vanish because%
    \begin{align*}
    	&\mathbf{E}_{Q}\left[ Y_{0,i}\left\{ -L_{1,w,i}+\frac{%
    		L_{0,w,i}p_{1}(V_{i},w)}{p_{0}(V_{i},w)}\right\} \right] \\
    	&=\mathbf{E}\left[ \beta _{0}(V_{i},w)\left\{
    	-1\{(D_{i,}W_{i})=(1,w)\}+\frac{1\{(D_{i,}W_{i})=(0,w)\}p_{1}(V_{i},w)}{%
    		p_{0}(V_{i},w)}\right\} \right] \\
    	&=\mathbf{E}\left[ \beta _{0}(V_{i},w)\left\{
    	p_{1}(V_{i},w)-p_{1}(V_{i},w)\right\} \right] =0,
    \end{align*}%
    and similarly,%
    \begin{align*}
    	\mathbf{E}_{Q}\left[Y_{1i}\left\{ L_{0,w,i}-\frac{%
    		L_{1,w,i}p_{0}(V_{i},w)}{p_{1}(V_{i},w)}\right\} \right] 
    	=\mathbf{E}\left[ \beta _{1}(V_{i},w)\left\{
    	p_{0}(V_{i},w)-p_{0}(V_{i},w)\right\} \right] =0.
    \end{align*}%
    Furthermore, observe that%
    \begin{align*}
    	\mathbf{E}_{Q}\left[ \left\{ -\{Y_{1i}-Y_{0i}\}L_{1,w,i}\right\} %
    	\right] 
    	&=-\mathbf{E}\left[ \{Y_{1i}-Y_{0i}\}1\{(D_{i,}W_{i})=(1,w)\}%
    	\right] \\
    	&=-\mathbf{E}\left[\{\beta _{1}(V_{i},w)-\beta
    	_{0}(V_{i},w)\}p_{1}(V_{i},w)\right] \\
    	&=-\mathbf{E}\left[ \tau (V_{i},w)p_{1}(V_{i},w)\right] ,
    \end{align*}%
    and similarly, 
    \begin{align*}
    	-\mathbf{E}_{Q}\left[\{Y_{1i}-Y_{0i}\}L_{0,w,i}\right] =-\mathbf{E%
    	}\left[\tau (V_{i},w)p_{0}(V_{i},w)\right] .
    \end{align*}%
    Hence, as for the last two terms in (\ref{develop}), we find that%
    \begin{align*}
    	\mathbf{E}_{Q}\left[Y_{i}\left\{ -L_{1,w,i}+\frac{%
    		L_{0,w,i}p_{1}(V_{i},w)}{p_{0}(V_{i},w)}\right\} \right] \frac{\hat{q}%
    		_{1,w}-q_{1,w}}{q_{1,w}} 
    	&=-\mathbf{E}_{Q}\left[\tau (V_{i},w)L_{1,w,i}\right] \frac{\hat{%
    			q}_{1,w}-q_{1,w}}{q_{1,w}} \\
    	&=-p_{1,w}\mathbf{E}_{1,w}\left[\tau (V_{i},w)\right] \frac{\hat{%
    			q}_{1,w}-q_{1,w}}{q_{1,w}} \\
    	&=-\mathbf{E}_{1,w}\left[ \tau (V_{i},w)\right] \frac{1}{n}%
    	\sum_{i=1}^{n}\left( L_{1,w,i}-p_{1,w}\right) ,
    \end{align*}%
    and%
    \begin{align*}
    	-\mathbf{E}_{Q}\left[Y_{i}\left\{ \frac{L_{1,w,i}p_{0}(V_{i},w)%
    	}{p_{1}(V_{i},w)}-L_{0,w,i}\right\} \right] \frac{\hat{q}_{0,w}-q_{0,w}}{%
    		q_{0,w}} 
    	&=-\mathbf{E}_{Q}\left[\tau (V_{i},w)L_{0,w,i}\right] \frac{\hat{%
    			q}_{0,w}-q_{0,w}}{q_{0,w}} \\
    	&=-p_{0,w}\mathbf{E}_{0,w}\left[\tau (V_{i},w)\right] \frac{\hat{%
    			q}_{0,w}-q_{0,w}}{q_{0,w}} \\
    	&=-\mathbf{E}_{0,w}\left[\tau (V_{i},w)\right] \frac{1}{n}%
    	\sum_{i=1}^{n}\left( L_{0,w,i}-p_{0,w}\right) .
    \end{align*}%
    Applying the result of (i) of this lemma to the first difference of (\ref%
    {develop}), we conclude that the difference in (ii) in this lemma is equal to%
    \begin{align*}
    	\frac{1}{n}\sum_{i=1}^{n}\frac{L_{1,w,i}\varepsilon _{1,w,i}}{%
    		p_{1}(V_{i},w)}-\frac{1}{n}\sum_{i=1}^{n}\frac{L_{0,w,i}%
    		\varepsilon _{0,w,i}}{p_{0}(V_{i},w)}+\Gamma _{n,w}+o_{P}(n^{-1/2}),
    \end{align*}%
    uniformly for $p\in A$, where%
    \begin{align*}
    	\Gamma _{n,w} &=\frac{1}{n}\sum_{i=1}^{n}\tau
    	(V_{i},w)L_{1,w,i}+\frac{1}{n}\sum_{i=1}^{n}\tau (V_{i},w)L_{0,w,i}
    	-\mathbf{E}_{1,w}\left[\tau (V_{i},w)\right] \frac{1}{n}%
    	\sum_{i=1}^{n}\left( L_{1,w,i}-p_{1,w}\right) \\
    	&\quad -\mathbf{E}_{0,w}\left[\tau (V_{i},w)\right] \frac{1}{n}%
    	\sum_{i=1}^{n}\left( L_{0,w,i}-p_{0,w}\right) .
    \end{align*}%
    The proof is complete because 
    \begin{align*}
    	\Gamma _{n,w} &=\frac{1}{n}\sum_{i=1}^{n}\left\{\tau (V_{i},w)-%
    	\mathbf{E}_{1,w}\left[\tau (V_{i},w)\right] \right\} L_{1,w,i} +\frac{1}{n}\sum_{i=1}^{n}\left\{\tau (V_{i},w)-\mathbf{E}_{0,w}%
    	\left[\tau (V_{i},w)\right] \right\} L_{0,w,i} \\
    	&\quad +\mathbf{E}_{1,w}\left[\tau (V_{i},w)\right] p_{1,w}+\mathbf{E}%
    	_{0,w}\left[\tau (V_{i},w)\right] p_{0,w}.
    \end{align*}%
    $\mathbf{\blacksquare }$\medskip
    
    \noindent \textsc{Proof of Lemma A1:} Let us consider the first
    statement in (\ref{convergences}). We write $\hat{\tau}_{ate}(p)-\tau
    _{ate}(p)$ as%
    \begin{align}
    	\sum_{w}\left\{ \frac{p_{1,w}}{%
    		n_{1,w}}\sum_{i\in S_{1,w}}\tilde{1}_{n,i}\frac{Y_{i}}{\tilde{p}%
    		_{1,i}(V_{i},w)}-\frac{p_{0,w}}{n_{0,w}}\sum_{i\in S_{0,w}}\tilde{1}_{n,i}%
    	\frac{Y_{i}}{\tilde{p}_{0,i}(V_{i},w)}\right\}-\tau
    	_{ate}(p).  \label{decomp2}
    \end{align}%
    
    Applying Lemma B3(ii) to term inside the bracket
    and recalling the definitions in (\ref{def4}), we obtain that $\hat{\tau}_{ate}(p)-\tau_{ate}(p)$ is asymptotically equivalent to (up to $%
    o_{P}(n^{-1/2})$ uniformly over all $p\in A$)%
    \begin{align*}
    	&\sum_{w}\left\{ \frac{1}{n}%
    	\sum_{i=1}^{n}\frac{L_{1,w,i}\varepsilon _{1,w,i}}{p_{1}(V_{i},w)}-%
    	\frac{1}{n}\sum_{i=1}^{n}\frac{L_{0,w,i}\varepsilon _{0,w,i}}{%
    		p_{0,w}(V_{i},w)}\right\} \\
    	&+\sum_{w}\frac{1}{n}%
    	\sum_{i=1}^{n}\left( \xi _{1,ate}(V_{i},w)L_{1,w,i}+\xi
    	_{0,ate}(V_{i},w)L_{0,w,i}\right) \\
    	&+\sum_{w}\left\{ \mathbf{E}%
    	_{1,w}[\tau (V_{i},w)]p_{1,w}+\mathbf{E}_{0,w}[\tau
    	(V_{i},w)]p_{0,w}\right\} -\tau _{ate}(p).
    \end{align*}%
    The second to the last term is actually $\tau _{ate}(p)$ canceling the last 
    $\tau _{ate}(p)$. This gives the first statement of Lemma A1.
    
    Now, we prove the second statement in (\ref{convergences}). Let 
    \begin{align*}
    	\mathbf{E}_{1}\left[ \beta _{0}(X_{i})\right] =\mathbf{E}\left[ \beta
    	_{0}(X_{i})|D_{i}=1\right]
    \end{align*}%
    and write $\hat{\tau}_{tet}(p)-\tau_{tet}(p)$ as%
    \begin{align}
    	&\frac{1}{p_1}\sum_{w%
    	}\left\{ \frac{p_{1,w}}{n_{1,w}}\sum_{i\in S_{1,w}}Y_{i}-\frac{%
    		p_{0,w}}{n_{0,w}}\sum_{i\in S_{0,w}}\tilde{1}_{n,i}\frac{\tilde{p}%
    		_{1,i}(V_{i},w)Y_{i}}{\tilde{p}_{0,i}(V_{i},w)}\right\} 
    	+\bar{R}_{n}-\tau _{tet}(p),  \label{decomp3}
    \end{align}%
    where%
    \begin{align*}
    	\bar{R}_{n}= M_{n}\sum_{w}\frac{p_{0,w}%
    	}{n_{0,w}}\sum_{i\in S_{0,w}}\tilde{1}_{n,i}\frac{\tilde{p}%
    		_{1,i}(V_{i},w)Y_{i}}{\tilde{p}_{0,i}(V_{i},w)},
    \end{align*}%
    with%
    \begin{align*}
    	M_{n} &=\frac{1}{p_{1}}-\left( \sum_{w}\frac{p_{0,w}}{n_{0,w}}\sum_{i\in S_{0,w}}\tilde{1}_{n,i}\tilde{p}_{1,i}(V_{i},w)/\tilde{p}%
    	_{0,i}(V_{i},w)\right) ^{-1}.
    \end{align*}%
    Note that
    \begin{align*}
    	\sum_{w}\frac{p_{0,w}}{n_{0,w}}\sum_{i\in S_{0,w}}\tilde{1}%
    	_{n,i}\frac{\tilde{p}_{1,i}(V_{i},w)}{\tilde{p}_{0,i}(V_{i},w)}
    	&=\sum_{w}p_{0,w}\mathbf{E}_{0,w}\left[ \frac{%
    		p_{1}(V_{i},w)}{p_{0}(V_{i},w)}\right] +O_{P}(n^{-1/2}) \\
    	&=\mathbf{E}_{0}\left[ \frac{p_{1}(X_{i})}{p_{0}(X_{i})}\right]
    	p_{0}+O_{P}(n^{-1/2}) =\mathbf{E}\left[ \frac{p_{1}(X_{i})(1-D_{i})}{p_{0}(X_{i})}\right]
    	+O_{P}(n^{-1/2}) \\
    	&=\mathbf{E}\left[ p_{1}(X_{i})\right] +O_{P}(n^{-1/2})=%
    	p_{1}+O_{P}(n^{-1/2}),
    \end{align*}%
    uniformly for all $p\in A$. The uniformity comes from the
    fact that 
    \begin{align*}
    	\frac{1}{n_{0,w}}\sum_{i\in S_{0,w}}\frac{p_{1}(V_{i},w)}{%
    		p_{0}(V_{i},w)}=\mathbf{E}_{0,w}\left[ \frac{p_{1}(V_{i},w)}{%
    		p_{0}(V_{i},w)}\right] +O_{P}(n^{-1/2}),
    \end{align*}%
    uniformly for $p\in A$. Also, 
    \begin{align*}
    	\sum_{w}\frac{p_{0,w}}{n_{0,w}}\sum_{i\in S_{0,w}}\tilde{1}%
    	_{n,i}\frac{\tilde{p}_{1,i}(V_{i},w)Y_{i}}{\tilde{p}_{0,i}(V_{i},w)%
    	} 
    	&=\sum_{w}p_{0,w}\mathbf{E}_{0,w}\left[ \frac{%
    		p_{1}(V_{i},w)Y_{i}}{p_{0}(V_{i},w)}\right] +O_{P}(n^{-1/2}) \\
    	&=\sum_{w}p_{0,w}\mathbf{E}_{0,w}\left[\frac{%
    		p_{1}(V_{i},w)\beta _{0}(V_{i},w)}{p_{0}(V_{i},w)}\right] +O_{P}(n^{-1/2}). \\
    	&=\mathbf{E}_{0}\left[ \frac{p_{1}(X_{i})\beta _{0}(X_{i})}{%
    		p_{0}(X_{i})}\right] p_{0}+O_{P}(n^{-1/2}),
    \end{align*}%
    uniformly for all $p\in A$. We can rewrite the leading term as%
    \begin{align*}
    	\mathbf{E}\left[\frac{p_{1}(X_{i})\beta _{0}(X_{i})(1-D_{i})}{%
    		p_{0}(X_{i})}\right] =\mathbf{E}\left[p_{1}(X_{i})\beta _{0}(X_{i})%
    	\right] =\mathbf{E}_{1}\left[\beta _{0}(X_{i})\right] p_{1}
    \end{align*}%
    Hence we can write $\bar{R}_{n}$ as (up to $o_{P}(n^{-1/2})\ $uniformly over 
    $p\in A$) 
    \begin{align*}
    	&\frac{1}{p_{1}}\left\{\sum_{w}\frac{p_{0,w}}{%
    		n_{0,w}}\sum_{i\in S_{0,w}}\tilde{1}_{n,i}\frac{\tilde{p}%
    		_{1,i}(V_{i},w)}{\tilde{p}_{0,i}(V_{i},w)}-p_{1}\right\} \mathbf{E}_{1}\left[
    	\beta _{0}(X_{i})\right]\\
    	&=\frac{1}{p_{1}}\left\{\sum_{w}\frac{p_{0,w}}{%
    		n_{0,w}}\sum_{i\in S_{0,w}}\tilde{1}_{n,i}\frac{\tilde{p}%
    		_{1,i}(V_{i},w)}{\tilde{p}_{0,i}(V_{i},w)}\mathbf{E}_{1}\left[
    	\beta _{0}(X_{i})\right]-p_{1}\mathbf{E}_{1}\left[
    	\beta _{1}(X_{i})\right]\right\}+\tau_{tet}(p). 
    \end{align*}%
    
    Plugging this result into (\ref{decomp3}) and defining 
    \begin{align*}
    	\tilde{\varepsilon}_{d,i}=Y_{di}-\mathbf{E}_{1}\left[\beta
    	_{d}(X_{i})\right] ,
    \end{align*}%
    we write $\hat{\tau}_{tet}(p)-\tau
    _{tet}(p)$ as (up to $o_{P}(n^{-1/2})\ $%
    uniformly over $p\in A$)%
    \begin{align}
    	&\frac{1}{p_{1}}\sum_{w}%
    	\frac{p_{1,w}}{n_{1,w}}\sum_{i\in S_{1,w}}\tilde{\varepsilon}_{1,i}
    	-\frac{1}{p_{1}}\sum_{w%
    	}\frac{p_{0,w}}{n_{0,w}}\sum_{i\in S_{0,w}}\frac{\tilde{1}_{n,i}
    		\tilde{p}_{1,i}(V_{i},w)\tilde{\varepsilon}_{0,i}}{\tilde{p}_{0,i}(V_{i},w)}
    	=\frac{1}{p_{1}}\left(
    	B_{n}-C_{n}-D_{n}\right) ,  \label{deve4}
    \end{align}%
    where%
    \begin{align*}
    	B_{n} &=\sum_{w}\frac{p_{1,w}}{n_{1,w}}\sum_{i\in
    		S_{1,w}}\tilde{\varepsilon}_{1,i}-\sum_{w}\frac{%
    		p_{0,w}}{n_{0,w}}\sum_{i\in S_{0,w}}\frac{p_{1}(V_{i},w)\tilde{%
    			\varepsilon}_{0,i}}{p_{0}(V_{i},w)}, \\
    	C_{n} &=\sum_{w}\frac{p_{0,w}}{n_{0,w}}\sum_{i\in
    		S_{0,w}}\tilde{\varepsilon}_{0,i}\left\{ \tilde{1}_{n,i}\frac{%
    		\tilde{p}_{1,i}(V_{i},w)}{\tilde{p}_{0,i}(V_{i},w)}-\hat{1}_{n,i}\frac{\hat{p%
    		}_{1,i}(V_{i},w)}{\hat{p}_{0,i}(V_{i},w)}\right\} ,\text{ and} \\
    	D_{n} &=\sum_{w}\frac{p_{0,w}}{n_{0,w}}\sum_{i\in
    		S_{0,w}}\tilde{\varepsilon}_{0,i}\left\{ \hat{1}_{n,i}\frac{\hat{p}%
    		_{1,i}(V_{i},w)}{\hat{p}_{0,i}(V_{i},w)}-\frac{p_{1}(V_{i},w)}{p_{0}(V_{i},w)%
    	}\right\} .
    \end{align*}%
    We consider $D_{n}$ first. By Lemma B1 and (\ref{cv4}), we write $D_{n}$ as
    (up to $o_{P}(n^{-1/2})\ $uniformly over $p\in A$)%
    \begin{align*}
    	&\sum_{w}\left\{ \frac{p_{0,w}}{n_{0,w}}\sum_{i\in
    		S_{0,w}}\tilde{\varepsilon}_{0,i}\hat{1}_{n,i}\left( \frac{\hat{p}%
    		_{1,i}(V_{i},w)p_{0}(V_{i},w)-p_{1}(V_{i},w)\hat{p}_{0,i}(V_{i},w)}{%
    		p_{0}^{2}(V_{i},w)}\right) \right\} \\
    	&=\sum_{w}\left\{ \frac{p_{0,w}}{q_{0,w}n}\sum_{i\in
    		S_{0,w}}\tilde{\varepsilon}_{0,i}\hat{1}_{n,i}\left( \frac{\hat{p}%
    		_{1,i}(V_{i},w)-p_{1}(V_{i},w)}{p_{0}(V_{i},w)}\right) \right\} \\
    	&\quad +\sum_{w}\left\{ \frac{p_{0,w}}{q_{0,w}n}\sum_{i\in
    		S_{0,w}}\tilde{\varepsilon}_{0,i}\hat{1}_{n,i}\left( \frac{%
    		p_{1}(V_{i},w)\{p_{0}(V_{i},w)-\hat{p}_{0,i}(V_{i},w)\}}{p_{0}^{2}(V_{i},w)}%
    	\right) \right\} \\
    	&=D_{1n}+D_{2n}.
    \end{align*}%
    Apply Lemma B2(i) to write $D_{1n}$ as (up to $o_{P}(n^{-1/2})$ uniformly
    for all $p\in A.$)%
    \begin{align*}
    	&\sum_{w}\left\{ \frac{1}{n}\sum_{i=1}^{n}\frac{%
    		\mathbf{E}_{Q,w,i}\left[ \tilde{\varepsilon}_{0,i}L_{0,w,i}\right] \mathcal{J%
    		}_{1,w,i}}{p_{0}(V_{i},w)\mathbf{E}_{Q,w,i}\left[ L_{w,i}\right] }\right\} -\sum_{w}\left\{ \frac{1}{n}\sum_{i=1}^{n}\frac{
    		\mathbf{E}_{Q,w,i}\left[ \tilde{\varepsilon}_{0,i}L_{0,w,i}\right]
    		p_{1}(V_{i},w)\mathcal{J}_{w,i}}{p_{0}(V_{i},w)\mathbf{E}_{Q,w,i}\left[
    		L_{w,i}\right] }\right\} .
    \end{align*}%
    Defining 
    \begin{align*}
    	\Delta _{d,w,i}= \beta _{d}(V_{i},w)-\mathbf{E}_{1}\left[
    	\beta _{d}(X_{i})\right],
    \end{align*}%
    we write the last difference as%
    \begin{align*}
    	\sum_{w}\left\{ \frac{1}{n}\sum_{i=1}^{n}\Delta
    	_{0,w,i}\mathcal{J}_{1,w,i}\right\} -\sum_{w}\left\{ \frac{1}{%
    		n}\sum_{i=1}^{n}p_{1}(V_{i},w)\Delta _{0,w,i}\mathcal{J}%
    	_{w,i}\right\} ,
    \end{align*}%
    because (using (\ref{eq1}) and (\ref{decomp1}))%
    \begin{align*}
    	\frac{\mathbf{E}_{Q,w,i}\left[ \tilde{\varepsilon}_{0,i}L_{0,w,i}\right] }{%
    		\mathbf{E}_{Q,w,i}\left[ L_{w,i}\right] }=p_{0}(V_{i},w)\left\{ \beta
    	_{0}(V_{i},w)-\mathbf{E}_{1}\left[\beta _{0}(X_{i})\right] \right\} =p_{0}(V_{i},w)\Delta
    	_{0,w,i},
    \end{align*}%
    and%
    \begin{align*}
    	&\frac{\mathbf{E}_{Q,w,i}\left[ \tilde{\varepsilon}_{0,i}L_{0,w,i}\right]
    		p_{1}(V_{i},w)}{\mathbf{E}_{Q,w,i}\left[ L_{w,i}\right] p_{0}(V_{i},w)}
    	= p_{1}(V_{i},w)\left\{ \beta _{0}(V_{i},w)-\mathbf{E}_{1}\left[
    	\beta _{0}(X_{i})\right] 
    	\right\} =p_{1}(V_{i},w)\Delta _{0,w,i}.
    \end{align*}%
    Applying Lemma B2(i), we write $D_{2n}$ as (up to $o_{P}(n^{-1/2})$
    uniformly for all $p\in A$)%
    \begin{align*}
    	&-\sum_{w}\frac{1}{n}\sum_{i=1}^{n}\frac{%
    		p_{1}(V_{i},w)\mathbf{E}_{Q,w,i}\left[ \tilde{\varepsilon}%
    		_{0,i}L_{0,w,i}\right] }{p_{0}^{2}(V_{i},w)\mathbf{E}_{Q,w,i}\left[ L_{w,i}%
    		\right] }\mathcal{J}_{0,w,i} +\sum_{w}\frac{1}{n}\sum_{i=1}^{n}\frac{%
    		p_{1}(V_{i},w)\mathbf{E}_{Q,w,i}\left[ \tilde{\varepsilon}%
    		_{0,i}L_{0,w,i}\right] }{p_{0}(V_{i},w)\mathbf{E}_{Q,w,i}\left[ L_{w,i}%
    		\right] }\mathcal{J}_{w,i} \\
    	&=-\sum_{w}\frac{1}{n}\sum_{i=1}^{n}\frac{%
    		p_{1}(V_{i},w)\Delta _{0,w,i}}{p_{0}(V_{i},w)}\mathcal{J}%
    	_{0,w,i}+\sum_{w}\frac{1}{n}%
    	\sum_{i=1}^{n}p_{1}(V_{i},w)\Delta _{0,w,i}\mathcal{J}_{w,i}.
    \end{align*}%
    Therefore, $D_{1n}+D_{2n}$ is equal to%
    \begin{align*}
    	&\sum_{w}\frac{1}{n}\sum_{i=1}^{n}\left\{\Delta
    	_{0,w,i}\mathcal{J}_{1,w,i}-p_{1}(V_{i},w)\Delta _{0,w,i}\mathcal{J%
    	}_{w,i}\right\}  -\sum_{w}\frac{1}{n}\sum_{i=1}^{n}\frac{%
    		p_{1}(V_{i},w)\Delta _{0,w,i}}{p_{0}(V_{i},w)}\mathcal{J}_{0,w,i}
    	\\
    	&\quad +\sum_{w}\frac{1}{n}\sum_{i=1}^{n}p_{1}(V_{i},w)%
    	\Delta _{0,w,i}\mathcal{J}_{w,i}+o_{P}(n^{-1/2}) \\
    	&=\sum_{w}\frac{1}{n}\sum_{i=1}^{n}\Delta _{0,w,i}%
    	\mathcal{J}_{1,w,i}-\sum_{w}\frac{1}{n}\sum_{i=1}^{n}\frac{%
    		p_{1}(V_{i},w)\Delta _{0,w,i}}{p_{0}(V_{i},w)}\mathcal{J}%
    	_{0,w,i}+o_{P}(n^{-1/2}),
    \end{align*}%
    uniformly for all $p\in A.$ As for the last difference, recall the
    definition $\mathcal{J}_{d,w,i}= L_{d,w,i}-\mathbf{E}_{Q,w,i}\left[
    L_{d,w,i}\right] $ and write it as%
    \begin{align*}
    	&\sum_{w}\frac{1}{n}\sum_{i=1}^{n}\Delta
    	_{0,w,i}L_{1,w,i} 
    	-\sum_{w}\frac{1}{n}\sum_{i=1}^{n}\frac{%
    		p_{1}(V_{i},w)\Delta _{0,w,i}}{p_{0}(V_{i},w)}L_{0,w,i} \\
    	&-\sum_{w}\frac{1}{n}\sum_{i=1}^{n}\Delta _{0,w,i}%
    	\mathbf{E}_{Q,w,i}\left[ L_{1,w,i}\right] 
    	+\sum_{w}\frac{1}{n}\sum_{i=1}^{n}\frac{%
    		p_{1}(V_{i},w)\Delta _{0,w,i}}{p_{0}(V_{i},w)}\mathbf{E}_{Q,w,i}%
    	\left[ L_{0,w,i}\right] .
    \end{align*}%
    Note that from (\ref{deve3}),%
    \begin{align}\label{qp}
    	&\mathbf{E}_{Q,w,i}\left[ L_{1,w,i}\right] -\frac{p_{1}(V_{i},w)}{%
    		p_{0}(V_{i},w)}\mathbf{E}_{Q,w,i}\left[ L_{0,w,i}\right]  =\frac{p_{1,w}}{q_{1,w}}q_{1}(V_{i},w)-\frac{p_{0,w}}{q_{0,w}}\frac{%
    		p_{1}(V_{i},w)q_{0}(V_{i},w)}{p_{0}(V_{i},w)}  \\
    	&=p_{1}(V_{i},w)\frac{f(V_{i},w)}{f_{Q}(V_{i},w)}-\frac{p_{0,w}}{q_{0,w}%
    	}\frac{q_{0}(V_{i},w)}{p_{0}(V_{i},w)}+\frac{p_{0,w}}{q_{0,w}}q_{0}(V_{i},w)
    	\notag \\
    	&=p_{1}(V_{i},w)\frac{f(V_{i},w)}{f_{Q}(V_{i},w)}-\frac{f(V_{i},w)}{%
    		f_{Q}(V_{i},w)}+p_{0}(V_{i},w)\frac{f(V_{i},w)}{f_{Q}(V_{i},w)}=0. 
    	\notag
    \end{align}%
    Therefore,%
    \begin{align*}
    	D_{n} &=D_{1n}+D_{2n} =\sum_{w}\frac{1}{n}\sum_{i=1}^{n}\Delta
    	_{0,w,i}L_{1,w,i}
    	-\sum_{w}\frac{1}{n}\sum_{i=1}^{n}\frac{%
    		p_{1}(V_{i},w)\Delta _{0,w,i}}{p_{0}(V_{i},w)}%
    	L_{0,w,i}+o_{P}(n^{-1/2}),
    \end{align*}%
    uniformly for all $p\in A$. 
    
    Now, we turn to $C_{n}$ (in (\ref{deve4})) which
    we write as%
    \begin{align*}
    	&\sum_{w}\frac{1}{n}\sum_{i=1}^{n}\tilde{%
    		\varepsilon}_{0,i}L_{0,w,i}\left\{ \tilde{1}_{n,i}\frac{\tilde{p}%
    		_{1,i}(V_{i},w)}{\tilde{p}_{0,i}(V_{i},w)}-\hat{1}_{n,i}\frac{\hat{p}%
    		_{1,i}(V_{i},w)}{\hat{p}_{0,i}(V_{i},w)}\right\} +o_{P}(n^{-1/2}) \\
    	&=\sum_{w}\frac{1}{n}\sum_{i=1}^{n}H_{i}\frac{\tilde{p}%
    		_{1,i}(V_{i},w)\{\hat{p}_{0,i}(V_{i},w)-\tilde{p}_{0,i}(V_{i},w)\}}{%
    		p_{0}^{2}(V_{i},w)} \\
    	&\quad +\sum_{w}\frac{1}{n}\sum_{i=1}^{n}H_{i}\frac{\{\tilde{p}%
    		_{1,i}(V_{i},w)-\hat{p}_{1,i}(V_{i},w)\}\tilde{p}_{0,i}(V_{i},w)}{%
    		p_{0}^{2}(V_{i},w)}+o_{P}(n^{-1/2}) \\
    	&=\sum_{w}\frac{1}{n}\sum_{i=1}^{n}H_{i}\frac{\tilde{p}%
    		_{1,i}(V_{i},w)\{\hat{p}_{0,i}(V_{i},w)-\tilde{p}_{0,i}(V_{i},w)\}}{%
    		p_{0}^{2}(V_{i},w)} \\
    	&\quad +\sum_{w}\frac{1}{n}\sum_{i=1}^{n}H_{i}\frac{\{\hat{p}%
    		_{0,i}(V_{i},w)-\tilde{p}_{0,i}(V_{i},w)\}\tilde{p}_{0,i}(V_{i},w)}{%
    		p_{0}^{2}(V_{i},w)}+o_{P}(n^{-1/2}) \\
    	&=\sum_{w}\frac{1}{n}\sum_{i=1}^{n}H_{i}\frac{\hat{p}%
    		_{0,i}(V_{i},w)-\tilde{p}_{0,i}(V_{i},w)}{p_{0}^{2}(V_{i},w)}%
    	+o_{P}(n^{-1/2}).
    \end{align*}%
    uniformly for all $p\in A$, where $H_{i}=\tilde{\varepsilon}%
    _{0,i}L_{0,w,i}$. The uniformity comes from the fact that the convergence
    rate of $\tilde{p}_{0,i}(V_{i},w)$ and $\hat{p}_{0,i}(V_{i},w)$ to $%
    p_{0}(V_{i},w)$ is uniform for $p$. The second equality follows from Lemma
    B2(ii). As for the last term, we apply Lemma B2(ii) to write it as (up to $%
    o_{P}(n^{-1/2})$, uniformly for all $p\in A$.)%
    \begin{align*}
    	&\sum_{w}\mathbf{E}_{Q}\left[ \frac{p_{1}(V_{i},w)%
    		\tilde{\varepsilon}_{0,i}L_{0,w,i}}{p_{0}(V_{i},w)}\right] \left( \frac{\hat{%
    			q}_{0,w}-q_{0,w}}{q_{0,w}}-\frac{\hat{q}_{1,w}-q_{1,w}}{q_{1,w}}\right) \\
    	&=\sum_{w}\mathbf{E}\left[p_{1}(V_{i},w)\Delta
    	_{0,w,i}\right] \left( \frac{\hat{q}_{0,w}-q_{0,w}}{q_{0,w}}-\frac{\hat{q}%
    		_{1,w}-q_{1,w}}{q_{1,w}}\right) +o_{P}(n^{-1/2}),
    \end{align*}%
    because%
    \begin{align*}
    	\mathbf{E}_{Q}\left[ \frac{p_{1}(V_{i},w)\tilde{\varepsilon}%
    		_{0,i}L_{0,w,i}}{p_{0}(V_{i},w)}\right] =\mathbf{E}\left[\frac{%
    		p_{1}(V_{i},w)}{p_{0}(V_{i},w)}\tilde{\varepsilon}_{0,i}1%
    	\{(D_{i},W_{i})=(0,w)\}\right]=\mathbf{E}\left[p_{1}(V_{i},w)\tilde{\varepsilon}_{0,i}\right]
    	=\mathbf{E}\left[p_{1}(V_{i},w)\Delta _{0,w,i}\right] .
    \end{align*}%
    Now, let us turn to $B_{n}$ (in (\ref{deve4})), which can be written as%
    \begin{align*}
    	\sum_{w}\frac{1}{n}\sum_{i=1}^{n}\tilde{\varepsilon}%
    	_{1,i}L_{1,w,i}-\sum_{w}\frac{1}{n}\sum_{i=1}^{n}\frac{%
    		p_{1}(V_{i},w)\tilde{\varepsilon}_{0,i}L_{0,w,i}}{p_{0}(V_{i},w)}%
    	+E_{n},
    \end{align*}%
    where%
    \begin{align*}
    	E_{n} =\sum_{w}\frac{1}{n}\sum_{i=1}^{n}
    	\tilde{\varepsilon}_{1,i}(\hat{L}_{1,w,i}-L_{1,w,i}) -\frac{1}{n}\sum_{i=1}^{n}\frac{p_{1}(V_{i},w)\tilde{\varepsilon}_{0,i}(\hat{L}_{0,w,i}-L_{0,w,i})}{p_{0}(V_{i},w)}.
    \end{align*}%
    Now, we focus on $E_{n}$. Observe that%
    \begin{align*}
    	&\frac{1}{n}\sum_{i=1}^{n}\tilde{\varepsilon}_{1,i}(\hat{L}%
    	_{1,w,i}-L_{1,w,i}) \\
    	&=\frac{1}{n}\sum_{i=1}^{n}\tilde{\varepsilon}_{1,i}p_{1,w}\left( 
    	\frac{q_{1,w}-\hat{q}_{1,w}}{q_{1,w}^{2}}\right)
    	1\{(D_{i},W_{i})=(1,w)\}+o_{P}(n^{-1/2}) \\
    	&=\frac{1}{n}\sum_{i=1}^{n}\tilde{\varepsilon}_{1,i}L_{1,w,i}%
    	\left( \frac{q_{1,w}-\hat{q}_{1,w}}{q_{1,w}}\right) +o_{P}(n^{-1/2})=\mathbf{E}_{Q}\left[ \tilde{\varepsilon}_{1,i}L_{1,w,i}\right]
    	\left( \frac{q_{1,w}-\hat{q}_{1,w}}{q_{1,w}}\right) +o_{P}(n^{-1/2}),
    \end{align*}%
    uniformly for all $p\in A.$ (Here the uniformity comes from that the
    convergence of $\hat{q}_{1,w}$ to $q_{1,w}$ does not depends on $p$). As for
    the last expectation,%
    \begin{align*}
    	\mathbf{E}_{Q}\left[\tilde{\varepsilon}_{1,i}L_{1,w,i}\right] 
    	&=(p_{1,w}/q_{1,w})\mathbf{E}_{Q}\left[\tilde{\varepsilon}%
    	_{1,i}1\{(D_{i},W_{i})=(1,w)\}\right] =\mathbf{E}\left[ \tilde{\varepsilon}_{1,i}1%
    	\{(D_{i},W_{i})=(1,w)\}\right] \\
    	&=\mathbf{E}\left[ p_{1}(V_{i},w)\left( \beta _{1}(V_{i},w)-%
    	\mathbf{E}_{1}\left[\beta _{0}(X_{i})\right] \right) \right] =\mathbf{E}\left[p_{1}(V_{i},w)\Delta _{1,w,i}\right] .
    \end{align*}%
    Hence%
    \begin{align*}
    	\frac{1}{n}\sum_{i=1}^{n}\tilde{\varepsilon}_{1,i}(\hat{L}%
    	_{1,w,i}-L_{1,w,i})=\mathbf{E}\left[ p_{1}(V_{i},w)\Delta _{1,w,i}%
    	\right] \frac{q_{1,w}-\hat{q}_{1,w}}{q_{1,w}}+o_{P}(n^{-1/2}),
    \end{align*}%
    uniformly for all $p\in A.$ Also,%
    \begin{align*}
    	&\frac{1}{n}\sum_{i=1}^{n}\frac{p_{1}(V_{i},w)\tilde{\varepsilon}%
    		_{0,i}(\hat{L}_{0,w,i}-L_{0,w,i})}{p_{0}(V_{i},w)} =\frac{1}{n}\sum_{i=1}^{n}\frac{p_{1}(V_{i},w)\tilde{\varepsilon}%
    		_{0,i}}{p_{0}(V_{i},w)}L_{0,w,i}\left( \frac{q_{0,w}-\hat{q}_{0,w}}{q_{0,w}}%
    	\right) L_{0,w,i} \\
    	&=\mathbf{E}\left[ p_{1}(V_{i},w)\Delta _{0,w,i}\right] \frac{%
    		q_{0,w}-\hat{q}_{0,w}}{q_{0,w}}+o_{P}(n^{-1/2}),
    \end{align*}%
    uniformly for all $p\in A.$ Therefore, we write $E_{n}$ as%
    \begin{align*}
    	\sum_{w}\mathbf{E}\left[ p_{1}(V_{i},w)\Delta
    	_{1,w,i}\right] \frac{q_{1,w}-\hat{q}_{1,w}}{q_{1,w}} -\sum_{w}\mathbf{E}\left[ p_{1}(V_{i},w)\Delta
    	_{0,w,i}\right] \frac{q_{0,w}-\hat{q}_{0,w}}{q_{0,w}}+o_{P}(n^{-1/2}),
    \end{align*}%
    uniformly for all $p\in A.$
    
    Now, we collect all the results for $B_{n},$ $C_{n},$ and $D_{n}$ and plug
    these into (\ref{deve4}) and to deduce that (up to $o_{P}(n^{-1/2})$
    uniformly for all $p\in A$)%
    \begin{align*}
    	\hat{\tau}_{tet}(p)-\tau _{tet}(p)=\frac{1}{p_{1}}\sum_{j=1}^{6}G_{jn}+o_{P}(n^{-1/2}),\ \text{uniformly over}%
    	\ p\in A,
    \end{align*}%
    where%
    \begin{align*}
    	G_{1n} &=\sum_{w}\frac{1}{n}\sum_{i=1}^{n}\left\{ 
    	\tilde{\varepsilon}_{1,i}L_{1,w,i}-\frac{p_{1}(V_{i},w)\tilde{%
    			\varepsilon}_{0,i}L_{0,w,i}}{p_{0}(V_{i},w)}\right\} , \\
    	G_{2n} &=\sum_{w}\mathbf{E}\left[ p_{1}(V_{i},w)%
    	\Delta _{1,w,i}\right] \frac{q_{1,w}-\hat{q}_{1,w}}{q_{1,w}}, \\
    	G_{3n} &=-\sum_{w}\mathbf{E}\left[ p_{1}(V_{i},w)%
    	\Delta _{0,w,i}\right] \frac{q_{0,w}-\hat{q}_{0,w}}{q_{0,w}}, \\
    	G_{4n} &=-\sum_{w}\mathbf{E}\left[ p_{1}(V_{i},w)
    	\Delta _{0,w,i}\right] \left( \frac{\hat{q}_{0,w}-q_{0,w}}{q_{0,w}}-\frac{%
    		\hat{q}_{1,w}-q_{1,w}}{q_{1,w}}\right) , \\
    	G_{5n} &=-\sum_{w}\frac{1}{n}\sum_{i=1}^{n}\Delta
    	_{0,w,i}L_{1,w,i},\text{ and} \\
    	G_{6n} &=\sum_{w}\frac{1}{n}\sum_{i=1}^{n}\frac{%
    		p_{1}(V_{i},w)\Delta _{0,w,i}}{p_{0}(V_{i},w)}L_{0,w,i}.
    \end{align*}%
    We rewrite $G_{2n}+G_{3n}+G_{4n}$ as 
    \begin{align*}
    	\sum_{w}\mathbf{E}\left[ p_{1}(V_{i},w)\left( \tau
    	(V_{i},w)-\mathbf{E}_{1}[\tau (X_{i})]\right) \right] \frac{q_{1,w}-\hat{q}_{1,w}}{q_{1,w}},
    \end{align*}%
    uniformly for all $p\in A.$
    By writing 
    \begin{align*}
    	\tilde{\varepsilon}_{d,i}=Y_{di}-\beta _{d}(V_{i},w)+\beta _{d}(V_{i},w)-\mathbf{E}_{1}\left[\beta _{d}(X_{i})\right] =\varepsilon _{d,w,i}+\Delta _{d,w,i},
    \end{align*}%
    $\ $and splitting the sums, we rewrite $\hat{\tau}_{tet}(p)-\tau
    _{tet}(p)$ as%
    \begin{align*}
    	\hat{\tau}_{tet}(p)-\tau_{tet}(p)=\frac{1}{p_{1}}\sum_{j=5}^{9}G_{jn}+o_{P}(n^{-1/2}),
    \end{align*}%
    uniformly for all $p\in A,$ where%
    \begin{align*}
    	G_{7n} &=\sum_{w}\frac{1}{n}\sum_{i=1}^{n}\left\{
    	\varepsilon _{1,w,i}L_{1,w,i}-\frac{p_{1}(V_{i},w)}{p_{0}(V_{i},w)}%
    	\varepsilon _{0,w,i}L_{0,w,i}\right\} , \\
    	G_{8n} &=\sum_{w}\frac{1}{n}\sum_{i=1}^{n}\left\{
    	\Delta _{1,w,i}L_{1,w,i}-\frac{p_{1}(V_{i},w)\Delta _{0,w,i}L_{0,w,i}}{%
    		p_{0}(V_{i},w)}\right\} , \\
    	G_{9n} &=\sum_{w}\mathbf{E}\left[ p_{1}(V_{i},w)%
    	\left( \tau (V_{i},w)-\mathbf{E}_{1}[\tau (X_{i})]\right) \right] \frac{q_{1,w}-\hat{q}_{1,w}}{q_{1,w}}.
    \end{align*}
    
    Noting that $\tau _{tet}(p)=\mathbf{E}_{1}\left[ \tau (X_{i})\right]$,
    we rewrite $G_{9n}$ as 
    \begin{align*}
    	&\sum_{w}\mathbf{E}\left[ p_{1}(V_{i},w)\left(
    	\tau (V_{i},w)-\tau _{tet}(p)\right) \right] \frac{q_{1,w}-\hat{q}_{1,w}}{%
    		q_{1,w}} \\
    	&=p_{1,w}\mathbf{E}_{1,w}\left[ \left( \tau (V_{i},w)-\mathbf{E}_{1}[\tau (X_{i})]\right) %
    	\right] \frac{q_{1,w}-\hat{q}_{1,w}}{q_{1,w}}=G_{10n}\text{, say.}
    \end{align*}%
    As for $G_{5n}+G_{6n}+G_{8n}$, we note that the part $G_{8n}$ that contains $%
    p_{1}(V_{i},w)\Delta _{0,w,i}L_{0,w,i}/p_{0}(V_{i},w)$ cancels with $G_{6n}$%
    , yielding that $G_{5n}+G_{6n}+G_{8n}$ is equal to 
    \begin{align*}
    	&\sum_{w}\frac{1}{n}\sum_{i=1}^{n}\left( \Delta
    	_{1,w,i}-\Delta _{0,w,i}\right) L_{1,w,i} =\sum_{w}\frac{1}{n}\sum_{i=1}^{n}\left( \tau
    	(V_{i},w)-\mathbf{E}_{1}[\tau (X_{i})]\right) L_{1,w,i} \\
    	&=\sum_{w}\frac{1}{n}\sum_{i=1}^{n}\left( \tau
    	(V_{i},w)-\tau _{tet}(p)\right) L_{1,w,i}=G_{11n},\text{ say,}
    \end{align*}%
    Thus, we can rewrite $\hat{\tau}_{tet}(p)-\tau_{tet}(p)$ as%
    \begin{align*}
    	\frac{1}{p_{1} }\left\{
    	G_{7n}+G_{10n}+G_{11n}\right\} .
    \end{align*}%
    However, as for $G_{10n}$, note that%
    \begin{align*}
    	G_{10n} &=\sum_{w}p_{1,w}\mathbf{E}_{1,w}\left[
    	\left( \tau (V_{i},w)-\tau _{tet}(p)\right) \right]
    	-\sum_{w}\frac{p_{1,w}\hat{q}_{1,w}}{q_{1,w}}\mathbf{E}%
    	_{1,w}\left[ \left( \tau (V_{i},w)-\tau _{tet}(p)\right) \right] =G_{12n}+G_{13n}.
    \end{align*}%
    Observe that%
    \begin{align*}
    	\frac{G_{12n}}{p_{1} } &=\frac{1}{p_{1}}\sum_{w}p_{1,w}\mathbf{E%
    	}_{1,w}\left[\left( \tau (V_{i},w)-\tau _{tet}(p)\right) \right]
    	\\
    	&=\frac{1}{p_{1}}\mathbf{E}\left[
    	\left( \tau (X_{i})-\tau _{tet}(p)\right) 1\{D_{i}=1\}\right] =\mathbf{E}_{1}[\tau (X_{i})]-\tau _{tet}(p)=0.
    \end{align*}%
    As for $G_{13n}$, note that
    \begin{align*}
    	\frac{G_{13n}}{p_{1}} =-\frac{1}{p_{1}}\sum_{w}\frac{p_{1,w}%
    		\hat{q}_{1,w}}{q_{1,w}}\mathbf{E}_{1,w}\left[\left( \tau
    	(V_{i},w)-\tau _{tet}(p)\right) \right] =-\frac{1}{p_{1}}\sum_{w\in \mathcal{%
    			W }}\frac{1}{n} \sum_{i=1}^{n}L_{1,w,i}\mathbf{E}_{1,w}\left[
    	\left( \tau (V_{i},w)-\tau _{tet}(p)\right) \right] .
    \end{align*}
    
    Therefore, we conclude that%
    \begin{align*}
    	&\frac{1}{p_{1}}\left\{
    	G_{7n}+G_{11n}+G_{13n}\right\} \\
    	&=\frac{1}{n}\sum_{i=1}^{n}\frac{1}{p_{1}}\sum_{w}\left\{ L_{1,w,i}\varepsilon
    	_{1,w,i}-\frac{L_{0,w,i}p_{1}(V_{i},w)\varepsilon _{0,w,i}}{%
    		p_{0}(V_{i},w)}\right\} +\frac{1}{n}\sum_{i=1}^{n}\frac{1}{p_{1}}\sum_{w}\left( \tau (V_{i},w)-\tau
    	_{tet}(p)\right) L_{1,w,i} \\
    	&\quad -\frac{1}{n}\sum_{i=1}^{n}\frac{1}{p_{1}}\sum_{w}\mathbf{E}_{1,w}\left[ \left( \tau
    	(V_{i},w)-\tau _{tet}(p)\right) \right] L_{1,w,i}+o_{P}(n^{-1/2}),
    \end{align*}%
    uniformly for all $p\in A.$ The wanted result follows immediately. $\mathbf{%
    	\blacksquare }$\medskip\

    The following lemma is used to prove Lemma B2(i) and may be useful for other
    purposes. Hence we make the notations and assumptions self-contained here.
    Let $(Z_{i},H_{i},X_{i})_{i=1}^{n}$ be an i.i.d. sample from $P,$ where $%
    Z_{i}$ and $H_{i}$ are random variables. Let $X_{i}=(X_{1i},X_{2i})\in 
    \mathbf{R}^{d_{1}+d_{2}}$ where $X_{1i}$ is continuous and $X_{2i}$ is
    discrete, and let $K_{ji}=K_{h}\left( X_{1j}-X_{1i}\right)
    1\{X_{2j}=X_{2i}\},$ $K_{h}(\cdot )=K(\cdot /h)/h^{d_{1}}.$ Let $\mathcal{X}$
    be the support of $X_{i}$ and $f(\cdot )$ be its density with respect to a $%
    \sigma $-finite measure.\medskip
    
    \noindent \textsc{Assumption D1 :} (i) For some $\sigma \geq 4,$ sup$_{x\in 
    	\mathcal{X}}||x_{1}||^{d_{1}}\mathbf{E}[|Z_{i}|^{\sigma
    }|X_{i}=(x_{1},x_{2})]<\infty ,\ \mathbf{E}[|H_{i}|^{\sigma }]<\infty ,$ and 
    $\mathbf{E}||X_{i}||^{\sigma }<\infty .$
    
    \noindent (ii) $f(\cdot ,x_{2}),$ $\mathbf{E}[Z_{i}|X_{1i}=\cdot
    ,X_{2i}=x_{2}]f(\cdot ,x_{2})$ and $\mathbf{E}[H_{i}|X_{1i}=\cdot
    ,X_{2i}=x_{2}]f(\cdot ,x_{2})$ are $L_{1}+1$ times continuously
    differentiable with bounded derivatives on $\mathbf{R}^{L_{1}}$ and their $%
    (L_{1}+1)$-th derivatives are uniformly continuous.
    
    \noindent (iii) $f$ is bounded and bounded away from zero on $\mathcal{X}$%
    .\medskip
    
    \noindent \textsc{Assumption D2 :}\textbf{\ }For the kernel $K\ $and the
    bandwidth $h$, Assumption \ref{ass:kernel} holds.\medskip
    
    \noindent \textsc{Lemma D1 :} \textit{Suppose that Assumptions D1-D2 hold}. 
    \textit{Let} $1_{n,i}=1\{||X_{i}||\geq \delta _{n}\}$. \textit{Then}%
    \begin{align*}
    	&\frac{1}{\sqrt{n}}\sum_{i=1}^{n}H_{i}\left\{ \mathbf{E}[Z_{i}|X_{i}]-\frac{%
    		1_{n,i}\frac{1}{n-1}\sum_{j=1,j\neq i}^{n}Z_{j}K_{ji}}{f(X_{i})}\right\} \\
    	&=\frac{1}{\sqrt{n}}\sum_{i=1}^{n}\left\{ \mathbf{E}\left[ \mathbf{E}\left[
    	H_{i}|X_{i}\right] Z_{i}\right] -\mathbf{E}\left[ H_{i}|X_{i}\right]
    	Z_{i}\right\} +o_{P}(1).
    \end{align*}
    
    \noindent \textsc{Proof :} For simplicity, we only prove the result for the
    case where $X_{i}=X_{1,i}$ so that $X_{i}$ is continuous. Write%
    \begin{align*}
    	&\frac{1}{\sqrt{n}}\sum_{i=1}^{n}H_{i}\left\{ \mathbf{E}[Z_{i}|X_{i}]-\frac{%
    		1_{n,i}}{(n-1)f(X_{i})}\sum_{j=1,j\neq i}^{n}Z_{j}K_{ji}\right\} \\
    	&=\frac{1}{\sqrt{n}}\sum_{i=1}^{n}H_{i}\left\{ \frac{\mathbf{E}[Z_{i}|X_{i}]%
    		\hat{f}(X_{i})}{f(X_{i})}-\frac{1_{n,i}}{(n-1)f(X_{i})}\sum_{j=1,j\neq
    		i}^{n}Z_{j}K_{ji}\right\} +\frac{1}{\sqrt{n}}\sum_{i=1}^{n}H_{i}\left\{ \frac{\mathbf{E}%
    		[Z_{i}|X_{i}]\{f(X_{i})-\hat{f}(X_{i})\}}{f(X_{i})}\right\} . \\
    	&=A_{1n}+A_{2n}.
    \end{align*}%
    It suffices to show that%
    \begin{align*}
    	A_{1n} &=\frac{1}{\sqrt{n}}\sum_{i=1}^{n}\mathbf{E}\left[ H_{i}|X_{i}\right]
    	\left\{ \mathbf{E}[Z_{i}|X_{i}]-Z_{i}\right\} +o_{P}(1),\text{ and} \\
    	A_{2n} &=\frac{1}{\sqrt{n}}\sum_{i=1}^{n}\left\{ \mathbf{E}\left[ \mathbf{E}%
    	\left[ H_{i}|X_{i}\right] \mathbf{E}[Z_{i}|X_{i}]\right] -\mathbf{E}\left[
    	H_{i}|X_{i}\right] \mathbf{E}[Z_{i}|X_{i}]\right\} +o_{P}(1).
    \end{align*}%
    Note that $\Pr \{||X_{i}||<\delta _{n}\}=\int_{||x||<\delta
    	_{n}}f_{X}(x)dx\leq C_{d_{1}}\delta _{n}^{d_{1}}\rightarrow 0,$ where $%
    C_{d_{1}}$ is a constant depending on $d_{1}$. With probability approaching
    one, we can write%
    \begin{align*}
    	A_{1n}=\frac{1}{(n-1)\sqrt{n}}\sum_{i=1}^{n}\sum_{j=1,j\neq
    		i}^{n}q_{h}(S_{i},S_{j})=\frac{1}{\sqrt{n}}\sum_{j=1}^{n}\mathbf{E}\left[
    	q_{h}(S_{i},S_{j})|S_{j}\right] +r_{1,n},
    \end{align*}%
    where $q_{h}(S_{i},S_{j})=H_{i}\left\{ \mathbf{E}[Z_{i}|X_{i}]-Z_{j}\right\}
    K_{ji}/f(X_{i})$ and $S_{i}=(X_{i},Z_{i},H_{i}),\ $and 
    \begin{align*}
    	r_{1,n}=\frac{1}{(n-1)\sqrt{n}}\sum_{i=1}^{n}\sum_{j=1,j\neq
    		i}^{n}\{q_{h}(S_{i},S_{j})-\mathbf{E}\left[ q_{h}(S_{i},S_{j})|S_{j}\right]
    	\}.
    \end{align*}%
    Observe that%
    \begin{align*}
    	n^{-1}\mathbf{E}\left( q_{h}(S_{i},S_{j})^{2}\right) =n^{-1}\mathbf{E}%
    	\left[ H_{i}^{2}\left\{ \mathbf{E}[Z_{i}|X_{i}]-Z_{j}\right\}
    	^{2}K_{ji}^{2}/f^{2}(X_{i})\right] 
    	\leq n^{-1}C\sqrt{\mathbf{E}\left[ K_{ji}^{4}\right] }%
    	=O(n^{-1}h^{-2d_{1}})=o(1)
    \end{align*}%
    by change of variables and by Assumptions D1(iii) and D2. Therefore, by
    Lemma 3.1 of \cite{Powell/Stock/Stoker:89:Eca}, $r_{1,n}=o_{P}(1).$ As for $%
    \mathbf{E}\left[ q_{h}(S_{i},S_{j})|S_{j}\right] $, we use change of
    variables, Taylor expansion, and deduce that%
    \begin{align*}
    	\mathbf{E}\left[ \left\vert \mathbf{E}\left[ q_{h}(S_{i},S_{j})|S_{j}\right]
    	-\mathbf{E}[H_{j}|X_{j}]\left\{ \mathbf{E}[Z_{j}|X_{j}]-Z_{j}\right\}
    	\right\vert \right] =o(n^{-1/2}).
    \end{align*}%
    The wanted representation follows from this.
    
    As for $A_{2n}$,%
    \begin{align*}
    	\frac{1}{\sqrt{n}}\sum_{i=1}^{n}\frac{H_{i}\mathbf{E}[Z_{i}|X_{i}]}{f(X_{i})}%
    	\left\{ f(X_{i})-\hat{f}(X_{i})\right\} =\frac{1}{(n-1)\sqrt{n}}%
    	\sum_{i=1}^{n}\sum_{j=1,j\neq i}^{n}s_{h}(S_{i},S_{j}),
    \end{align*}%
    where%
    \begin{align*}
    	s_{h}(S_{i},S_{j})=\frac{H_{i}\mathbf{E}[Z_{i}|X_{i}]}{f(X_{i})}\left\{
    	f(X_{i})-K_{ji}\right\} .
    \end{align*}%
    Since we can write $\mathbf{E}\left[ K_{ji}|X_{i}\right]
    =f(X_{i})+O_{P}(h^{L_{1}+1})$ uniformly over $1\leq i\leq n$, we find that%
    \begin{align*}
    	\mathbf{E}\left[ s_{h}(S_{i},S_{j})|S_{i}\right] =\frac{H_{i}\mathbf{E}%
    		[Z_{i}|X_{i}]}{f(X_{i})}\left\{ f(X_{i})-\mathbf{E}\left[ K_{ji}|X_{i}\right]
    	\right\} =o_{P}(n^{-1/2}),
    \end{align*}%
    uniformly over $1\leq i\leq n.$ Hence we can write%
    \begin{align*}
    	A_{2n}=\frac{1}{\sqrt{n}}\sum_{j=1}^{n}\mathbf{E}\left[
    	s_{h}(S_{i},S_{j})|S_{j}\right] +r_{2,n}+o_{P}(1),
    \end{align*}%
    where%
    \begin{align*}
    	r_{2,n}=\frac{1}{(n-1)\sqrt{n}}\sum_{i=1}^{n}\sum_{j=1,j\neq
    		i}^{n}\{s_{h}(S_{i},S_{j})-\mathbf{E}\left[ s_{h}(S_{i},S_{j})|S_{j}\right]
    	\}.
    \end{align*}%
    Note that\ $n^{-1}\mathbf{E}\left[ s_{h}(S_{i},S_{j})^{2}\right] =o(1)$ and
    that%
    \begin{align*}
    	&\mathbf{E}\left[ s_{h}(S_{i},S_{j})|S_{j}\right] =\mathbf{E}\left[ H_{i}%
    	\mathbf{E}[Z_{i}|X_{i}]-\frac{\mathbf{E}\left[ H_{i}|X_{i}\right] \mathbf{E}%
    		[Z_{i}|X_{i}]}{f(X_{i})}K_{ji}|S_{j}\right] \\
    	&=\mathbf{E}\left[ H_{j}\mathbf{E}[Z_{j}|X_{j}]\right] -\mathbf{E}\left[
    	H_{j}|X_{j}\right] \mathbf{E}[Z_{j}|X_{j}]+o_{P}(n^{-1/2}),
    \end{align*}%
    uniformly over $1\leq j\leq n$, yielding the desired representation for $%
    A_{2n}$. $\mathbf{\blacksquare }$
    
    \bibliographystyle{chicago}
    \bibliography{treatment_based_sampling}

\end{document}